\documentclass[10pt, aps,twocolumn,superscriptaddress]{revtex4-2}
\usepackage{pifont}
\usepackage{graphicx}
\usepackage{amsmath}
\usepackage{bm}
\usepackage{color}
\usepackage{array} 
\usepackage[percent]{overpic}
\usepackage[bookmarks=true,colorlinks=true,urlcolor=blue,linkcolor=blue,citecolor=blue,breaklinks]{hyperref}
\usepackage{amssymb}
\usepackage{booktabs}
\usepackage{physics}
\usepackage{comment}
\usepackage[toc]{appendix}
\usepackage[normalem]{ulem}

\begin{document}

\sloppy

\renewcommand{\arraystretch}{1.4}

\newcommand{\mycheckmark}{\ding{51}}
\newcommand{\xmark}{\ding{55}}

\definecolor{oiBlue}{RGB}{0,114,178}
\definecolor{oiRed}{RGB}{213,94,0}
\definecolor{oiGreen}{RGB}{0,158,115}
\definecolor{oiCyan}{RGB}{86,180,233}

\newcommand{\change}[1]{\textcolor{oiBlue}{#1}}
\newcommand{\hr}[1]{\textcolor{oiRed}{#1}}
\newcommand{\jlp}[1]{\textcolor{oiGreen}{#1}}

\newcommand{\rsout}[1]{\textcolor{oiRed}{\sout{#1}}}
\newcommand{\gsout}[1]{\textcolor{oiGreen}{\sout{#1}}}

\newcommand{\rev}[1]{\textcolor{oiBlue}{#1}}

\newcommand{\naturesection}[1]{%
  \par
  \vspace{1.0\baselineskip}%
  \noindent{\large\bfseries #1}\par
  \nopagebreak[4]%
  \vspace{0.35\baselineskip}%
}

\newcommand{\naturesubsection}[1]{%
  \par
  \vspace{0.65\baselineskip}%
  \noindent{\bfseries #1}\par
  \nopagebreak[4]%
  \vspace{0.20\baselineskip}%
}

\title{Quantum Geometric Injection and Shift Optical Forces Drive Coherent Phonons}

\author{J. Luke Pimlott}
\email{jlp62@bath.ac.uk}
\affiliation{Department of Physics, University of Bath, Claverton Down, Bath BA2 7AY, United Kingdom}

\author{Habib Rostami}
\email{hr745@bath.ac.uk}
\affiliation{Department of Physics, University of Bath, Claverton Down, Bath BA2 7AY, United Kingdom}

\begin{abstract}
We develop a gauge-invariant quantum-geometric theory of light-induced Raman forces driving coherent lattice dynamics in solids. We show that the resonant interband response separates into injection and shift forces, providing the lattice analogue of injection and shift photocurrents. Crucially, the electronic quantum metric and Berry curvature directly govern these forces, weighted by distinct electron--phonon quantities. The injection force is controlled by the quantum geometric tensor and differences between band-diagonal electron--phonon couplings, producing the resonant displacive response. The shift force is a resonant impulsive response governed by quantum geometry and a phononic shift vector arising from interband electron--phonon coherence. We derive polarization selection rules, symmetry constraints, and temporal signatures. Applied to interlayer shear phonons in a bilayer Haldane model, the theory predicts tunability with optical frequency, topology, electronic temperature, and current-induced symmetry breaking. These results establish a quantum-geometric framework for coherent lattice dynamics and identify coherent phonons as experimentally accessible signatures of electronic quantum geometry in quantum materials.
\end{abstract}
\maketitle

\section*{Introduction}Coherent optical manipulation of quantum materials involves light-induced forces that act by exciting lattice vibration modes. These drive ultrafast phase transitions to transient phases, such as from a normal metal to a topological insulator, or to superconducting, ferroelectric, or ferromagnetic phases \cite{Disa2021,Torre_rmp_2021,Sie2019,Soranzio_prr_2019,Basini2024DynamicalMultiferro}, and impact nonlinear photocurrent generation \cite{Okamura2022SoftPhononShiftCur,Huang2025PhononMedShiftCur}.  
Coherent excitation of Raman-active phonons is driven by light-induced
\mbox{\em Raman forces} through {\em impulsive} or {\em displacive} mechanisms  \cite{Zeiger1992,Dhar1994,Garrett1996,Merlin1997,Stevens2002,Udina2019CoherentOscillTheory,rostami2022,rostami2023}, enabling ultrafast non-thermal control of quantum materials. In the impulsive mechanism, ions vibrate coherently around their equilibrium positions, typical in the transparent regime. In contrast, the displacive mechanism, prevalent in the absorbing regime, involves ion displacement to new equilibrium positions. 

In this work, we develop a general theory of the rectified Raman force and show that its distinct components are governed by the quantum geometry of the electronic states, encoded in the \emph{quantum geometric tensor} (QGT) \cite{ProvostValleeQMetric,Berryorig,Paivi2022}. The band-resolved QGT is defined as
\begin{equation}
Q^{mn}_{ab}
=
g^{mn}_{ab}
-\frac{i}{2}\Omega^{mn}_{ab},
\label{eq:QGT}
\end{equation}
where $m$ and $n$ denote electronic bands, and $a$ and $b$ label Cartesian directions. The real part, $g^{mn}_{ab}$, is the band-resolved \emph{quantum metric}, while $\Omega^{mn}_{ab}=-2{\rm Im}[Q^{mn}_{ab}]$ corresponds to the band-resolved \emph{Berry curvature}.

Quantum-geometric effects have been extensively studied in contexts such as the anomalous Hall effect \cite{Nagaosa2010AHE}, flat-band superfluidity \cite{Paivi2022,Peotta2015}, nonlinear photocurrents \cite{Holder_TRS&QG,Ahn_prx_2020,Watanabe_prx_2021,Slager2025}, nonlinear Hall effects in ${\cal PT}$-symmetric antiferromagnets \cite{Wang2021NLHEQuantumMetric,Naizhou2023QMNonlinearTransport,Anyuan2023}, and electron--phonon coupling \cite{Yu2024QGandEPC,Hu_prl_2025,Hu_prl_256404_2025,Pellitteri2025}. Despite this progress, it remains unresolved how electronic quantum geometry governs the light-induced Raman forces that generate coherent phonons capable of driving nonequilibrium phase transitions, and whether these forces decompose into distinct microscopic and temporal mechanisms. Conventional theories of coherent-phonon generation \cite{Stevens2002} are formulated primarily in terms of dielectric response and electron--phonon coupling (EPC), and do not expose the underlying quantum-geometric structure of the Raman-force response.

Here, we present a quantum-geometric framework for unconventional light-induced atomic forces, introducing the terms {\em injection force} and {\em shift force} by analogy with established second-order photocurrent processes~\cite{SipeSOOR}.
\begin{figure}[t]
    \centering
\includegraphics[width=0.99\linewidth]{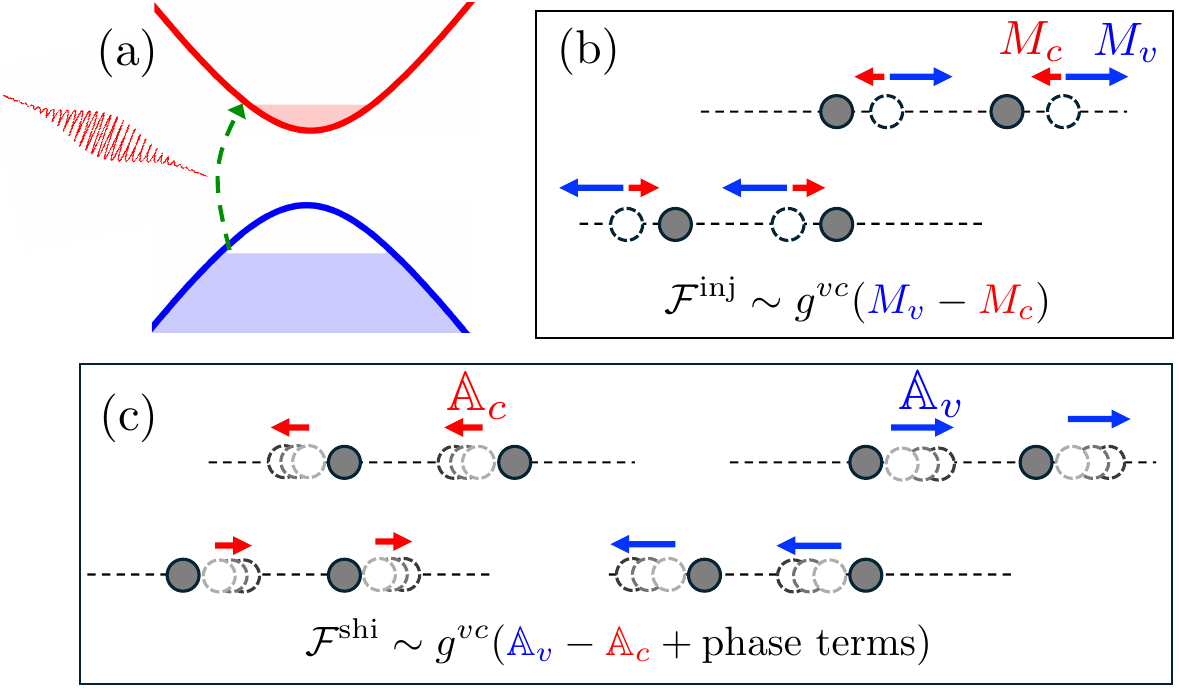}
\caption{
(a) Light-driven interband transitions generate a 
Raman force (RF) in layered materials via electron–phonon coupling (EPC).
(b) Injection (displacive) force arises from the difference in EPC-induced atomic forces between two bands.
(c) Shift (impulsive) force arises from a gauge-invariant geometric shift in phonon conjugate momentum generated by interband coherence.
\label{fig:1}
}
\end{figure}
We show that, within the clean, noninteracting limit of the electron-mediated Raman response considered here, the injection force is the sole displacive component and arises only in the absorbing (resonant) regime, where it is governed by the QGT and weighted by the mismatch in EPC strength between optically-connected bands (Fig.~\ref{fig:1}a). It therefore originates from the distinct atomic forces exerted on ions by photoexcited electrons in each band (Fig.~\ref{fig:1}b), analogous to how the injection current depends on the difference in electron group velocities between bands~\cite{SipeSOOR}.

The shift force, meanwhile, acts as an unconventional resonant impulsive force in the absorbing regime, unlike the conventional impulsive force which typically arises in the transparent (off-resonant) regime. We formulate the shift force in terms of the QGT and a {\em phononic shift vector}, which quantifies the gauge-invariant geometric shift in the momentum conjugate to the phonon coordinate during an interband optical transition (Fig.~\ref{fig:1}c). This is the phononic analogue of the electronic shift vector underlying the shift current, which describes the non-diffusive real-space displacement of an electron wavepacket during an interband transition~\cite{SipeSOOR,YoungRappeShiftCurr}. 
This parameter is defined via the {\em Mead–Truhlar 
connection}~\cite{Mead_Truhlar_1979,Resta2000}: an emergent gauge field over the atomic displacement manifold, generated by the adiabatic evolution of the electronic states. It acts as an effective vector potential on phonons (shifting canonical phonon momentum $\mathbf{P}\to\mathbf{P}-\hbar\mathbb{A}$) due to electron-phonon interactions.

We apply the general quantum-geometric theory of Raman forces to quantify the forces associated with the coherent excitation of interlayer shear phonons \cite{Tan_nm_2012,Boschetto_nl_2013,Zhao_nl_2013,Bartram2022TMD_IL-Phonon}, which correspond to the out-of-phase sliding of atomic layers in van der Waals materials. Such coherent phonon excitation is central to optically controlling stacking order in layered materials like bilayer and trilayer graphene \cite{Zhang_lsa_2020} and transition metal dichalcogenides (TMDs) such as MoS$_2$, WSe$_2$, MoTe$_2$, and WTe$_2$ \cite{Sie2019,Zhang_prx_2019,Fukuda_apl_2020}, driven by ultrashort, intense laser pulses.
The coherent dynamics of shear phonons play a key role in light-induced switching of electronic topology \cite{Sie2019,Rodriguez-Vega2022MagTopPhaseTran} and magnetism \cite{Rodriguez-Vega2022MagTopPhaseTran}, modulation of Kondo screening in heterobilayers \cite{Zhong2026}, and control of interlayer ferroelectric polarization in h-BN \cite{Yang_prl_2024,Yasuda2024SlidingFerrohBN}, and TMDs \cite{Wang2022TMDFerroelec,Bian2024ResistantSlidingFerroMoS2}. 
These effects have applications in field-effect transistors and ferroelectric memory devices \cite{Sun2025SlidingFerroelecReview} and offer a promising route for controlling superconducting transitions in twisted moiré graphene systems \cite{qin_prl_2021}.

\section*{Results and discussion} \subsection*{Raman-force rectification: photogalvanic analogy}
Light-induced Raman force is 
a second-order response to the incident light field $\mathbf{E}(t)$: 
\begin{equation}\label{eq:1}
{\cal F}_a(\omega_\Sigma)=    \int \frac{d\omega}{2\pi}
 \chi_{abc}(\omega,\omega_\Sigma-\omega) E^b(\omega) E^c(\omega_\Sigma-\omega),
 \vspace{-0.1cm}
\end{equation}
assuming the Einstein convention for summation on the repeated Cartesian indices, and with $\chi_{abc}$ the Raman-force  response function, which is the correlation of EPC with light-matter couplings. 

The rectified Raman-force response \(\chi_{abc}(\omega,-\omega)\) was previously shown to describe rigid atomic displacement and phonon splitting and softening under monochromatic driving \cite{rostami2022,rostami2023}. Those works, however, did not identify the resonant response in terms of distinct injection and shift mechanisms, nor formulate their quantum-geometric structure or contrasting temporal dynamics under pulsed excitation. 

Using the Kadanoff--Baym formalism, we derive the complete Raman-force
susceptibility in Supplementary Note~1. The resulting diagrams are
evaluated explicitly in the band basis in Supplementary Note~2
(see also Methods). We show that
the Raman-force response separates into resonant injection and shift
channels, together with an off-resonant contribution. Under pulsed excitation, the injection channel contains both a singular term leading to a displacive response and a regular term with impulsive character, whereas the shift force
is intrinsically regular and impulsive.
In the nearly-rectified regime
($\hbar|\omega_\Sigma|\ll\varepsilon_g$, with $\varepsilon_g$ the band gap),
the susceptibility separates into a pole and a regular part:
\begin{align}
\chi_{abc}(\omega,\omega_\Sigma-\omega)
=
\frac{\alpha_{abc}(\omega)}{i\omega_\Sigma}
+\beta_{abc}(\omega)
+\mathcal{O}(\omega_\Sigma).
\label{eq:2}
\end{align}
The first term has a pole at $\omega_\Sigma=0$, describing a force
that accumulates under illumination in the absence of relaxation.
The force-generation rate $d\mathcal{F}/dt$ nevertheless remains
finite, analogous to the {\em injection photocurrent}
\cite{SipeSOOR,deJuan2017}. We therefore call this contribution
the {\em injection force}, denote its coefficient by
$\alpha^{\text{inj}}_{abc}(\omega)\equiv\alpha_{abc}(\omega)$,
and show below that it produces a displacive lattice response.
The regular part is analytic in $\omega_\Sigma$ and approaches
$\beta_{abc}(\omega)$ as $\omega_\Sigma\to0$. It comprises
three contributions:
\begin{align}
\beta_{abc}(\omega)
&=
\beta^{\text{reg-inj}}_{abc}(\omega)
+\beta^{\text{shi}}_{abc}(\omega)
+\beta^{\text{off}}_{abc}(\omega).
\end{align}
The first is a resonant term,
$\beta^{\text{reg-inj}}_{abc}
=(i/2)\partial_\omega\alpha^{\text{inj}}_{abc}(\omega)$,
which we call the {\em regular injection} response.
The second, also resonant, is governed by a phononic shift vector,
as shown below; we call it the {\em shift force} by analogy with
the shift photocurrent. The third collects all off-resonant
contributions. 

\begin{figure}[t]
    \centering
    \includegraphics[width=0.9\linewidth]{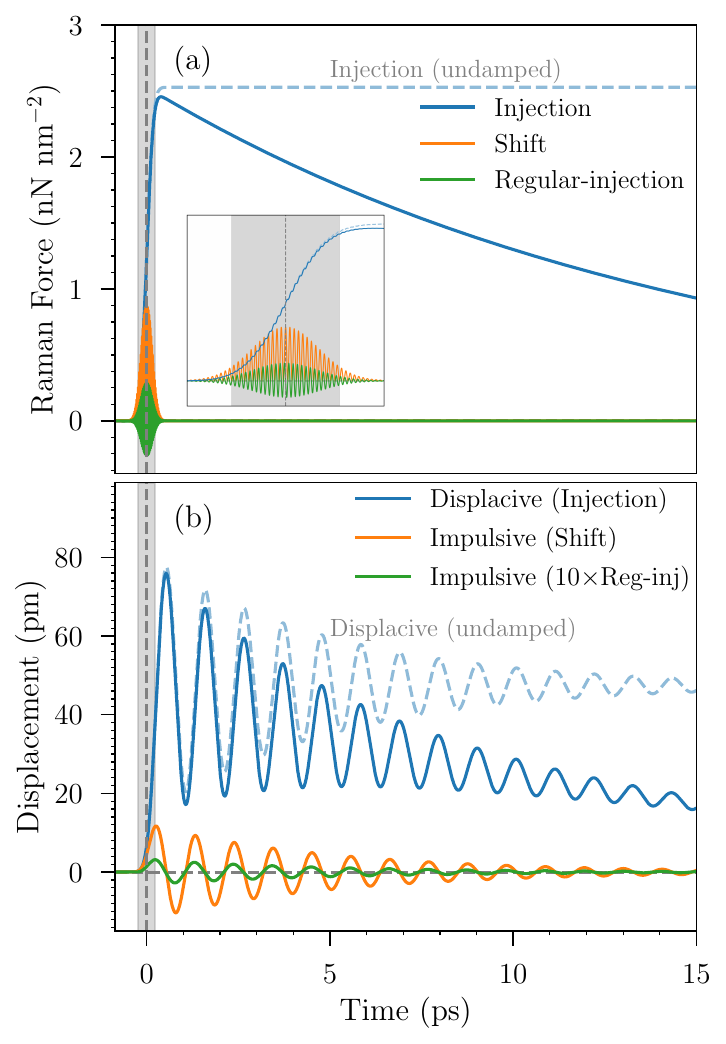}
    \caption{Resonant Raman forces and phonon dynamics.
(a) Injection, shift, and regular injection forces calculated from
Eq.~\eqref{eq:Ft} and Eqs.~\eqref{eq:alpha_analytic}–\eqref{eq:beta_analytic}
for a Gaussian pulse
$E(t)=E_0\cos(\omega_0t)e^{-t^2/(2\sigma^2)}$,
with \mbox{$E_0 \approx 100~\mathrm{MV/m}$}, \mbox{$\sigma=200~\mathrm{fs}$},
\mbox{$\hbar\omega_0=2.2\delta_{\rm H}$}, Haldane gap, \mbox{$\delta_{\rm H}=50~\mathrm{meV}$},
and \mbox{$\tau=\Gamma^{-1}=15~\mathrm{ps}$}.
Gray shading indicates the pulse width; the inset resolves the forces
during the pulse.
(b) Shear-phonon displacements driven separately by the injection
(displacive), shift (impulsive), and regular injection (impulsive)
forces, with phonon energy $\hbar\omega_p=3.9~\mathrm{meV}$,
lifetime $\tau_p=\gamma_p^{-1}=5~\mathrm{ps}$, and
areal mass density $\rho\approx1.5~\mathrm{mg/m^2}$.
The regular injection displacement is magnified by a factor of ten.
Dashed curves show the injection response in the clean limit
($\Gamma\to0$); the shift and regular injection responses are
independent of $\Gamma$ at this order.}
    \label{fig:2}
\end{figure}

We focus here on the resonant forces, deferring the off-resonant response to the Supplementary Material. In the time domain, the injection, regular injection, and shift forces take the form   
\begin{align}
    &{\cal F}^{\mathrm{inj}}_a(t) \!=-\!\!\int_{\!-\infty}^t\! dt'\!\int\! dt''\, \widetilde{\alpha}^{\text{inj}}_{abc}(t'') E^b(t'-t'')E^c(t') e^{\Gamma(t'-t)}, \nonumber\\
    &{\cal F}^{\text{shi+reg}}_a(t) \!=\!\int\! dt' \Big[\widetilde{\beta}^{\mathrm{shi}}_{abc}(t')-\frac{t'}{2}\widetilde{\alpha}^{\mathrm{inj}}_{abc}(t')\Big] E^b(t-t')E^c(t),\label{eq:Ft}
\end{align}
where $\Gamma$ is the electronic relaxation rate.
All integrals extend over the real axis, and tildes denote
inverse Fourier transforms with respect to the optical
frequency. For a Gaussian pulse,
\mbox{$\mathbf E(t)=\mathbf E_0\cos(\omega_0t)e^{-t^2/(2\sigma^2)}$},
the singular injection contribution accumulates during
excitation and relaxes on the timescale $\Gamma^{-1}$,
producing a long-lived displacive response. 
The regular injection term provides an impulsive correction during the pulse, governed by the spectral dispersion of
$\alpha^{\mathrm{inj}}$. The shift force is also impulsive but arises from a distinct mechanism involving interband electronic coherence.
For frequency-independent $\alpha^{\mathrm{inj}}_{abc}$,
the regular injection correction vanishes, and
${\cal F}^{\mathrm{inj}}_a(t)
=-\alpha^{\mathrm{inj}}_{abc}
\int_{-\infty}^{t}dt'\,e^{-\Gamma(t-t')}E^b(t')E^c(t')$. Likewise, frequency-independent $\beta^{\mathrm{shi}}_{abc}$
gives ${\cal F}^{\mathrm{shi}}_a(t)
= \beta^{\mathrm{shi}}_{abc}E^b(t)E^c(t)$.
For fixed polarization, these forces follow the accumulated
and instantaneous pulse intensity, respectively, reproducing
the conventional displacive and impulsive temporal profiles
\cite{Stevens2002}.
Using the frequency-dependent Raman response tensors of Eqs.~\eqref{eq:alpha_analytic} and \eqref{eq:beta_analytic}, we compute the real-time resonant Raman driving force via Eq.~\eqref{eq:Ft} and obtain the lattice displacement, $u(t)$ shown in Fig.~\ref{fig:2} by solving
\begin{equation}
    \ddot u(t) + 2\gamma_p \dot u(t) + \omega_p^2 u(t) = {\cal F}(t)/\rho,
\end{equation}
where $\rho$ is the mass density, $\omega_p$ the phonon frequency, and $\gamma_p$ the phonon decay rate.
The injection term $\alpha^{\text{inj}}_{abc}$ produces a \emph{displacive} force that, in the clean limit ($\Gamma\to0$), approaches a finite constant after the pulse, whereas for finite electronic relaxation it decays on the timescale $\Gamma^{-1}$. In contrast, the shift term $\beta^{\text{shi}}_{abc}$ and the regular injection term $\beta^{\text{reg-inj}}_{abc}$ generate \emph{impulsive} forces confined to the pulse duration and independent of $\Gamma$ at this order. This distinction in the force profiles translates directly into the coherent phonon dynamics: the numerical results reveal an almost $\pi/2$ phase difference between the oscillations generated by the injection and shift forces, consistent with the ideal cosine- and sine-like responses expected for displacive and impulsive excitation, respectively~\cite{Merlin1997,Stevens2002}.

This temporal structure closely parallels that of injection and shift photocurrents driven by an ultrashort pulse~\cite{Laman_jap_2005}. In particular, disorder scattering at rate $\Gamma$ causes the injection contribution to relax, while the shift contribution remains localized to the optical excitation and does not develop an injection-like relaxation tail, reflecting the same non-diffusive character as the \emph{shift photocurrent}~\cite{SipeSOOR}. Table~\ref{tab:analogy} summarizes this correspondence between injection and shift mechanisms in photocurrents and Raman forces. This analogy, in turn, motivates extending the quantum-geometric description developed for photocurrents~\cite{Watanabe_prx_2021,Ahn_prx_2020} to Raman forces.

{\setlength{\tabcolsep}{3pt}%
\renewcommand{\arraystretch}{1}%
\begin{table*}[t]
\caption{Analogy between coherent photocurrent and coherent Raman force
in the rectified ($\omega_\Sigma \approx 0$) sector of the second-order response.
Here $I(t)\propto |{\bf E}(t)|^2$ is the pulse intensity envelope, and $\tau=1/\Gamma$ is the
electronic relaxation time. Here we compare the conventional singular injection response; the regular injection term arising from the finite-bandwidth correction is not included in the photocurrent comparison.}\label{tab:analogy}
\begin{tabular}{|c|c|c|c|}
\hline\hline
Type & Regime & Photocurrent & Raman force (this work) \\
\hline\hline
 & clean limit
   & $J(\omega_\Sigma\!\to\!0) \sim 1/(i\omega_\Sigma)$ (singular) \cite{SipeSOOR}
   & $\mathcal{F}(\omega_\Sigma\!\to\!0) \sim 1/(i\omega_\Sigma)$ (singular) \\
Injection
 & monochromatic
   & $J \propto \tau \times$ interband velocity asymm. \cite{deJuan2017,Ahn_prx_2020}
   & $\mathcal{F} \propto \tau \times$ interband EPC asymm. \\
 & pulsed
   & long-lived (memoryful), relaxation-limited \cite{Laman_jap_2005}
   & displacive:
     $\displaystyle \mathcal{F}(t)\propto
       \!\int_{-\infty}^{t}\!dt'\,I(t')\,e^{(t'-t)/\tau}$ \\
\hline
 & clean limit
   & $J(\omega_\Sigma\!\to\!0) = \text{const.}$ \cite{vonBaltzKrautPhotogalvOrig,SipeSOOR}
   & $\mathcal{F}(\omega_\Sigma\!\to\!0) = \text{const.}$ \\
Shift
 & monochromatic
   & $J \propto$ electronic shift vector, $\tau$-indep. \cite{vonBaltzKrautPhotogalvOrig,SipeSOOR}
   & $\mathcal{F} \propto$ phononic shift vector, $\tau$-indep. \\
 & pulsed
   & instantaneous (memoryless), $\tau$-indep. \cite{Laman_jap_2005,NastosSipe2006}
   & impulsive:
     $\mathcal{F}(t)\propto I(t)$ (vanishes after pulse) \\
     \hline
\end{tabular}
\end{table*}
}

In Supplementary Note~3, we develop the quantum-geometric formulation
of the resonant injection and shift forces. The resonant response
separates into a band-diagonal injection channel, containing both the
singular population pole and its regular finite-bandwidth correction,
and an interband-coherence channel underlying the shift force. The cancellation of the remaining resonant counterterms is demonstrated
in Supplementary Note~4, while the complete off-resonant response, consisting of three distinct contributions, is derived and numerically
analyzed in Supplementary Note~5.  
Hereafter, we focus on the resonant injection and shift Raman-force 
channels and their quantum-geometric structure under linearly and
circularly polarized optical excitation.

\subsection*{Quantum-geometric injection and shift forces}
For linearly polarized (LP) excitation, the injection and shift forces can be expressed in terms of the quantum metric as (see Supplementary Note~3)
\begin{equation} \label{eq:RF_LP}
  \begin{bmatrix}
      \alpha^{\text{inj-LP}}_{abb}(\omega)
      \\[2pt]
      \beta^{\text{shi-LP}}_{abb} (\omega)
  \end{bmatrix}
   =
\frac{g_s\pi e^2}{\hbar} \sum_{\mathbf p} \sum_{mn} f_{mn} g^{nm}_{bb}\!
\begin{bmatrix}
\frac{\Delta^{nm}_{a}}{\hbar}  
\\[2pt]
  {\mathbb R}^{mn}_{a;b}  
\end{bmatrix}
 \!\delta(\omega-\omega_{nm}) 
\end{equation}
with spin degeneracy factor \mbox{$g_s=2$}. 
Here, the band-resolved quantum metric reads
\begin{equation}
    g^{nm}_{bc} = \mathrm{Re}[{\cal A}^{nm}_b {\cal A}^{mn}_c],
\end{equation} 
with ${\cal A}^{nm}_b = i\hbar \langle \phi_n | \partial_{p_b} \phi_m \rangle$ the Berry connection and $|\phi_n\rangle$ the Bloch state of band $n$. Additionally, 
\begin{equation}
    \Delta^{mn}_a = \langle \phi_m | \hat M_a | \phi_m \rangle - \langle \phi_n | \hat M_a | \phi_n \rangle =[\partial_{u_a}\varepsilon_{mn}]_{\mathbf{u}\to0}
\end{equation}
is the EPC-difference between bands, $\hbar\omega_{mn}=\varepsilon_{mn} = \varepsilon_m - \varepsilon_n$, and $f_{mn} = f_m - f_n$, with $f_n$ the Fermi-Dirac distribution at energy $\varepsilon_n$.
Remarkably, the shift Raman force is given in terms of the {\em phononic shift vector}:  
\begin{equation}
    {\mathbb R}^{mn}_{a;\ell}
    = \left [\partial_{u_a} \vartheta^{mn}_\ell + {\mathbb A}^{mm}_{a}-{\mathbb A}^{nn}_{a}
    \right ]_{{\mathbf u}\to 0}, \label{eq.PhononicSV}
\end{equation}
with electronic velocity operator $\hat v_\ell=\partial_{p_\ell}\hat H$, whose elements can be written in polar form $v_\ell^{mn} = \abs{v_\ell^{mn}} e^{-i \vartheta_\ell^{mn}}$, and
Mead--Truhlar connection $\mathbb A_a^{nm} =i\langle\phi_n|\partial_{u_a}\phi_m\rangle$ \cite{Mead_Truhlar_1979,Resta2000}. The phononic shift vector is invariant under independent, band-dependent \(U(1)\) gauge transformations, \( |\phi_n\rangle \rightarrow e^{i\theta_n(\mathbf p,\mathbf u)}|\phi_n\rangle \).
The phononic shift vector quantifies the gauge-invariant geometric shift in the momentum conjugate to the phonon coordinate during an interband optical transition, with $\hbar{\mathbb R}$ setting the associated momentum shift, as illustrated in Fig.~\ref{fig:1}c. It encodes the geometric contribution to the impulsive lattice response generated by optically induced interband coherence and is the phononic analogue of the electronic shift vector.
A numerically-accessible expression is derived in
Supplementary Note~6, in terms of the interband electron--phonon coupling (EPC) matrix elements $M^{mn}_a$, together with \mbox{$M^{mn}_{a\ell}=\mel{\phi_m}{\partial_{p_\ell}\hat{M}_a}{\phi_n}$} which is the electron--phonon--photon coupling (EPPC).

For a circularly polarized (CP) light field, the imaginary part of the QGT, corresponding to the band-resolved Berry curvature, 
\begin{equation}
    \Omega^{nm}_{bc}= -2{\rm Im}[{\cal A}^{nm}_{b} {\cal A}^{mn}_{c}],
\end{equation}
also contributes to the injection-force response. Accordingly, the CP injection force is given by
\begin{equation}\label{eq:RF_CPinj}
\alpha^{\text{inj-CP}}_{abc}(\omega)
=\frac{g_s\pi e^2}{2\hbar^2} \sum_{\mathbf p} \sum_{mn} f_{mn}  \Omega^{nm}_{bc} \Delta^{mn}_{a}  \delta(\omega-\omega_{nm}),
\end{equation}
and similarly the CP shift force reads
\begin{align}\label{eq:RF_CPshift}
   &\beta^{\text{ shi-CP}}_{abc} (\omega) = \frac{g_s\pi e^2}{4\hbar} \sum_{\mathbf p} \sum_{mn} f_{mn} \Big [ \Omega_{bc}^{mn} \big(\mathbb{R}_{a;+}^{mn} + \mathbb{R}_{a;-}^{mn} \big) - \nonumber \\
    &\hspace{1.3cm} \big(g_{bb}^{mn} + g_{cc}^{mn} \big) \big(\mathbb{R}_{a;+}^{mn} - \mathbb{R}_{a;-}^{mn} \big ) \Big ] \delta(\omega - \omega_{nm}),
\end{align}
notably depending on the {\em chiral phononic shift vector} $\mathbb{R}_{a;\pm}^{mn}$. This quantity is obtained from Eq. \eqref{eq.PhononicSV} by setting $\ell = \pm$ for the chiral components (with $p_\pm = \frac{p_b \pm i p_c}{\sqrt{2}}$), and it is the phononic equivalent of the chiral shift vector describing the gyration photocurrent as discussed in detail in Ref. \cite{Watanabe_prx_2021}. 
The analytical expressions for the injection and shift-force susceptibilities in Eqs.~\eqref{eq:RF_LP}--\eqref{eq:RF_CPshift} summarize the main results of this work. 
These expressions are not model-specific; rather they apply to generic Bloch-band Hamiltonians in which a zone-center Raman-active phonon couples to the electronic system as in Eq.~\eqref{eq.Hamiltonian}.

\subsection*{Equivalent velocity- and length-gauge formulations}
The quantum-geometric expressions for the injection and shift Raman forces
were obtained above using the velocity gauge, where the optical field enters
through the vector potential. Using a density-matrix formulation with the dipole light--matter
interaction, an independent length-gauge derivation reproduces the
velocity-gauge injection and shift susceptibilities, thereby
establishing their gauge invariance (see Supplementary Note~7).

In the resonant interband regime, the length-gauge calculation reproduces
the same injection and shift susceptibilities obtained from the diagrammatic
velocity-gauge approach. The band-diagonal EPC sector yields the injection
response governed by the quantum geometric tensor and the band-resolved EPC
difference, whereas the interband-coherence sector yields the shift response
governed by the electronic quantum geometry weighted by the phononic shift vector. The agreement between the two formulations demonstrates that the resonant injection and shift Raman-force susceptibilities are gauge independent and that their quantum-geometric interpretation does not depend on the representation chosen for the light--matter coupling.

\subsection*{Conventional Raman-force theory as a limit} A useful connection with conventional descriptions of coherent-phonon
generation emerges by reducing the general Raman-force response to a
resonant diagonal-EPC limit. In this limit, the conventional displacive
response is identified with the injection Raman force, whereas the
interband-coherence channel responsible for the shift force is absent.

We take a two-band insulator with a filled valence band, $v$ and empty conduction band, $c$. For light linearly polarized along $\hat{\bm e}_b$ and force projected along $\hat{\bm e}_a$, we additionally approximate the EPC difference between optically connected bands $\Delta_a^{cv}=M_a^{cc}-M_a^{vv}\simeq\overline{\Delta}_a$ as being independent of momentum. Within this system the linear electric susceptibility reads
\begin{equation}
\hspace{-3mm}\chi^{(1)}_{bb}(\omega)
\!=\!
\frac{g_s e^2}{\epsilon_0}
\sum_{\mathbf p}
\left[
\frac{g^{cv}_{bb}}{\hbar\omega+\varepsilon_{cv}+i0^+}
-
\frac{g^{cv}_{bb}}{\hbar\omega-\varepsilon_{cv}+i0^+}
\right],
\end{equation}
where dielectric function, $\epsilon_{bb}(\omega)=
\epsilon_0[1+\chi^{(1)}_{bb}(\omega)]$, with $\epsilon_0$ being the vacuum permittivity, and where $\varepsilon_{cv}$ is the energy gap between bands.
For an approximately constant gap
$\varepsilon_{cv}(\mathbf p)\simeq\varepsilon_g$, the
quantum-geometric injection response reduces to
\begin{equation}
\alpha^{\mathrm{inj}}_{abb}(\omega) \simeq    \frac{\overline{\Delta}_a}{\varepsilon_g}\, \omega \, {\rm Im}\epsilon_{bb}(\omega),
\end{equation}
and therefore the regular injection term follows as
\begin{equation}
\beta^{\text{reg-inj}}_{abb}(\omega)
\simeq 
 \frac{i}{2}\frac{\overline{\Delta}_a}{\varepsilon_g}\,
\frac{d}{d\omega}\left[\omega \, {\rm Im}\epsilon_{bb}(\omega)\right],
\end{equation}
Similarly, the shift response in this system can be written in terms of the dielectric function as
\begin{equation}
    \beta^{\mathrm{shi}}_{abb}(\omega) \simeq \frac{\hbar\overline{\mathbb{R}}_{a;b}}{\varepsilon_g}\, \omega \, {\rm Im}\epsilon_{bb}(\omega), \label{eq.dielec_shiftF}
\end{equation}
with $\overline{\mathbb{R}}_{a;b}$ being the momentum-averaged interband component of the phononic shift vector, where for a two-band model with a momentum-independent EPC operator, it takes the form (see Supplementary Note~6)
\begin{equation}
\mathbb{R}_{a;b}^{vc}
\simeq
\frac{1}{\varepsilon_{g}}\operatorname{Im}
\left[
\frac{M_a^{vc}(v_b^{cc}-v_b^{vv})}
{v_b^{vc}}
\right].
\end{equation}
We see that nonzero interband EPC matrix elements, such as $M_a^{vc}$, are therefore necessary, but not sufficient, for a finite shift force, as discussed later in regard to the time reversal symmetry constraint.

Our framework recovers the conventional dielectric-function
description in the reduced limit
where, for the two-band insulator with momentum-independent and band-diagonal EPC, and taking the clean limit, we find (see Supplementary Note~8)
\begin{equation}
 \chi_{abb}\simeq
\frac{\overline{\Delta}_a}{\varepsilon_g}
\left[\frac{\omega\operatorname{Im}\epsilon_{bb}(\omega)}
{i\omega_\Sigma} -
\frac{1}{2}\partial_\omega
\{\omega(\epsilon_{bb}^{*}(\omega)-\epsilon_0)\}
\right].
\label{eq:chi_stevens_limit}
\end{equation}
The first term is the singular injection response, $\alpha^{\mathrm{inj}}_{abb}/(i\omega_\Sigma)$, which identifies the conventional displacive force with the population-driven response governed by the diagonal EPC difference.
The imaginary part of the second term gives the regular injection contribution, $\beta^{\text{reg-inj}}_{abb} =(i/2)\partial_\omega\alpha^{\mathrm{inj}}_{abb}$. Its real part gives the off-resonant response, which can remain
finite in the transparent sub-gap regime. As shown in the
Supplementary Note 5, the three off-resonant contributions
combine under the two-band approximation specified above as
\begin{equation}
    \sum_{i=1}^{3}\beta^{\rm off,i}_{abb}(\omega)
    \simeq
    -\frac{\overline{\Delta}_{a}}{2\varepsilon_g}
    \partial_\omega
    \left[\omega\bigl(\operatorname{Re}\epsilon_{bb}(\omega)
    -\epsilon_0\bigr)\right].
\end{equation}
This expression recovers the conventional dispersive term
$\omega\partial_\omega\operatorname{Re}\epsilon_{bb}$ and
retains the additional contribution
$\operatorname{Re}\epsilon_{bb}-\epsilon_0$.
Equation~\eqref{eq:chi_stevens_limit} thus reproduces the absorptive and dispersive terms shown by Stevens \textit{et al.}~\cite{Stevens2002}, while retaining the complete finite part of the expansion.
Near resonance, $\hbar\omega\simeq\varepsilon_g$, the factor
$\overline{\Delta}_a\omega/\varepsilon_g$ approaches
$\overline{\Delta}_a/\hbar$, enabling comparison with the
conventional result after matching coupling and normalization
conventions.

The shift response (Eq.~\eqref{eq.dielec_shiftF}) vanishes in this diagonal-EPC limit. It requires interband EPC and the associated electronic
coherences, providing a distinct resonant impulsive
contribution beyond the population-driven response.
A related population-based description was developed by
O'Mahony \textit{et al.}~\cite{OMahoney_2019}, who obtain
photoinduced lattice forces from nonequilibrium occupations
weighted by diagonal EPC matrix elements. Such a treatment
captures the population-driven physics underlying
injection, whereas a description restricted to populations
omits the interband coherences responsible for the shift
force.

\subsection*{Symmetry and polarization selection rules}
The Raman force exhibits distinct symmetry characteristics relative to the second-order photocurrent
\cite{vonBaltzKrautPhotogalvOrig,SipeSOOR,YoungRappeShiftCurr}.
Before evaluating the numerical values of the shift and injection Raman-force responses, we first examine their symmetry properties under parity ($\mathcal{P}$), time-reversal ($\mathcal{T}$), and combined $\mathcal{PT}$ symmetry operations. Berry curvature is odd under $\mathcal{T}$ and even under $\mathcal{P}$, whereas quantum metric is even under both $\mathcal{P}$ and $\mathcal{T}$. For the inversion-even Raman mode considered here, the phonon displacement—and consequently the EPC elements and the components of $\Delta^{mn}_{a}$—are even under both $\mathcal{P}$ and $\mathcal{T}$.
Additionally, the linear and chiral phononic shift vectors exhibit distinct symmetry properties: the linear phononic shift vector is \emph{odd} under $\mathcal{T}$, while the chiral phononic shift vector changes sign and chirality under ${\cal T}$, that is, ${\cal T} {\mathbb R}_{\pm}{\cal T}^{-1}=-\mathbb{R}_{\mp}$. Both linear and chiral phononic shift vectors, however, are {\em even} under ${\cal P}$. The impact of $\mathcal{PT}$-symmetry emerges when $\mathcal{P}$ and $\mathcal{T}$ operations are applied concurrently. Berry curvature vanishes in $\mathcal{PT}$-symmetric contexts \cite{Watanabe_prx_2021,Holder_TRS&QG}, leaving only quantum metric contributions. Considering the outlined symmetries and the analytical forms presented in Eqs.~\eqref{eq:RF_LP}–\eqref{eq:RF_CPshift}, we summarize the main symmetry attributes of LP and CP shift and injection Raman forces within $\mathcal{P}$-, $\mathcal{T}$-, and $\mathcal{PT}$-symmetric systems in Table \ref{Tab.Symmetry}. Permitted and prohibited contributions are marked with \mycheckmark{} and \xmark{}, respectively.

To illustrate these general results, we consider interlayer shear phonons in a bilayer Haldane model~\cite{Haldaneorig,BLH_Sorn}, which provides a proof-of-principle model in which inversion $\mathcal{P}$ and time-reversal $\mathcal{T}$ symmetry can be controlled independently. The tight-binding Hamiltonian and electron--phonon coupling are
specified in Methods. 

\begin{table}[t]
    \centering
    \caption{Symmetry properties of the resonant Raman-force susceptibility for the inversion-even Raman mode considered here under linearly polarized (LP) and circularly polarized (CP) excitation. Here, ${\mathbb R}$ and ${\mathbb R}_\pm$ denote the linear and chiral phononic shift vectors, respectively.}
    \label{Tab.Symmetry}
    \begin{tabular}{|c|c|c|c|c|}
            \hline \hline
        Sym. & $\alpha^{\text{inj-LP}}_{abc}$ & $\beta^{\text{shi-LP}}_{abc}$ & $\alpha^{\text{inj-CP}}_{abc}$ & $\beta^{\text{shi-CP}}_{abc}$ \\
        \hline \hline
        $\mathcal{P}$  & $g \otimes \Delta$: \mycheckmark & $g \otimes {\mathbb R}$: \mycheckmark & $\Omega \otimes \Delta$: \mycheckmark & $g \otimes {\mathbb R}_{\pm}$: \mycheckmark \ $\Omega \otimes {\mathbb R}_{\pm}$: \mycheckmark \\
        \hline
        $\mathcal{T}$  & $g \otimes \Delta$: \mycheckmark & $g \otimes {\mathbb R}$: \xmark & $\Omega \otimes \Delta$: \xmark & $g \otimes {\mathbb R}_{\pm}$: \mycheckmark \ $\Omega \otimes {\mathbb R}_{\pm}$: \mycheckmark \\
        \hline
        $\mathcal{PT}$ & $g \otimes \Delta$: \mycheckmark & $g \otimes {\mathbb R}$: \xmark & $\Omega \otimes \Delta$: \xmark & $g \otimes {\mathbb R}_{\pm}$: \mycheckmark \ $\Omega \otimes {\mathbb R}_{\pm}$: \xmark \\
        \hline
    \end{tabular}
\vspace{-0.25cm}
\end{table}

At equilibrium, the injection and shift tensors are respectively
even and odd in the $\mathcal{T}$-breaking Haldane gap $\delta_\text{H}$,
$\alpha^{\rm inj}_{abc}(\omega,\delta_{\rm H})
=\alpha^{\rm inj}_{abc}(\omega,-\delta_{\rm H})$ and
$\beta^{\rm shi}_{abc}(\omega,\delta_{\rm H})
=-\beta^{\rm shi}_{abc}(\omega,-\delta_{\rm H})$.
Combining these parities with threefold rotational symmetry
and mirror covariance, under which
$(x,\delta_{\rm H})\to(-x,-\delta_{\rm H})$ because the Haldane
mass is axial and reflection-odd, yields
(see Supplementary Note~9)
\begin{align}
\alpha^{\rm inj}_{yxx} &= \alpha^{\rm inj}_{xxy} = \alpha^{\rm inj}_{xyx} = -\alpha^{\rm inj}_{yyy},
\nonumber\\
\beta^{\rm shi}_{xyy} &= \beta^{\rm shi}_{yyx} = \beta^{\rm shi}_{yxy} = -\beta^{\rm shi}_{xxx}.
\label{eq:sym}
\end{align}
All remaining components vanish.
All ``parallel-field'' components $\chi_{abb}$ are purely real at $\omega_1+\omega_2=0$ due to the
permutation symmetry $\chi_{abc}(\omega_1,\omega_2)=\chi_{acb}(\omega_2,\omega_1)$ and the
reality condition $\chi_{abc}(\omega_1,\omega_2)=\chi^{\ast}_{abc}(-\omega_1,-\omega_2)$
\cite{Butcher_Cotter}.

\begin{figure}
    \centering
    \includegraphics[width=0.9\linewidth]{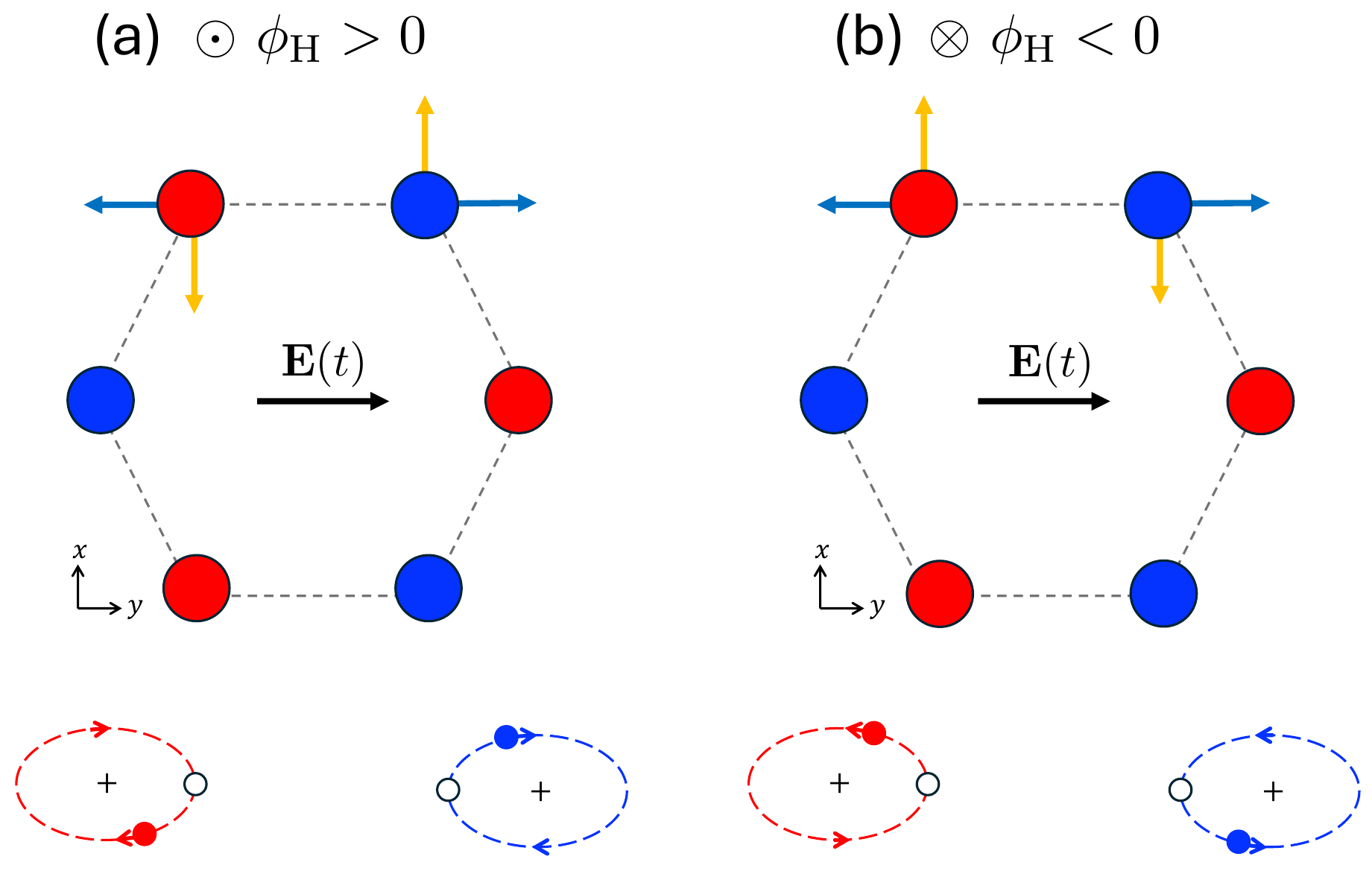}
    \caption{Rotational motion of a zone-center $E_{2g}$ mode
driven by linearly polarized light for
(a) $\phi_{\rm H}>0$ ($\odot$) and
(b) $\phi_{\rm H}<0$ ($\otimes$), where $\phi_\text{H}$ is the Haldane phase.
Blue and gold arrows denote the displacive injection and
impulsive shift forces, respectively.
The perpendicular drives produce elliptical motion, becoming
circular for equal oscillation amplitudes in the ideal limit
of degenerate modes and negligible relaxation.
Open circles mark the original equilibrium positions;
crosses mark the photoinduced equilibria.
The two sublattices oscillate out of phase but rotate in the
same sense, which reverses when $\phi_{\rm H}$ changes sign.}
    \label{fig:3}
\end{figure}

A dynamical consequence of Eq.~\eqref{eq:sym} is that
linearly polarized light can launch rotational motion of a
two-component Raman mode, as illustrated for an $E_{2g}$ mode
in Fig.~\ref{fig:3}. For a $y$-polarized pump,
the shift and injection forces act along $x$ and $y$,
respectively. In the ultrashort-pulse limit, neglecting phonon damping and
electronic relaxation, the ideal impulsive and displacive
responses (see Fig.~\ref{fig:2}b and Ref.~\cite{Stevens2002})
of initially stationary, degenerate modes yield
\begin{equation}
u_x(t)=u^{\rm shi}_0\sin(\omega_p t),\quad
u_y(t)=u^{\rm inj}_0[1-\cos(\omega_p t)].
\end{equation}
The trajectory is elliptical about the photoinduced equilibrium
$(0,u^{\rm inj}_0)$ and becomes circular when
$|u^{\rm shi}_0|=|u^{\rm inj}_0|$.
Reversing Haldane phase, $\phi_{\rm H}$, hence reversing $\delta_{\rm H}$ reverses the shift response while
preserving injection, thereby switching the sense of rotation
at fixed pump polarization. Together, the directional,
polarization, and reversal properties provide a more specific
signature of the shift mechanism than a sine-like
coherent-phonon oscillation alone. We now present the numerical results for both LP and CP injection and shift forces in this model.

\begin{figure*}[t]
    \centering
\includegraphics[width=0.9\linewidth]{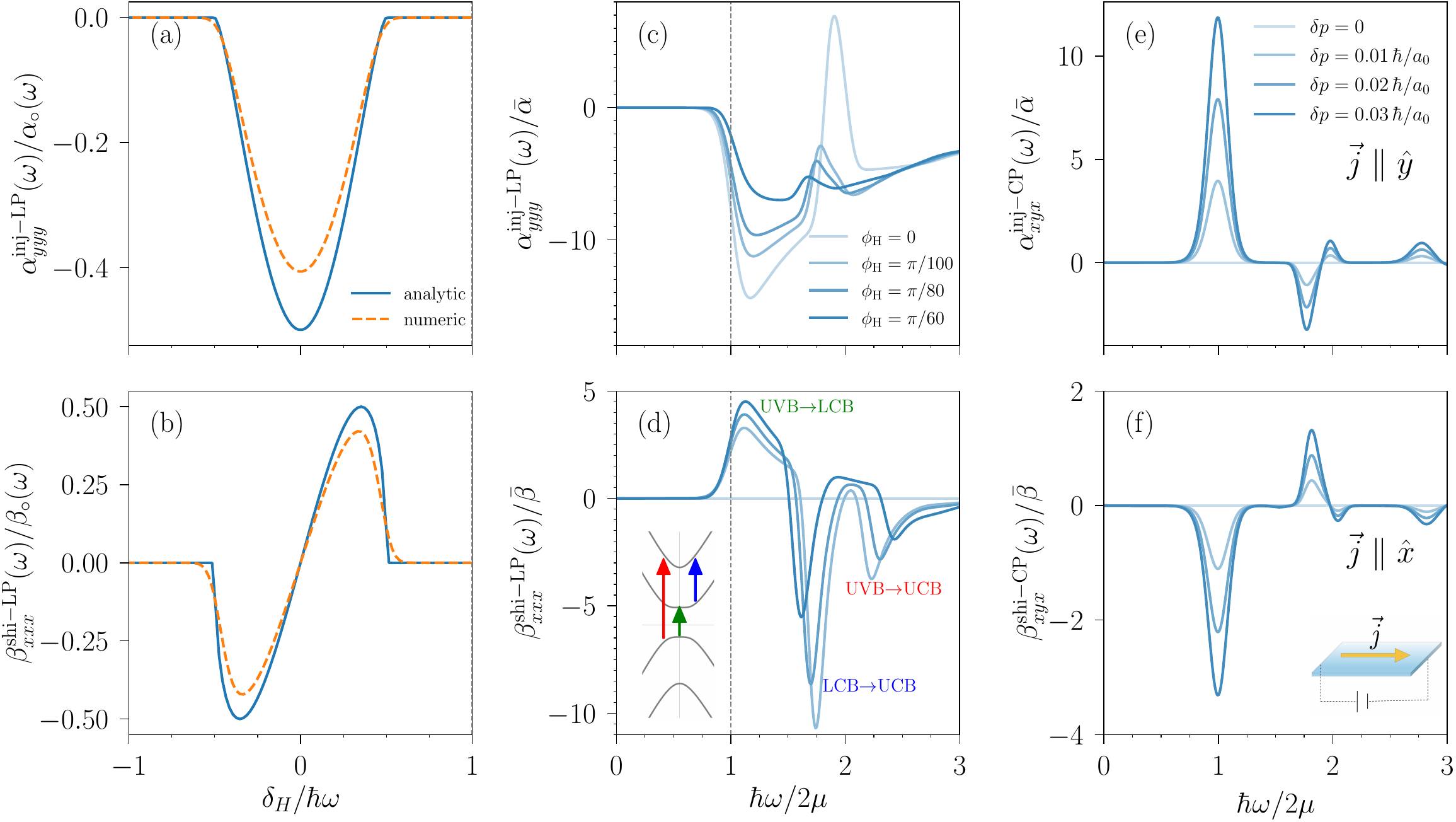}
\caption{
Injection and shift shear Raman-force susceptibilities for linearly and circularly polarized light in the bilayer Haldane model.
(a,b) LP injection and shift responses as functions of the $\mathcal{T}$-breaking Haldane gap for $\hbar\omega=0.1$ eV, comparing analytical and numerical results and demonstrating even and odd gap dependence, respectively.
(c,d) Corresponding LP spectra at $T=50\,\mathrm{K}$ for different Haldane fluxes $\phi_{\rm H}$. The inset in (d) identifies the interband transitions responsible for the spectral features: UVB-to-LCB, LCB-to-UCB, and UVB-to-UCB.
(e,f) CP injection and shift spectra at $T=50\,\mathrm{K}$ for different current-induced momentum shifts $\delta\mathbf{p}$ of the electronic distribution. The current and momentum shift are directed along $y$ in (e) and $x$ in (f), as indicated. Both responses vanish at zero current and become finite when the current breaks the in-plane symmetry. We use $\phi_{\rm H}=0$ in (e), $\phi_{\rm H}=\pi/80$ in (f), and a sublattice potential $\delta=0.05\,\mathrm{eV}$ in both panels. We set chemical potential, $\mu=0.1$ eV throughout, which is within the LCB.
The normalization factors are
$\alpha_\circ(\omega)=e^2\mathcal{M}/(\hbar^2\omega)$,
$\beta_\circ(\omega)=e^2\mathcal{M}/(\hbar\omega)^2$,
$\bar{\alpha}=100\,\mathrm{nN}/(\mathrm{V}^2\,\mathrm{ps})$, and
$\bar{\beta}=1\,\mathrm{nN}/\mathrm{V}^2$.
The model parameters are
$\gamma_0=3.16\,\mathrm{eV}$,
$\gamma_1=0.381\,\mathrm{eV}$, and
$\gamma_3=0.38\,\mathrm{eV}$ \cite{BLGHoppingParams};
$\gamma'=0.3\,\mathrm{eV}$ \cite{PeresGraphene}; and
$a_0=1.42\,\text{\AA}$,
$b=2.59\,\text{\AA}$, and
$\beta_3=5.24$ \cite{CappelutiBLGPhonon}.}
    \label{fig:4}
\end{figure*}

\subsection*{Frequency and topology control of linearly polarized injection and shift forces}
The numerical results for the injection and shift shear Raman-force responses to LP light (described by Eq. \eqref{eq:RF_LP}) in the bilayer Haldane model are depicted in Fig.~\ref{fig:4}. In panels (a) and (b), we show the injection and shift responses as functions of the $\mathcal{T}$-breaking Haldane gap, scaled to the light frequency, $\hbar\omega$. As seen, the injection force is largest in the gapless regime and diminishes as the Haldane gap increases. It retains the same sign for both positive and negative values of the Haldane gap. On the other hand, panel (b) displays the shift response, which shows an odd-parity dependence on the Haldane gap: it vanishes in the gapless regime (corresponding to the $\mathcal{T}$-symmetric case) as expected from above, rises, and then falls again as the Haldane gap increases. 

We analyze the system using a low-energy \mbox{${\bf k} \cdot {\bf p}$} model near the hexagonal Brillouin zone corners (valleys \mbox{$\tau = \pm$}), defined by \mbox{${\cal H} = -p^2 (\cos 2\varphi \hat\sigma_x + \tau \sin 2\varphi \hat\sigma_y)/2m - \tau \delta_{\rm H} \hat\sigma_z$}, where $\varphi$ is the polar angle of momentum ${\mathbf p}$. Here, Pauli matrices represent the layer pseudo-spin, $m={2 \hbar^2 \gamma_1}/{9 \gamma_0^2 a_0^2}$, and $\delta_{\rm H}$ denotes the Haldane gap. Moreover, close to the valley point, the EPC reads   $(M_x,M_y)= {\cal M}(\tau\hat\sigma_y,\hat \sigma_x)$, with \mbox{${\cal M}= 
{3 a_0 \gamma_3 \beta_3}/{\sqrt{2} b^2}$} being the electron coupling constant to shear phonons. 
Using the two-band model, we analytically evaluate the relevant quantum-geometric quantities (see Methods) and obtain the injection and shift Raman-force responses for linearly polarized excitation:
\begin{align}\label{eq:alpha_analytic}
   & \alpha^{\rm inj-LP}_{yyy}= -4\alpha_\circ(\omega) \left({1}/{4}-z^2\right)^{3/2}
    \Theta\left({1}/{2}-|z|\right),
\\ \label{eq:beta_analytic}
   & \beta^{\rm shi-LP}_{xxx} = + 4\beta_\circ(\omega) z\left({1}/{4}-z^2\right)^{1/2}\Theta\left({1}/{2}-|z|\right),
\end{align}
where \mbox{$z=\delta_{\rm H}/\hbar\omega$}, $\Theta(\dots)$ is the Heaviside step function, and by including the spin degeneracy factor, we have \mbox{$\alpha_\circ(\omega)=e^2{\cal M}/\hbar^2\omega$} and \mbox{$\beta_\circ(\omega)=e^2 {\cal M}/(\hbar\omega)^2$}.  
These analytical results are compared with full Brillouin zone numerical calculations from the four-band model, as shown in Fig.~\ref{fig:4}a,b. The two-band model reproduces the qualitative dependence of the full four-band calculation, while quantitative differences likely
arise from trigonal warping and multiband transitions present in
the full tight-binding model.

To further investigate the effect of multiband transitions, we compute the injection and shift response functions as a function of light frequency scaled by $2\mu$. As shown in Fig.~\ref{fig:4}c, in the absence of magnetic flux ($\phi_{\rm H} = 0$), corresponding to standard bilayer graphene, the injection response susceptibility $\alpha_{yyy}^{\rm inj\text{-}LP}$ is finite and exhibits a step at the interband optical gap $\hbar\omega = 2\mu$, associated with transitions between the upper valence band (UVB) and lower conduction band (LCB). 
At higher frequencies, a peak emerges due to transitions from the LCB to the upper conduction band (UCB). Notably, this transition induces a $\pi$-phase (direction) shift in the injection force, revealing a strong frequency dependence in the phase of coherent phonon oscillations and the direction of the rectified displacement.
As magnetic flux $\phi_{\rm H}$ is introduced and increased, a Haldane gap opens between the UVB and LCB, resulting in a reduced step height in the low-frequency response, consistent with the trend shown in panel~(a).
Similarly, Fig.~\ref{fig:4}d depicts the shift response as a function of frequency. The response shows an interband transition step corresponding to transitions from the UVB to LCB. The step height increases from zero to a finite value and exhibits an increasing trend with increasing magnetic flux, consistent with Fig.~\ref{fig:4}b. At higher energies, additional spectral features arise from LCB-to-UCB and UVB-to-UCB transitions (with LVB-to-LCB contributions yielding a subdominant response), with the former producing the pronounced sign-changing feature near the split-band threshold. The corresponding UVB-to-UCB feature is considerably weaker in the injection response and does not appear as a distinct spectral feature.

Hot-electron effects are generally unavoidable in nonlinear optical experiments on small-gap materials, and they can significantly modify the Raman-force response through the temperature dependence of the electronic occupations. We therefore analyze the effect of electronic temperature on the LP Raman force in Supplementary Note~10. The main effect is not simply thermal broadening: increasing temperature redistributes carriers between valence and conduction bands and shifts the chemical potential downward at fixed carrier density. These changes reduce Pauli blocking near the low-energy absorption edge, activate transitions closer to the Haldane gap, and modify the relative weight of UVB-to-LCB, LCB-to-UCB, and LVB-to-UVB transitions. As a result, the LP injection and shift forces acquire a pronounced temperature dependence, with the injection response enhanced by thermally activated conduction-band transitions, while the shift response reflects a competition between population spreading and the temperature-induced shift of the chemical potential. These effects are shown in Supplementary Fig. 3.

\subsection*{Current-induced circular Raman forces from Fermi-surface symmetry breaking}
Here, we discuss the resonant Raman shear-force response to circularly polarized (CP) light. This is motivated by the growing interest in optically exciting coherent chiral or axial phonons and controlling phonon angular momentum with circularly polarized light fields~\cite{Juraschek2025}.

For the second-order force response in Eq.~\eqref{eq:1}, the susceptibility can be decomposed into components that are symmetric and antisymmetric under exchange of the two optical indices,
$
\chi^{\rm S/A}_{abc}=\frac{1}{2}\left(\chi_{abc}\pm\chi_{acb}\right)$.
The symmetric part governs the LP-response, while the antisymmetric part governs the CP-response, with the relevant circular component in two dimensions given by \(\chi^{\rm A}_{axy}\). 
Using intrinsic permutation symmetry, \(\chi_{abc}(\omega_1,\omega_2)=\chi_{acb}(\omega_2,\omega_1)\), and the reality condition, \(\chi_{abc}(\omega_1,\omega_2)=\chi^*_{abc}(-\omega_1,-\omega_2)\)~\cite{Butcher_Cotter}, the rectified CP response satisfies $\chi^{\rm A}_{axy}(\omega,-\omega)
    =
    i\,{\rm Im}\,\chi_{axy}(\omega,-\omega)$. 
Therefore, the CP Raman force is governed by the imaginary, antisymmetric part of the rectified Raman susceptibility, in direct analogy with helicity-dependent nonlinear photocurrents {\cite{Watanabe_prx_2021}, with the corresponding resonant CP injection and shift susceptibilities given by Eqs.~\eqref{eq:RF_CPinj} and \eqref{eq:RF_CPshift}.

For the equilibrium bilayer Haldane model, the CP Raman
shear-force response vanishes despite the finite Haldane flux
because threefold rotational symmetry remains intact.
This is because of the symmetry constraints outlined in Eq.~\eqref{eq:sym}, which enforce $\chi_{axy}$ to be real, thus making its antisymmetric (circular) component vanish.
A finite CP Raman shear force therefore requires breaking
$C_3$ symmetry. A source--drain dc current provides a natural
route by producing a stationary drift distortion of the
electronic distribution that selects an in-plane direction. We model this drift by a momentum-shifted Fermi distribution,
\begin{equation}
    f_m^{\rm neq}(\mathbf p) =    f_m(\mathbf p+\delta\mathbf p),
    \qquad
    \delta\mathbf p=e\tau\mathbf E_{\rm dc},
\end{equation}
where $\tau=1/\Gamma$ is the intraband transport relaxation time and $\mathbf E_{\rm dc}$ is the static bias electric field. The current-induced distortion,
$\delta f_m(\mathbf p)=f_m^{\rm neq}(\mathbf p)-f_m(\mathbf p)$,
reduces in the small-shift limit to
$\delta f_m(\mathbf p)\simeq\delta\mathbf p\cdot\nabla_{\mathbf p}f_m(\mathbf p)$.
Since $\delta\mathbf p$ changes sign under both $\mathcal{T}$ and $\mathcal{P}$,
a finite drift shift breaks the corresponding equilibrium symmetries and allows
response components forbidden in the equilibrium case summarized in
Table~\ref{Tab.Symmetry}.

The resulting CP spectra are shown in Fig.~\ref{fig:4}e,f for both shift and injection channels. At \(\delta p=0\), both CP responses vanish, as required by symmetry. For finite \(\delta p\), the responses become nonzero and increase with the magnitude of the current-induced distribution shift. Since the current-induced occupation change $\delta f_m$ is concentrated near the Fermi surface at low temperature, the CP Raman force is dominated by optical transitions involving the partially occupied low-energy conduction band (LCB) containing the Fermi level. As a result, the CP response is highly sensitive to current-induced distortions of the Fermi surface. This contrasts with the LP response, which is already symmetry-allowed in equilibrium and therefore does not rely on such nonequilibrium Fermi-surface distortions.

Three spectral features appear in both CP injection and shift shear Raman forces in Fig.~\ref{fig:3}e,f: a UVB-to-LCB resonance at \(\hbar\omega = 2\mu\), a sign-changing LCB-to-UCB resonance, and a weaker LVB-to-LCB resonance.  All three features are Fermi-surface effects involving transitions to or from the LCB, as expected. The sign change of the LCB-to-UCB resonance relative to the other two spectral features reflects the variation of the relevant interband EPC matrix element over the momentum shell sampled across the transition threshold.  

 \section*{Conclusion}
Taken together, the quantum-geometric decomposition, its gauge-invariant
formulation, and the recovery of the conventional dielectric-function limit
establish a unified picture of rectified Raman forces.
Within this picture,
the resonant response separates into two mechanisms  with
distinct microscopic origins and temporal dynamics. The conventional
displacive force is identified with the injection contribution and is
governed by the quantum geometric tensor together with the band-resolved
asymmetry of the diagonal EPC. The shift force, by contrast, originates from interband EPC and
electronic coherences and provides a distinct impulsive mechanism in the
absorbing regime governed by the quantum geometric tensor and the phononic
shift vector. 

The polarization dependence is strongly constrained by time-reversal symmetry. For LP excitation, the injection force is allowed in a time-reversal-symmetric system, whereas the shift force vanishes and requires broken time-reversal symmetry. For CP excitation, the injection force vanishes under time reversal, while the shift force can remain finite. Thus, in a time-reversal-symmetric system, circularly polarized
light provides a direct route to generating the resonant shift
Raman force without magnetic time-reversal-symmetry breaking,
provided that the relevant response is allowed by crystalline
symmetry.

For the bilayer system considered here, these mechanisms provide distinct
routes to coherently control interlayer shear motion. Their magnitude and
direction can be controlled through the optical frequency and Haldane flux,
while circularly polarized responses can additionally be activated through
current-induced Fermi-surface symmetry breaking. Their temporal signatures
are also qualitatively different: the injection contribution produces a
displacive force that persists on the electronic relaxation timescale,
whereas the shift contribution acts impulsively during the optical pulse, as
illustrated in Fig.~\ref{fig:2}.

Experimentally, these dynamics can be probed through coherent phonon
displacements in pump--probe measurements~
\cite{Udina2019CoherentOscillTheory}, including time-resolved
reflectivity~\cite{Zeiger1992,Soranzio_prr_2019,Bartram2022TMD_IL-Phonon},
second-harmonic generation~
\cite{Mankowsky2017ReversalFerroelecPol,Zhang_prx_2019},
Raman spectroscopy~\cite{Kuramochi_JACS_2021}, coherent band shifts in
time-resolved ARPES~\cite{Shi2019ARPES_CDW,Sayers2023ARPES_CDW}, and
Bragg-peak intensity changes in electron or x-ray diffraction~
\cite{Sie2019,
Soranzio2022XRDLatticeManip}. Substrate effects can be reduced through lattice mismatch, free suspension~
\cite{Boschetto_nl_2013}, or by studying bulk layered materials. In time-resolved reflectivity~
\cite{Zeiger1992}, for example, the differential signal follows the coherent
displacement, \(\Delta R(t)/R\propto u(t)\), providing a direct route to
distinguishing the displacive injection and impulsive shift dynamics.

Our results therefore extend quantum geometry from electronic transport and optical response to the generation and control of coherent lattice motion, providing a quantum-geometric route toward light-induced phase control in quantum materials. Coherent-phonon amplitudes, phases, polarization dependence, and temporal dynamics thereby offer experimentally accessible signatures of the quantum metric, Berry curvature, and electron--phonon geometric quantities such as the phononic shift vector.

\section*{Methods}
\subsection*{Kadanoff--Baym Raman-force formalism}
We consider a generic electronic Bloch Hamiltonian coupled to a zone-center Raman-active phonon displacement ${\bf u}$ with frequency $\omega_p$. Expanding the electronic Hamiltonian to linear order in the phonon displacement, we write~\cite{Giustino_rmp_2017}
\begin{align}
   \hat {\cal H}
   = \sum_{\mathbf p} \hat\psi^\dagger_{\mathbf p}
   \left[
   \hat H({\mathbf p}) + \hat{\mathbf M}({\mathbf p})\cdot {\mathbf u}
   \right]
   \hat\psi_{\mathbf p},
   \label{eq.Hamiltonian}
\end{align}
where $\hat H({\mathbf p})$ is the equilibrium electronic Bloch Hamiltonian and
\begin{align}
    \hat M_a({\mathbf p})
    =
    \frac{\partial \hat H({\mathbf p},{\mathbf u})}
    {\partial u_a}
    \bigg|_{{\mathbf u}=0} \label{eq.EPC_operator}
\end{align}
is the electron--phonon coupling (EPC) operator associated with the phonon coordinate $u_a$.  The electronic field operator at momentum $\mathbf p$ is denoted by $\hat{\psi}_{\mathbf p}$. 
The light--matter interaction Hamiltonian is obtained by coupling light to the electrons via the minimal coupling substitution $\mathbf{p}\to\mathbf{p}+e\mathbf{A}(t)$, where $\mathbf{A}(t)$ is a homogeneous vector potential. Here, we focus on the electron-mediated Raman force and neglect direct ionic coupling to the optical field.
Four relevant light--matter couplings are given by
$\partial_{p_a}\hat H$, $\partial_{p_a}\partial_{p_b}\hat H$,
$\partial_{p_b}\hat M_a$, and
$\partial_{p_c}\partial_{p_b}\hat M_a$, corresponding respectively
to one-photon, two-photon, one-photon--phonon, and
two-photon--phonon couplings \cite{Basko_2009}.

Using the Kadanoff–Baym variational formalism~\cite{KadanoffBaymPaperPR,KadanoffBaym,Rostami2021} (summarized below; full derivation in Supplementary Note 1), we
derive the Raman force perturbatively to second order in the electric
field. The susceptibility $\chi_{abc}$ contains resonant injection
and shift contributions together with three off-resonant contributions.
The complete set of diagrams is shown in Supplementary Fig.~1, while
Fig.~\ref{fig:5} shows the triangle and oval diagrams that contain
the resonant terms. The formal evaluation of these contributions in the band basis and their quantum-geometric decomposition are presented in Supplementary Note 2 and Supplementary Note 3, respectively. 
 
In the band basis, the triangle diagram yields the singular injection response, its regular correction, a first off-resonant contribution, $\beta^{\text{off,1}}_{abc}$, and a counterterm, $\gamma^\Delta_{abc}$.
The oval diagram yields the resonant shift force, $\beta^{\rm shi}_{abc}$; a second off-resonant contribution, $\beta^{\text{off,2}}_{abc}$, which remains finite in the transparent regime; and a counterterm, $\gamma^{\rm O}_{abc}$. The latter is equal in magnitude and opposite in sign to the triangle-diagram counterterm $\gamma^\Delta_{abc}$, so that in the clean, noninteracting limit
\begin{equation}
\gamma^\Delta_{abc}+\gamma^{\rm O}_{abc}=0.
\end{equation}
A detailed proof of this cancellation is provided in Supplementary Note~4.

\subsection*{Shear phonons in a bilayer Haldane model}
Having developed the Raman-force formalism for a generic multiband
Bloch Hamiltonian, we now specialize to a two-dimensional bilayer
Haldane model and its interlayer shear phonon.

\begin{figure}[t]
    \centering
        \begin{overpic}[width=\linewidth]{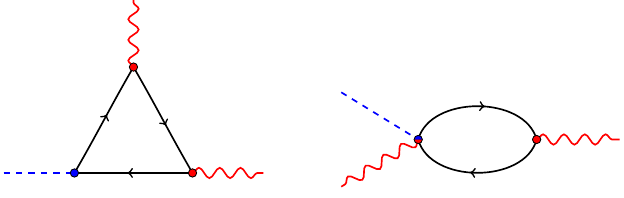}
        \put(18,-2){\large (a)}
        \put(75,-2){\large (b)}
    \end{overpic}
\caption{Feynman diagrams contributing to the resonant Raman-force response. Solid, dashed, and wavy lines denote electron, phonon, and photon propagators, respectively. Red, blue, and combined red-blue vertices
represent velocity, electron--phonon, and mixed electron--phonon--photon couplings. (a) The triangle diagram
yields the singular and regular injection contributions, an off-resonant term, and a counterterm. (b) The oval diagram
yields the shift contribution, a second off-resonant term, and the opposite counterterm. The counterterms cancel in the clean, noninteracting limit.
} 
\label{fig:5}
\end{figure}

We quantitatively compute the injection and shift forces that excite interlayer shear modes in a bilayer system. To incorporate $\mathcal{P}$- and $\mathcal{T}$-symmetry breaking, we analyze injection and shift responses in the bilayer Haldane model \cite{Haldaneorig,BLH_Sorn}, which breaks $\mathcal{T}$-symmetry via magnetic flux and \(\mathcal{P}\)-symmetry via sublattice staggering. 
We employ a four-band Hamiltonian with hexagonal symmetry, where block-diagonal terms describe an individual Haldane model \cite{Haldaneorig} in each layer with flux $\phi_{\rm H}$, as given by
\vspace{-0.2cm}
\begin{equation}
\hat{\mathcal H}
= \hat H_H \otimes \hat\tau_0
  + \hat H_{12} \otimes \hat\tau_{-}
  + \hat H_{12}^\dagger \otimes \hat\tau_{+},
\label{eq:H_compact}
\end{equation}
where $\hat\tau_i$ and $\hat\sigma_i$ \mbox{($i=x,y,z$)} are Pauli matrices in layer and sublattice space, 
\mbox{$\hat\tau_\pm=(\hat\tau_x\pm i\hat\tau_y)/2$}, \mbox{$\hat\sigma_\pm=(\hat\sigma_x\pm i\hat\sigma_y)/2$},
\mbox{$\hat H_H = d_0 \hat\sigma_0 + \mathbf d\cdot\hat{\boldsymbol\sigma}$}, with 
\begin{align}
     &d_0 = -2 \gamma' \cos\phi_{\rm H} \left( \text{Re}[\zeta_{\mathbf{p}}] + 3/2 \right),
    \nonumber \\
    &d_x - i d_y = -\gamma_0 \xi_{\mathbf{p}},
    \nonumber \\
    &d_z = \delta + 2 \gamma' \sin\phi_{\rm H} \ \text{Im}[\zeta_{\mathbf{p}}].
\end{align}
The interlayer term is $  \hat H_{12} = \gamma_1 \hat\sigma_+ - \gamma_3 \xi_{\mathbf p}\hat\sigma_-$. 
Here, the nearest and next-nearest neighbor form factors are 
\begin{align}
\xi_{\mathbf{p}} &= e^{i p_y} +2 e^{-i p_y/2} \cos\left( {\sqrt{3} p_x}/{2} \right),
\\
\zeta_{\mathbf{p}} &= e^{i \sqrt{3} p_x} + 2 e^{-i \sqrt{3} p_x/2} \cos\left( {3 p_y}/{2} \right).
\end{align}
Note that $\mathbf{p}$ is measured in units of $\hbar/a_0$. The parameters $\gamma_0$ and $\gamma'$ describe the nearest and next-nearest neighbor intralayer hopping energies, while $\gamma_1$ and $\gamma_3$ correspond to the interlayer hopping terms \cite{McCannBLG}. The parameter $\delta$ represents the sublattice symmetry-breaking term that opens a trivial gap and breaks $\mathcal{P}$ symmetry, while the Haldane magnetic flux $\phi_{\rm H}$ generates the $\mathcal{T}$-breaking Haldane mass
$\delta_{\rm H}=3\sqrt{3}\gamma'\sin\phi_{\rm H}$, corresponding to a gap $2|\delta_{\rm H}|$ when $\delta=0$.
Since $\phi_{\rm H}$ affects only intralayer hopping phases, the interlayer EPC is the same as in Bernal-stacked bilayer graphene~\cite{CappelutiBLGPhonon}, namely
\begin{equation}
\hat{\mathbf M}
= \hat M_{12}\otimes\tau_{-} + \hat M_{12}^\dagger\otimes\tau_{+}, ~~~ \hat M_{12} = -i\hbar\,\mathcal{C}\partial_{\mathbf p}\xi_{\mathbf p}\,\sigma_{-},
\end{equation}
where ${\cal C}={\sqrt{2}\gamma_3 \beta_3}/{b^2}$, with Grüneisen parameter, \mbox{$\beta_3 = -{\partial \ln \gamma_3}/{\partial \ln b}$} quantifying the strength of the EPC, and $b$ the interatomic distance for $\gamma_3$-hopping.

Projecting the four-band model onto its low-energy two-band sector allows us to obtain analytical expressions for the relevant quantum-geometric quantities. We find
\begin{equation}
    {\bm \Delta}^{cv} = -2\mathcal{M}(1+X^2)^{-1/2}\,
(\sin 2\varphi,\;\cos 2\varphi),
\end{equation} 
\begin{align}
\label{eq:g_and_delta_2band}
\mathbf{g}^{cv}
&= \frac{\hbar^2}{p^2(1+X^2)^2}
\begin{bmatrix}
X_s^2(\varphi) & -\tfrac{1}{2}\sin 2\varphi \\
-\tfrac{1}{2}\sin 2\varphi & X_c^2(\varphi)
\end{bmatrix},
\end{align}
and
\begin{align}
\mathbb{R}^{cv}
&= \frac{\mathcal{M}X}{\epsilon_p\sqrt{1+X^2}}
\begin{bmatrix}
-{\cos^2\varphi}/{X_s^2(\varphi)} &
{\sin^2\varphi}/{X_c^2(\varphi)} \\[2pt]
{\sin 2\varphi}/{2X_s^2(\varphi)} &
{\sin 2\varphi}/{2X_c^2(\varphi)}
\end{bmatrix}, 
\end{align}
where $X=\delta_{\rm H}/\epsilon_p$ with $\epsilon_p=p^2/2m$,  $X_s^2(\varphi) = X^2 + \sin^2\varphi$, and $X_c^2(\varphi) = X^2 + \cos^2\varphi$. From the $\varphi$-dependence of the Cartesian components of $\boldsymbol{\Delta}^{cv}$, ${\bf g}^{cv}$, and $\mathbb{R}^{cv}$, we see that the non-vanishing injection and shift components obey Eq.~\eqref{eq:sym}.
Moreover, for small $X\ll1$, the shift vector rises linearly with $X$, whereas for larger $X\gg1$, it diminishes with increasing magnetic flux. 
Substituting the quantum-geometric quantities derived above yields
the analytical LP injection- and shift-force susceptibilities given
in Eqs.~\eqref{eq:alpha_analytic}
and~\eqref{eq:beta_analytic}, respectively. 
 
\section*{Acknowledgments} 
We acknowledge support from the Engineering and Physical Sciences Research Council (Grant No. UKRI122) and the Royal Society (Grant No. IES\textbackslash R2\textbackslash 242309). We thank R. Stanier, S. Patel, E. Da Como, D. Wolverson, S. Bending, E. Cappelluti, N. Nagaosa and L. Benfatto for helpful comments. 
 
\section*{Correspondence} 
Habib Rostami (hr745@bath.ac.uk)
 
\section*{Author contributions}
H.R. conceived and supervised the project. J.L.P. carried out the analytical and numerical calculations. J.L.P. and H.R. developed the theoretical framework, analyzed the results, and wrote and revised the manuscript.
 
\section*{Data availability}
The data supporting the findings of this study are available from the corresponding author upon reasonable request.
 
\section*{Code availability}
The custom code used to generate the numerical results and figures is available from the corresponding author upon reasonable request.
\section*{Competing interests}
The authors declare no competing interests.
 
\bibliography{refs}
\end{document}


\bibliographystyle{apsrev4-2}

\renewcommand{\arraystretch}{1.4}

\title{Supplementary Information for ``Quantum Geometric Injection and Shift Optical Forces Drive Coherent Phonons''}

\author{J. Luke Pimlott}
\email{jlp62@bath.ac.uk}
\affiliation{Department of Physics, University of Bath, Claverton Down, Bath BA2 7AY, United Kingdom}

\author{Habib Rostami}
\email{hr745@bath.ac.uk}
\affiliation{Department of Physics, University of Bath, Claverton Down, Bath BA2 7AY, United Kingdom}


\maketitle
\tableofcontents

\begin{figure}
    \centering
    \begin{overpic}[width=\linewidth]{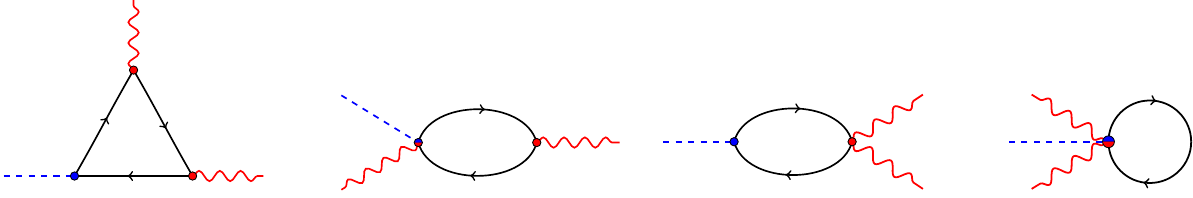}
        \put(10,-2){\large (a)}
        \put(38.5,-2){\large (b)}
        \put(65.5,-2){\large (c)}
        \put(95,-2){\large (d)}
    \end{overpic}
    \vspace{1mm}
    \caption{The four topologically-unique Feynman diagrams for the Raman-force susceptibility: solid, dashed, and wavy lines respectively denote electrons, phonons, and photons. The diagrams from (a)–(d) respectively represent Eqs.~\eqref{eq.pia}–\eqref{eq.pid}. Diagrams (a) and (b) have two permutations.}
    \label{fig:feyn}
\end{figure}

\suppnote{Kadanoff--Baym derivation of the Raman-force susceptibility} 
\label{app.KB_propagator}
In the presence of light–matter interaction with a light field with vector potential $\mathbf{A}$ and electron-phonon interaction operator $\hat{\cal M}_a$, phonon displacement field component $ u_a$, the thermodynamical average of the atomic force is expressed in terms of the field-dependent Green's function \cite{KadanoffBaymPaperPR,KadanoffBaym,Rostami2021}
\begin{align}\label{eq:F}
    {\cal F}_a(1) = i\int_{1'}\int_{1''} {\rm tr}[\hat {\cal M}_a(1',1'';1) \hat G(1'',1')] \eval_{\bf{u}\rightarrow0}.
\end{align}
Here, $1 = (\mathbf{r}_1, t_1)$ is a shorthand for the space-time coordinate, $1'=1''^{+}=(\mathbf{r}_{1''}, t_{1''}+0^+)$, ${\rm tr}[\dots]$ denotes the trace over internal spinor degrees of freedom, and the Green's function is defined as 
\begin{align}
    \hat G(1,1') = -i \langle {\cal T}_t[\hat\psi_{H}(1) \hat \psi^\dagger_{H}(1')] \rangle 
\end{align}
in which $\langle \cdots \rangle$ stands for the thermodynamic average, ${\cal T}_t$ denotes the time-ordering operation, and $\hat \psi_H(\mathbf{r}, t)$ is the field operator in the Heisenberg picture, expressed in the basis of the full Hamiltonian $\hat{\mathcal{H}}$, which includes kinetic, light–matter, and electron-phonon interaction terms. The electron-phonon coupling, $\hat {\cal M}_a$ reads 
\begin{align}
    \hat {\cal M}_a(1',1'';1) = - \frac{\delta \hat G^{-1}(1',1'')}{\delta u^a(1)}\Bigg|_{{\bf u}\to 0},
\end{align}
where the inverse Green's function obeys
\begin{align}
    \hat G^{-1}(1',1'') = \left\{i\partial_{t_{1'}} -\hat H(-i\hbar \nabla_{1'} + e{\bf A}(1')) - \hat M_a(-i\hbar \nabla_{1'} + e{\bf A}(1'))  u^a (1')\right\}\delta(1'-1''),
\end{align}
Here, $\hat{\mathcal H}(\mathbf p)
=\hat H(\mathbf p)+\hat M_a(\mathbf p)u_a$ is the
momentum-resolved form of Eq.~(26) of the main text, while
$\hat H(\mathbf p)\equiv\hat H(-i\hbar\nabla)$ is the
noninteracting electronic Hamiltonian. The light--matter interaction is included via the minimal-coupling substitution $\mathbf{p}\to\mathbf{p}+e\mathbf{A}$, which can be expanded for small $\mathbf{A}$, giving paramagnetic (one-photon coupling) and diamagnetic (two-photon coupling) current operators, as 
\begin{align}
    \hat{H}({\bf p}) \rightarrow 
    \hat{H} ({\bf p} + e {\bf A}) \approx \hat{H}({\bf p}) + e [\nabla_{p_{b_1}} \hat{H}({\bf p})] A^{b_1} + \frac{e^2}{2} [\nabla_{p_{b_1}}\nabla_{p_{b_2}} \hat{H}({\bf p})] A^{b_1} A^{b_2} + \order{{\bf A}^3}. \label{eq:H-minimalcoupling}
\end{align}
Similarly, part of the light–matter coupling emerges as a mixed photon–phonon–electron interaction, originating from the (minimally-coupled) momentum dependence of the electron–phonon coupling, expanded to 
\begin{align}
    \hat{M}_a({\bf p}) \rightarrow 
    \hat{M}_a ({\bf p} + e {\bf A}) \approx \hat{M}_a({\bf p}) + e [\nabla_{p_{b_1}} \hat{M}_a ({\bf p})] A^{b_1} +  \frac{e^2}{2} [\nabla_{p_{b_1}} \nabla_{p_{b_2}} \hat{M}_a ({\bf p})] A^{b_1} A^{b_2} + \order{{\bf A}^3}, \label{eq:M-minimalcoupling}
\end{align}
with repeated indices summed according to the Einstein summation convention. Following the Taylor expansion, we define the one-photon and two-photon current operators, as well as photon-phonon-electron coupling terms in the space-time coordinate basis, based on the derivatives of the inverse Green's function,
\begin{align}
    &\nabla_{p_b}  \hat {H} 
    ~~~\to~~~ \hat j_b(1',1'';2) = \frac{\delta \hat G^{-1}(1',1'')}{\delta A^b(2)}
    \\
    &\nabla_{p_b} \nabla_{p_c} \hat {H}
    ~~~\to ~~~
    \hat j_{bc}(1',1'';2,3) = \frac{\delta^2 \hat G^{-1}(1',1'')}{\delta A^c(3) \delta A^b(2)}
    \\
    & \nabla_{p_b} \hat {M}_a
    ~~~\to~~~ \hat {\cal M}_{ab}(1',1'';1,2) = -\frac{\delta \hat G^{-1}(1',1'')}{\delta A^b(2) \delta u^{a}(1)}
    \\
    & \nabla_{p_b} \nabla_{p_c} \hat {M}_a
    ~~~\to~~~ \hat {\cal M}_{abc}(1',1'';1,2,3) = -\frac{\delta \hat G^{-1}(1',1'')}{\delta A^c(3) \delta A^b(2) \delta u^{a}(1)}
\end{align}
(noting that ${\delta G^{-1}}/{\delta A}=-{\delta \mathcal{H}}/{\delta A}$ and ${\delta G^{-1}}/{\delta u}=-{\delta\mathcal{H}}/{\delta u}$). With the light–matter coupling term expressed in terms of the Green's function, the Raman susceptibility can be obtained by applying a variational approach and differentiating Eq.~\eqref{eq:F} with respect to the light field:
\begin{align}
    \pi_{abc} (1;2,3) = \frac{1}{2} \frac{\delta^2 {\cal F}_a(1)}{\delta A_b(2)\delta A_c(3)} \Bigg|_{\bf A\to 0}.
\end{align}
The lengthy but straightforward variational 
calculus generates four distinct terms for the Raman response given by
\begin{align}
    \pi_{abc}(1;2,3) =  
    \frac{i}{2}(\pi^{\rm A}_{abc}(1;2,3) + \pi^{\rm B}_{abc}(1;2,3) + \pi^{\rm C}_{abc}(1;2,3) + \pi^{\rm D}_{abc}(1;2,3))
\end{align}
where each term explicitly reads 
\begin{align}
  \pi^{\rm A}_{abc}(1;2,3) &= -\int_{1',1''}   \int_{2',2''} \int_{3',3''} \bigg \{  \Tr[\hat{\cal M}_a(1',1'';1) G(1'',3') \hat{j}_c(3',3'';3) G(3'',2') \hat{j}_b(2',2'';2) G(2'',1')]  \nonumber \\
    &\hspace{3cm} + \Tr[\hat{\cal M}_a(1',1'';1) G(1'',2') \hat{j}_b(2',2'';2) G(2'',3') \hat{j}_c(3',3'';3) G(3'',1')] \bigg\}, \label{eq.pia}
\end{align}
\begin{align}
  \pi^{\rm B}_{abc}(1;2,3) &=  \int_{1',1''} \int_{2',2''} \bigg \{  \Tr[\hat{\cal M}_{ab}(1',1'';1,2) G(1'',2') \hat{j}_c(2',2'';3) G(2'',1')] \nonumber \\
    &\hspace{2cm} + \Tr[\hat{\cal M}_{ac}(1',1'';1,3) G(1'',2') \hat{j}_b(2',2'';2) G(2'',1')] \bigg \}, \label{eq.pib}
\end{align}
\begin{align}
    \pi^{\rm C}_{abc}(1;2,3) =  \int_{1',1''} \int_{2',2''} \Tr[\hat{\cal M}_a(1',1'';1) G(1'',2') \hat{j}_{bc}(2',2'';2,3) G(2'',1')], \label{eq.pic}
\end{align}
and finally 
\begin{align}
     \pi^{\rm D}_{abc}(1;2,3) = -\int_{1',1''}  \Tr[\hat{\cal M}_{abc}(1',1'';1,2,3) G(1'',1')].  \label{eq.pid}
\end{align}
These terms can be illustrated using four distinct Feynman diagrams depicted in \suppfigref{fig:feyn}. Finally, the gauge-invariant response function to the external electric fields, $\chi_{abc}$, is obtained by relating the electric field to a homogeneous vector potential, ${\bf E}(t) = -\partial_t {\bf A}(t)$, which in the frequency domain becomes ${\bf A}(\omega) = {\bf E}(\omega) / (i\omega)$, leading to 
\begin{align}
    \chi_{abc}(\omega_\Sigma;\omega_1,\omega_2) = \frac{\pi_{abc}(\omega_\Sigma;\omega_1,\omega_2)}{(i\omega_1)(i\omega_2)},
\end{align}
having Fourier transformed the susceptibility to the frequency domain, where our electric fields are $E^b(\omega_1)$ and $E^c(\omega_2)$, and with $\omega_\Sigma=\omega_1+\omega_2$ the frequency of the Raman force generated. Note that experimentally, one would observe both $\chi_{abc}(\omega_\Sigma;\omega_1,\omega_2)$ and $\chi_{acb}(\omega_\Sigma;\omega_2,\omega_1)$ simultaneously, which as they are equal means that the total observed response would be twice that implied by a single susceptibility component.
\suppnote{Band-basis expressions for the Raman-force diagrams} \label{app.band_basis}
We move from the real to imaginary-time formalism \cite{Mahan}, allowing us to use a Matsubara summation approach in the band basis to formally evaluate each diagram in terms of electronic dispersion, and the matrix elements of light-matter and electron-phonon coupling terms. First, we fix the notation as
\begin{align}
    \hat{j}_{b_1 b_2 \dots b_n} &= -e^n \hat{v}_{b_1 b_2 \dots b_n}, 
    & 
    \hat{\mathcal{M}}_{a b_1 b_2 \dots b_n} &= e^n \hat{M}_{a b_1 b_2 \dots b_n},
    \label{eq:nphotonvertex}
\end{align}
where $\hat{v}_b = \partial_{p_b} \hat{\mathcal{H}}$ is the velocity operator, 
$\hat{M}_a = \partial_{u_a} \hat{\mathcal{H}}$ is the electron-phonon coupling operator, and 
$\hat{M}_{ab} = \partial_{p_b} \hat{M}_a$ is the electron-phonon-photon coupling operator. 
In compact notation, the eigenvalue problem of the electronic Hamiltonian reads ${\cal H}|\phi_n\rangle = \varepsilon_n |\phi_n\rangle$, where $\varepsilon_n$ is the energy of band $n$ with eigenvector $|\phi_n\rangle$. The Matsubara Green's function can then be written in the band basis as
\begin{align}
    \langle \phi_n|\hat G (i\varepsilon_\nu)|\phi_m\rangle = \delta_{nm} G_n= \frac{ \delta_{nm} }{ i\varepsilon_\nu-\varepsilon_n}, 
\end{align}
where $\varepsilon_\nu= (2\nu+1) \pi k_B T$ stands for the fermionic Matsubara energy and $\nu$ is an integer. We use the standard Matsubara sums \cite{Mahan} for one, two and three propagators, given by
\begin{align}
    \frac{1}{\beta} \sum_\nu G_{m}(i \varepsilon_\nu) =& f(\varepsilon_m) = f_m \label{eq.1GMatsSum} 
    \\
    \frac{1}{\beta} \sum_\nu G_{n}(i\varepsilon_\nu + i\hbar\omega) G_{m}(i\varepsilon_\nu) =& \frac{1}{\hbar} \frac{f_{mn}}{\omega + \omega_{mn}} \label{eq.2GMatsSum} 
    \\
    \frac{1}{\beta} \sum_\nu G_l(i\varepsilon_\nu+i\hbar\omega) G_n(i\varepsilon_\nu+i\hbar\omega_1) G_m(i\varepsilon_\nu) =& \frac{1}{\hbar^2} \left\{  
    \frac{f_{mn}}{(\omega_1-\omega_{nm})(\omega-\omega_{lm})} + \frac{f_{ln}}{(\omega_2-\omega_{ln})(\omega-\omega_{lm})} \right\}
    , \label{eq.3GMatsSum}
\end{align}
where $\beta = 1/k_B T$ is the inverse temperature and $f_m$ denotes the Fermi–Dirac distribution for band $m$. We define $f_{mn} = f_m - f_n$, and $\omega_{nm} = (\varepsilon_n - \varepsilon_m)/\hbar$ is the energy difference between bands $n$ and $m$, expressed in frequency units.
With these sums, we compute the explicit values of each diagram, which are
\begin{align}
    \chi^{\rm A}_{abc} (\omega_\Sigma;\omega_1,\omega_2) =&\frac{-e^2}{2\hbar^2 (i\omega_1)(i\omega_2)}\sum_{\bf p} 
    \sum_{mnl} \bigg \{\frac{f_{mn} M^{ml}_a v^{ln}_c v^{nm}_b}{(\omega_1 + i\gamma-\omega_{nm})(\omega_\Sigma+i\Gamma-\omega_{lm})} + \frac{f_{mn} M^{lm}_a v^{mn}_c v^{nl}_b}{(\omega_2 + i\gamma-\omega_{mn})(\omega_\Sigma+i\Gamma-\omega_{ml})} 
    \nonumber\\&+ [b,\omega_1 \leftrightarrow c,\omega_2] \bigg \} 
    \label{eq:chi-A}
    \\
    \chi^{\rm B}_{abc} (\omega_\Sigma;\omega_1,\omega_2) =& \frac{-e^2}{2\hbar (i\omega_1)(i\omega_2)} \sum_{\bf p} \sum_{mn} \bigg\{\frac{f_{mn} M^{mn}_{ac} v^{nm}_b}{\omega_1 + i\gamma + \omega_{mn}} + [b,\omega_1 \leftrightarrow c,\omega_2] \bigg\}
    \label{eq:chi-B}
    \\
  \chi^{\rm C}_{abc} (\omega_\Sigma;\omega_1,\omega_2) =& \frac{-e^2}{2\hbar (i\omega_1)(i\omega_2)}\sum_{\bf p} \sum_{mn} \frac{f_{mn}M^{mn}_a v^{nm}_{bc}}{\omega_\Sigma+i\Gamma + \omega_{mn}}   
    \label{eq:chi-C}
    \\
    \chi^{\rm D}_{abc} (\omega_\Sigma;\omega_1,\omega_2) =& \frac{-e^2}{2(i\omega_1)(i\omega_2)} \sum_{\bf p} \sum_m f_m M^{mm}_{abc} 
    \label{eq:chi-D}
\end{align}
where $O^{mn}=\langle \phi_m|\hat O|\phi_n\rangle$ is the matrix element of operator $\hat O$. Here, we include a small imaginary part in the frequency to account for phenomenological damping. Specifically, we add $\gamma$ to the frequencies $\omega_1$ and $\omega_2$, and $\Gamma$ to their sum, $\omega_\Sigma$. Physically, in the case of rectified Raman force we are interested in, $\gamma$ represents interband dephasing, while $\Gamma=\tau^{-1}$ is the intraband transport relaxation rate \cite{Holder_TRS&QG}.

Unless otherwise stated, analytical expressions in the
Supplementary Information are given per spin. The spin
degeneracy $g_s=2$ is included in the main-text microscopic
response formulas and in all numerical results.
\suppnote{Low-output-frequency decomposition of the Raman-force response}
\label{app.rect_force}
The rectified Raman-force response is obtained in the
low-output-frequency regime,
$\omega_\Sigma=\omega_1+\omega_2\to0$, with
$\omega_1\approx-\omega_2$. In this limit, the triangle
response in Eq.~\eqref{eq:chi-A} can be separated according to whether
the EPC vertex is band diagonal, $M_a^{nn}$, or band off-diagonal,
$M_a^{mn}$ ($m\neq n$). The band-diagonal EPC sector contains three
contributions: the singular injection pole, proportional to
$1/\omega_\Sigma$; a regular resonant correction to the injection
channel; and a regular off-resonant contribution. By contrast, the
band-off-diagonal EPC sector generates a counterterm, $\gamma^\triangle_{abc}$.
We may therefore write
\begin{align}
\chi^{\rm A}_{abc}
=
\chi^{{\rm EPC\text{-}intra}}_{abc}
+
\chi^{{\rm EPC\text{-}inter}}_{abc}
=
\chi^{\rm inj,sing}_{abc}
+
\chi^{\rm inj,reg}_{abc}
+
\chi^{\rm off,1}_{abc}
+
\gamma^\triangle_{abc}.
\label{eq:chiA-decomp}
\end{align}

Similarly, the susceptibility in Eq.~\eqref{eq:chi-B} can be separated
according to whether it depends on the Abelian, $\mathbb{A}^{nn}_a$,
or non-Abelian, $\mathbb{A}^{mn}_a$ ($m\neq n$), components of the
Mead--Truhlar (MT) gauge connection,
\begin{equation}
\mathbb{A}^{mn}_a
=
i\left\langle \phi_m \left| \partial_{u_a}\phi_n \right. \right\rangle_{\mathbf{u}=0}.
\end{equation}
The corresponding low-output-frequency decomposition of diagram (b)
is
\begin{align}
\chi^{\rm B}_{abc}
=
\chi^{\rm Abelian}_{abc}
+
\chi^{\rm non\text{-}Abelian}_{abc}
=
\chi^{\rm shi}_{abc}
+
\chi^{\rm off,2}_{abc}
+
\gamma^{\rm O}_{abc}.
\label{eq:chiB-decomp}
\end{align}
The Abelian sector gives the resonant shift response together with an
off-resonant contribution, whereas the non-Abelian sector generates
the counterterm $\gamma^{\rm O}_{abc}$, which exactly cancels
$\gamma^\triangle_{abc}$ in the clean, noninteracting limit.

Finally, the remaining contributions from Eqs.~\eqref{eq:chi-C} and
\eqref{eq:chi-D} combine, in the low-output-frequency limit, into a
single off-resonant contribution,
\begin{align}
\chi^{\rm C}_{abc}
+
\chi^{\rm D}_{abc}
=
\chi^{\rm off,3}_{abc}.
\label{eq:chiCD-decomp}
\end{align}
Consequently, after cancellation of the counterterms,
$\gamma^\triangle_{abc}+\gamma^{\rm O}_{abc}=0$, as demonstrated in
Supplementary Note~\ref{app.cancel}, the complete low-output-frequency
response separates into resonant injection and shift channels together
with three off-resonant contributions. It can therefore be written as
\begin{align}
\chi_{abc}
\approx
\chi^{\rm inj,sing}_{abc}
+
\chi^{\rm inj,reg}_{abc}
+
\chi^{\rm shi}_{abc}
+
\sum_{j=1}^{3}\chi^{\rm off,j}_{abc}
+
\mathcal{O}(\omega_\Sigma).
\label{eq:full-lowfreq-decomp}
\end{align}
The resonant injection and shift responses constitute the main focus of
this work, while the complete off-resonant response is analyzed in
Supplementary Note~\ref{app.off_res}. In the remainder of this section,
we develop the quantum-geometric decomposition of the resonant injection
and shift channels, with the injection channel comprising both singular
and regular contributions. We further note that the
responses to linearly polarized (LP) and circularly polarized (CP)
light are obtained from the corresponding symmetric and antisymmetric
optical-index sectors, which reduce to the relevant real and imaginary
parts of the response tensor at rectification \cite{rostami2022}.
\subsection{Quantum geometry of injection contribution}\label{app.inj}

For the band-diagonal EPC sector, $l=m$, and hence
$\omega_{lm}=\omega_{ml}=0$. Setting
$\omega_1=\omega$ and $\omega_2=\omega_\Sigma-\omega$, and relabelling
$m\leftrightarrow n$ in the appropriate terms, we obtain
\begin{align}
\chi^{{\rm EPC-intra}}_{abc}
(\omega_\Sigma;\omega,\omega_\Sigma-\omega)
=&
-\frac{e^2}{2\hbar^2}
\frac{1}{\omega(\omega-\omega_\Sigma)}
\frac{1}{\omega_\Sigma+i\Gamma}
\sum_{\mathbf p}\sum_{mn}
f_{mn}
\left(M_a^{mm}-M_a^{nn}\right)
v_b^{nm}v_c^{mn}
\nonumber\\
&\times
\left[
\frac{1}{\omega+i\gamma-\omega_{nm}}
-
\frac{1}{\omega-\omega_\Sigma-i\gamma-\omega_{nm}}
\right].
\label{eq:EPCintra-low}
\end{align}
The band-off-diagonal EPC sector, meanwhile, reduces in the
low-output-frequency limit to
\begin{align}
\lim_{\omega_\Sigma\to0} \chi^{{\rm EPC-inter}}_{abc}
(\omega_\Sigma;\omega,\omega_\Sigma-\omega)= \gamma^{\triangle}_{abc}(\omega)
&=
-\frac{e^2}{2\hbar^2\omega^2}
\sum_{\mathbf p}\sum_{mnl}
\Bigg\{
\frac{
f_{mn} M_a^{ml} v_c^{ln} v_b^{nm}
}{
(\omega+i\gamma-\omega_{nm})
(i\Gamma-\omega_{lm})
}
\bigg|_{l\neq m}
\nonumber\\
&+
\frac{
f_{mn} M_a^{lm} v_c^{mn} v_b^{nl}
}{
(-\omega+i\gamma-\omega_{mn})
(i\Gamma-\omega_{ml})
}
\bigg|_{l\neq m}
+
[b,\omega\leftrightarrow c,-\omega]
\Bigg\}.
\label{eq:gammaTriangle}
\end{align}
This term is exactly cancelled, in the clean noninteracting limit, by
the counterterm $\gamma^{\rm O}_{abc}$ arising from diagram (b).
The details of this cancellation are presented in Supplementary
Note~\ref{app.cancel}.

We now focus on the band-diagonal EPC sector and take the intrinsic clean limit at fixed nonzero
$\omega_\Sigma$, $\gamma,\Gamma\to0^+$, consistent with the
prescription used throughout the preceding sections, and only then
expand for small $\omega_\Sigma$. The factors entering
Eq.~\eqref{eq:EPCintra-low} therefore admit the expansions
\begin{align}
\frac{1}{\omega(\omega-\omega_\Sigma)}
&= \frac{1}{\omega^2}+\frac{\omega_\Sigma}{\omega^3}+O(\omega_\Sigma^2),
\label{eq.w1w2} \\
\frac{1}{\omega-\omega_{nm}+i0^+}-\frac{1}{\omega-\omega_{nm}-\omega_\Sigma-i0^+} 
&= -2i\pi\delta(\omega-\omega_{nm}) -\omega_\Sigma \left[\mathcal{P}\frac{1}{(\omega-\omega_{nm})^2}-i\pi\delta'(\omega-\omega_{nm})\right]+O(\omega_\Sigma^2).
\label{eq.d1d2}
\end{align}

Substituting Eqs.~\eqref{eq.w1w2} and \eqref{eq.d1d2} into
Eq.~\eqref{eq:EPCintra-low}, the leading terms in the
$\omega_\Sigma$ expansion reproduce the singular injection response.
Introducing the band-resolved quantum geometric tensor
\begin{align}
Q_{bc}^{nm}
=
\frac{v_b^{nm}v_c^{mn}}{\omega_{nm}^2}
=
g_{bc}^{nm}
-\frac{i}{2}\Omega_{bc}^{nm},
\end{align}
where $g_{bc}^{nm}$ and $\Omega_{bc}^{nm}$ denote the band-resolved
(interband) quantum metric and Berry curvature, respectively, we obtain
\begin{equation}
\chi^{\rm inj,sing}_{abc}(\omega_\Sigma\to 0;\omega,-\omega)
=
\frac{\alpha^{\rm inj}_{abc}(\omega)}{i\omega_\Sigma},
\end{equation}
with
\begin{equation}
\alpha^{\rm inj}_{abc}(\omega)
=
-\frac{\pi e^2}{\hbar^2}
\sum_{\mathbf p}\sum_{mn}
f_{mn}\,
\Delta_a^{mn}
Q_{bc}^{nm}\,
\delta(\omega-\omega_{nm}),
\end{equation}
where
$\Delta_a^{mn}=M_a^{mm}-M_a^{nn}$.
The injection susceptibility $\alpha^{\rm inj}_{abc}(\omega)$ enters
the total low-output-frequency response through the singular term in
Eq.~(3) of the main text. Its compact quantum-geometric form closely
parallels the corresponding expression for the injection photocurrent
\cite{Ahn_prx_2020,Watanabe_prx_2021}. The LP and CP injection
responses are obtained from the real and imaginary parts of
$Q_{bc}^{nm}$, respectively.

The remaining resonant contribution is regular in
$\omega_\Sigma$ and arises from the $\delta(\omega-\omega_{nm})$
and $\delta'(\omega-\omega_{nm})$ terms generated by
Eqs.~\eqref{eq.w1w2} and \eqref{eq.d1d2}. Their frequency dependence
can be simplified using the distributional identity
\begin{align}
\frac{\omega_{nm}^{2}}{\omega^{3}}
\delta(\omega-\omega_{nm})
-
\frac{\omega_{nm}^{2}}{2\omega^{2}}
\delta'(\omega-\omega_{nm})
=
-\frac{1}{2}\delta'(\omega-\omega_{nm}),
\label{eq:injRegDistribution}
\end{align}
which follows from the general distributional identity
\begin{equation}
f(x)\delta'(x-a)+ f'(a)\delta(x-a)
=
f(a)\delta'(x-a).
\label{eq:deltaPrimeIdentity}
\end{equation} 
Using the definition of $\alpha^{\rm inj}_{abc}(\omega)$ above, the
regular resonant contribution to the injection channel can therefore
be written compactly as
\begin{align}
\chi^{\rm inj,reg}_{abc}(\omega)
\equiv
\beta^{\text{reg-inj}}_{abc}(\omega)
=
\frac{i}{2}
\frac{\partial\alpha^{\rm inj}_{abc}(\omega)}{\partial\omega}.
\label{eq:injReg}
\end{align}
Thus, the band-diagonal EPC sector of the triangle diagram contains
both the singular injection pole and a regular resonant correction,
with the latter entirely determined by the spectral dispersion of the
injection susceptibility.

The principal-value term in Eq.~\eqref{eq.d1d2} gives rise to an
off-resonant contribution, which is discussed separately in
Supplementary Note~\ref{app.off_res}.

\subsection{Quantum geometry of shift contribution}\label{app.shi}
%
Using the covariant derivative approach \cite{VenturaGaugeCov,Parker} with respect to phonon displacement in the $a$-direction,
\begin{align}
    M_{ab}^{mn} = \langle \phi_m|\hat M_{ab}|\phi_n\rangle =& \partial_{u_a} v_{b}^{mn} - i\comm{\hat{\mathbb{A}}_a}{\hat{v}_b}^{mn} = \partial_{u_a} v_{b}^{mn} -i \sum_l \big(\mathbb{A}_a^{ml} v_b^{ln} - v_b^{ml} \mathbb{A}_a^{ln} \big), \label{eq.PhononCovDev}
\end{align}
where the Mead-Truhlar (MT) connection is defined as
\begin{align}
    \mathbb{A}^{mn}_a 
    = \langle \phi_m | \hat{\mathbb{A}}_a | \phi_n \rangle 
    = i \langle \phi_m | \partial_{u_a} | \phi_n \rangle,
\end{align}
Therefore, the Abelian contribution, whose resonant component gives the shift-force susceptibility, reads  
\begin{align}
    \lim_{\omega_\Sigma\to 0}\chi^{\text{Abelian}}_{abc} = \chi^{\text{shi}}_{abc}(\omega) + \chi^{\text{off,2}}_{abc}(\omega) =
    \frac{-e^2}{2\hbar \omega^2} \sum_{\mathbf{p}} \sum_{mn} f_{mn} \Bigg \{ \frac{\Big(\partial_{u_a} v_c^{mn} -i \big(\mathbb{A}_a^{mm} v_c^{mn} - v_c^{mn} \mathbb{A}_a^{nn} \big) \Big) v^{nm}_b}{\omega + i\gamma + \omega_{mn}} + [b,\omega \leftrightarrow c,-\omega] \Bigg\}\label{eq.InitShift}.
\end{align}
and the non-Abelian counterpart, meanwhile, is given by
\begin{align}
     \lim_{\omega_\Sigma\to 0}\chi^{\text{non-Abelian}}_{abc}=   \gamma^{\rm O}_{abc}(\omega) = \frac{-e^2}{2\hbar\omega^2} \sum_{\mathbf{p}} \sum_{mnl} f_{mn}\Bigg\{ \frac{- i  \mathbb{A}_a^{ml} v_c^{ln} v^{nm}_b \eval_{l \neq m}}{\omega + i\gamma + \omega_{mn}} + \frac{-i \mathbb{A}_a^{lm} v_c^{nl} v^{mn}_b \eval_{l \neq m}}{\omega + i\gamma - \omega_{mn}} + [b,\omega \leftrightarrow c,-\omega] \Bigg\},
\end{align}
which cancels $\gamma^{\triangle}_{abc}(\omega)$ in the clean limit, as discussed in the following Supplementary Note~\ref{app.cancel}. 
Turning back to the Abelian part, we represent velocity matrix elements in polar form, as $v_c^{mn} = \abs{v_c^{mn}} e^{-i \vartheta_c^{mn}}$,
which after some algebra gives us
\begin{align}\label{eq:abel}
\chi^{\text{shi}}_{abc}(\omega) + \chi^{\text{off,2}}_{abc}(\omega)=
     \frac{-e^2}{2\hbar \omega^2} \sum_{\mathbf{p}} \sum_{mn} f_{mn} \left \{ \frac{ \Big( \abs{v^{nm}_b} \partial_{u_a} \abs{v_c^{mn}} e^{-i (\vartheta_c^{mn} + \vartheta^{nm}_b)} -i \omega_{nm}^2 \mathbb{R}_{a;c}^{mn} Q_{bc}^{nm} \Big)}{\omega + i\gamma + \omega_{mn}} + [b,\omega \leftrightarrow c,-\omega] \right \} ,
\end{align}
where the \textit{phononic shift vector}, evaluated at
$\mathbf u=\mathbf 0$, is defined as 
\begin{align}
    \mathbb{R}_{a;c}^{mn} = \partial_{u_a} \vartheta_{c}^{mn} + \mathbb{A}_a^{mm} - \mathbb{A}_a^{nn}, \label{eq.PhSV}
\end{align}
with $\vartheta_{c}^{mn} = -\Im \ln v_{c}^{mn}$.
This expression is invariant under independent
band-dependent $U(1)$ transformations
$|\phi_n\rangle\rightarrow e^{i\theta_n(\mathbf p,\mathbf u)}
|\phi_n\rangle$. Indeed,
\begin{align}
v_c^{mn} = \langle \phi_m|v_c|\phi_n\rangle
\rightarrow e^{i(\theta_n-\theta_m)}v_c^{mn},
\qquad
\mathbb A_a^{nn}=i\langle \phi_n|\partial_{u_a}\phi_n\rangle
\rightarrow\mathbb A_a^{nn}
-\partial_{u_a}\theta_n.
\end{align}
Consequently,
\begin{align}
\mathbb R_{a;c}^{mn}
\rightarrow\mathbb R_{a;c}^{mn}
-\partial_{u_a}(\theta_n-\theta_m)
-\partial_{u_a}\theta_m
+\partial_{u_a}\theta_n
=\mathbb R_{a;c}^{mn}.
\end{align} 
Taking the resonant contribution in the clean limit, and following some basic algebra and use of the relation \mbox{$ v_b^{mn} = i \omega_{mn} \mathcal{A}_b^{mn} $}, which expresses the interband velocity matrix element in terms of the electronic Berry connection, \mbox{$\mathcal{A}_b^{mn} = i\hbar \langle \phi_m | \partial_{p_b} | \phi_n \rangle $}, we obtain the shift-force susceptibility as \mbox{$\beta^{\rm shi}_{abc}(\omega) \equiv \chi^{\rm shi}_{abc}(\omega)$} as 
\begin{align}
    \beta^{\rm shi}_{abc}(\omega)
    = \frac{i\pi e^2}{2\hbar} \sum_{\mathbf{p}} \sum_{mn} f_{mn} \left\{\big( \abs{\mathcal{A}^{nm}_b} \partial_{u_a} \abs{\mathcal{A}_c^{nm}} - \abs{\mathcal{A}^{nm}_c} \partial_{u_a} \abs{\mathcal{A}_b^{nm}} \big) e^{-i (\vartheta^{nm}_b - \vartheta_c^{nm})} -i \big(\mathbb{R}_{a;c}^{mn} + \mathbb{R}_{a;b}^{mn} \big) Q_{bc}^{nm} \right\} \delta(\omega - \omega_{nm}) \label{eq.full_shift_term}.
\end{align}
This expression is rather complicated and opaque, and to make progress, we consider specific cases for the response to LP and CP light separately (in the way outlined above). For LP incoming light, we can, without loss of generality, assume that the input polarizations are identical, i.e., $c=b.$ In this case, the expression simplifies significantly to
\begin{align}
    \beta^{\mathrm{shi}-\mathrm{LP}}_{abb}(\omega)
    = \frac{\pi e^2}{\hbar} \sum_{\mathbf{p}} \sum_{mn} f_{mn} \mathbb{R}_{a;b}^{mn} \Re \left[Q_{bb}^{nm}\right] \delta(\omega - \omega_{nm}),
\end{align}
which resembles the well-known shift current formula expressed in terms of the quantum geometric tensor~\cite{Ahn_prx_2020, Watanabe_prx_2021}, whose real part corresponds to the quantum metric.

For CP light, we reformulate the expression using the chiral Berry connection and the chiral shift vector, as introduced in the context of the gyration (circular shift) current \cite{Watanabe_prx_2021}. The chiral Berry connections are defined as
\begin{align}
    \mathcal{A}_{\pm}^{mn} 
    = \frac{1}{\sqrt{2}} (\mathcal{A}_{b}^{mn} \pm i \mathcal{A}_{c}^{mn}) 
    = |\mathcal{A}_{\pm}^{mn}| e^{-i\vartheta_{\pm}^{mn}} 
    = [\mathcal{A}_{\mp}^{nm}]^*,
\end{align}
where the $\pm$ indices denote the two circular polarizations, and where $\vartheta_\ell^{mn}$ describes the phase of both velocity and Berry connection elements beyond a constant term of $\pi/2$. The corresponding chiral phononic shift vector is given by
\begin{align}
    \mathbb{R}_{a;\pm}^{mn} 
    = \partial_{u_a} \vartheta_{\pm}^{mn} + \mathbb{A}_a^{mm} - \mathbb{A}_a^{nn}, 
    \label{eq.ChiPhSV}
\end{align}
which captures the phase winding of the chiral Berry connection along with the difference in diagonal MT connections.
Using these definitions, the shift-force susceptibility to CP light becomes
\begin{align}
    \beta^{\mathrm{shi}-\mathrm{CP}}_{abc}(\omega)
    = -\frac{\pi e^2}{2\hbar} \sum_{\mathbf{p}} \sum_{mn} f_{mn} 
    \left( |\mathcal{A}_{+}^{mn}|^2 \mathbb{R}_{a;+}^{mn} 
         - |\mathcal{A}_{-}^{mn}|^2 \mathbb{R}_{a;-}^{mn} 
    \right) 
    \delta(\omega - \omega_{nm}).
\end{align}
The squared chiral Berry amplitudes can be further expressed in terms of quantum-geometric quantities as~\cite{Watanabe_prx_2021}
\begin{align}
    |\mathcal{A}_{\pm}^{mn}|^2 
    = \frac{g_{bb}^{mn} + g_{cc}^{mn} \mp \Omega_{bc}^{mn}}{2}.
\end{align}
Substituting this back into the previous expression yields
\begin{align}
    \beta^{\mathrm{shi}-\mathrm{CP}}_{abc}(\omega)
    = -\frac{\pi e^2}{4\hbar} \sum_{\mathbf{p}} \sum_{mn} f_{mn} 
    \left\{ 
        (g_{bb}^{mn} + g_{cc}^{mn}) (\mathbb{R}_{a;+}^{mn} - \mathbb{R}_{a;-}^{mn}) 
        - \Omega_{bc}^{mn} (\mathbb{R}_{a;+}^{mn} + \mathbb{R}_{a;-}^{mn})
    \right\} 
    \delta(\omega - \omega_{nm}),
\end{align}
where we see contributions from both the symmetric (quantum metric) and antisymmetric (Berry curvature) parts of the quantum geometric tensor. This structure highlights the geometric origin of the Raman-force response under circularly polarized excitation, which closely resembles that of the nonlinear photocurrent reported in Ref.~\cite{Watanabe_prx_2021}.

The off-resonant contribution $\chi^{\text{off,2}}_{abc}(\omega)$ originates from the principal-value term
in Eq.~\eqref{eq:abel} and is discussed in Supplementary
Note~\ref{app.off_res}. 
 
\suppnote{Cancellation of counterterms between the triangle and oval diagrams}\label{app.cancel}
In the contributions from the injection and shift diagrams (\suppfigref{fig:feyn}a,b), after removing the parts that contribute to the injection and shift-force susceptibilities, we are left with:
\begin{align}
    \gamma^{\triangle}_{abc}(\omega) + \gamma^{\rm O}_{abc}(\omega)
    = \frac{-e^2}{2\hbar^2\omega^2} \sum_{\mathbf{p}} \sum_{mnl} f_{mn}\Bigg\{&\frac{ M^{ml}_a v^{ln}_c v^{nm}_b \eval_{l \neq m}}{(\omega + i\gamma-\omega_{nm})(i\Gamma-\omega_{lm})} + \frac{ M^{lm}_a v^{mn}_c v^{nl}_b \eval_{l \neq m}}{(-\omega + i\gamma-\omega_{mn})(i\Gamma-\omega_{ml})} 
    \nonumber \\
    +& \frac{- i\hbar \mathbb{A}_a^{ml} v_c^{ln} v^{nm}_b \eval_{l \neq m}}{\omega + i\gamma + \omega_{mn}} + \frac{-i\hbar \mathbb{A}_a^{lm} v_c^{nl} v^{mn}_b \eval_{l \neq m} )}{\omega + i\gamma - \omega_{mn}} + [b,\omega \leftrightarrow c,-\omega] \Bigg\}.
\end{align}
We treat the response in the intrinsic geometric limit, where impurity, disorder, and many-body scattering (dephasing) rates are negligible. The intrinsic response is obtained by first taking the clean limit, 
$\lim_{\{\Gamma,\gamma\} \to 0^+}$, at fixed band gap, and for the gapless case, the gap is sent to zero at the end: 
$\lim_{\text{gap} \to 0} \lim_{\{\Gamma,\gamma\} \to 0^+}$. 
This protocol captures the clean-system cancellation of the first and second lines in the above expression, independent of the magnitude of the band gap---that is, it remains valid even for an infinitesimal gap, as long as the gap exceeds the negligible scattering rates.  Therefore, in this limit, we can neglect the $\Gamma$ term and replace $i\Gamma - \omega_{lm} \to -\omega_{lm}$ in the denominators on the first line, yielding:
\begin{align}
    \gamma^{\triangle}_{abc}(\omega) + \gamma^{\rm O}_{abc}(\omega)
    = \frac{-e^2}{2\hbar^2\omega^2} \sum_{\mathbf{p}} \sum_{mnl} f_{mn}\Bigg\{& \frac{ M^{ml}_a v^{ln}_c v^{nm}_b \eval_{l \neq m}}{(\omega + i\gamma+\omega_{mn})(\omega_{ml})} + \frac{ M^{lm}_a v^{mn}_c v^{nl}_b \eval_{l \neq m}}{(-\omega + i\gamma-\omega_{mn})(\omega_{lm})} 
     \nonumber \\
    +& \frac{-i\hbar \mathbb{A}_a^{ml} v_c^{ln} v^{nm}_b \eval_{l \neq m}}{\omega + i\gamma + \omega_{mn}} + \frac{-i\hbar  \mathbb{A}_a^{lm} v_c^{nl} v^{mn}_b \eval_{l \neq m} }{\omega + i\gamma - \omega_{mn}} + [b,\omega \leftrightarrow c,-\omega] \Bigg\}.
\end{align}
Now, using the identity $M_a^{ml} = i \hbar\omega_{ml} \mathbb{A}_a^{ml}$ for $m \neq l$ in the second line, and then simplifying and explicitly showing all terms, we get
\begin{align}
    \gamma^{\triangle}_{abc}(\omega) + \gamma^{\rm O}_{abc}(\omega)  
    =& \frac{e^2}{2\hbar^2\omega^2} \sum_{\mathbf{p}} \sum_{mnl} \frac{f_{mn}}{\omega_{lm}} \Bigg\{ \frac{M^{ml}_a v^{ln}_c v^{nm}_b}{\omega + i\gamma+\omega_{mn}} - \frac{M^{lm}_a v^{mn}_b v^{nl}_c}{\omega + i\gamma-\omega_{mn}} + \frac{M^{ml}_a v^{ln}_b v^{nm}_c}{-\omega + i\gamma+\omega_{mn}} - \frac{M^{lm}_a v^{mn}_c v^{nl}_b }{-\omega + i\gamma-\omega_{mn}}  
    \nonumber \\
    &\hspace{2.4cm}- \frac{M_a^{ml} v_c^{ln} v^{nm}_b}{\omega + i\gamma + \omega_{mn}} + \frac{M_a^{lm} v_c^{nl} v^{mn}_b}{\omega + i\gamma - \omega_{mn}} - \frac{M_a^{ml} v_b^{ln} v^{nm}_c}{-\omega + i\gamma + \omega_{mn}} + \frac{M_a^{lm} v_b^{nl} v^{mn}_c}{-\omega + i\gamma - \omega_{mn}} \Bigg\}\eval_{l \neq m},
\end{align}
and thus in the clean limit we obtain an identically exact cancellation:
\begin{align}
      \gamma^{\triangle}_{abc}(\omega) + \gamma^{\rm O}_{abc}(\omega) = 0,
\end{align}
as the terms on the top row clearly each cancel with the term directly below it. This reasoning can be compared to the cancellation of the counter terms for the injection and shift photocurrents, following a similar logic to that used in Sec.~III~D of Ref.~\cite{Watanabe_prx_2021}, for example.

\suppnote{Quantum geometry of off-resonant contributions}\label{app.off_res}
Here we examine the complete off-resonant Raman-force response, which
consists of three distinct contributions. The first,
$\beta^{\rm off,1}_{abc}$, originates from the band-diagonal EPC sector
of the triangle diagram in Eq.~\eqref{eq:chi-A}, represented by
\suppfigref{fig:feyn}a, and arises from the principal-value part of its
low-output-frequency expansion. The second,
$\beta^{\rm off,2}_{abc}$, comes from the Abelian sector of
Eq.~\eqref{eq:chi-B}, represented by \suppfigref{fig:feyn}b, which also
generates the resonant shift force. The third,
$\beta^{\rm off,3}_{abc}$, is obtained from the remaining
susceptibilities in Eqs.~\eqref{eq:chi-C} and \eqref{eq:chi-D},
represented by the diagrams in \suppfigref{fig:feyn}c and d.
Together,
\begin{align}
\beta^{\rm off}_{abc}
=
\beta^{\rm off,1}_{abc}
+
\beta^{\rm off,2}_{abc}
+
\beta^{\rm off,3}_{abc}.
\end{align}
These contributions are regular in the low-output-frequency limit and
are dispersive rather than on shell, being governed by principal-value
terms rather than the Dirac delta functions associated with the
resonant injection and shift responses. Except for the nonlinear Drude
term contained in $\beta^{\rm off,3}_{abc}$, they can be expressed in
terms of the quantum geometric tensor, closely paralleling the
geometric structure of the resonant response.

\subsection{Contribution from diagram (a): 
$\beta^{\mathrm{off},1}_{abc}$}
\label{app.off_res3}

We begin with the off-resonant contribution from the triangle diagram
in \suppfigref{fig:feyn}a, whose band-diagonal EPC sector also generates
the resonant injection response. At leading order, taking the
principal-value part of the clean-limit in the singular injection term gives
\begin{align}
\alpha^{\rm off}_{abc}(\omega)
=
-\frac{e^2}{2\hbar^2\omega^2}
\sum_{\mathbf p}\sum_{mn}
f_{mn}\,
\pv\Bigg\{
\frac{
\left(M_a^{mm}-M_a^{nn}\right)
\omega_{nm}^2 Q_{bc}^{nm}
}{
\omega-\omega_{nm}
}
-
\frac{
\left(M_a^{mm}-M_a^{nn}\right)
\omega_{nm}^2 Q_{bc}^{nm}
}{
\omega-\omega_{nm}
}
\Bigg\}
=0 .
\label{eq.InjPVLeadingCancel}
\end{align}
Thus, the leading principal-value counterpart of the singular injection
pole cancels identically. This cancellation, however, does not exclude
a regular off-resonant contribution arising at the next order in the
low-output-frequency expansion.
Indeed, the principal-value term in Eq.~\eqref{eq.d1d2}, which arises
from the same band-diagonal EPC sector responsible for the regular injection
response, gives
\begin{equation}
\beta^{\rm off,1}_{abc}(\omega)
=
\frac{e^2}{2\hbar^2}
\sum_{\mathbf p}\sum_{mn}
f_{mn}\,
\Delta_a^{mn}Q_{bc}^{nm}
\left(\frac{\omega_{nm}}{\omega}\right)^2
\mathcal{P}
\frac{1}{(\omega-\omega_{nm})^2}.
\label{eq:OffRes1}
\end{equation}
Eq.~\eqref{eq:OffRes1} is a regular impulsive contribution arising
from the same band-diagonal EPC sector of the triangle diagram that
generates the singular and regular resonant injection responses.
Unlike those resonant contributions, however,
$\beta^{\rm off,1}_{abc}$ belongs to the off-resonant sector that is governed by a principal-value double pole.
\subsection{Contributions from diagram (b): ${\beta}^{\rm off,2}_{abc}$}
Next, we consider the off-resonant component of the oval diagram in \suppfigref{fig:feyn}b, which serves as the off-resonant analogue to the shift force. Taking the principal value part of the clean limit of Eq.~\eqref{eq.InitShift} yields
\begin{align}
    {\beta}^{\rm off,2}_{abc}(\omega)
    = \frac{-e^2}{2\hbar \omega^2} \sum_{\mathbf{p}} \sum_{mn} f_{mn}\pv\Bigg \{\frac{v^{nm}_b \partial_{u_a} v_c^{mn} -i \big(\mathbb{A}_a^{mm} - \mathbb{A}_a^{nn} \big) v^{nm}_b v_c^{mn} }{\omega - \omega_{nm}} + [b,\omega \leftrightarrow c,-\omega] \Bigg\}.
\end{align}
Writing explicitly the second term and exchanging $m \leftrightarrow n$ in this term yields, after some algebra, an expression in terms of the quantum geometric tensor $Q_{bc}^{nm}$:
\begin{align}
    {\beta}^{\rm off,2}_{abc}(\omega)
    = \frac{-e^2}{2\hbar^2 \omega^2} \sum_{\mathbf{p}} \sum_{mn} f_{mn} \pv\Bigg \{\frac{2 \omega_{nm} \Delta_a^{nm} Q_{bc}^{nm} + \hbar\omega_{nm}^2 \partial_{u_a} Q_{bc}^{nm}}{\omega - \omega_{nm}} \Bigg\}, \label{eq.PVSusc}
\end{align}
where we use that EPC-difference, $\Delta_a^{nm} = \hbar\partial_{u_a} \omega_{nm}$. In the second term, it enters via a displacement derivative, $\partial_{u_a} Q^{mn}_{bc}$: the phononic counterpart to the momentum-derivatives of $Q^{mn}_{bc}$ that define quantum-geometric \emph{dipole moments} in the photogalvanic effect \cite{Watanabe_prx_2021,Anyuan2023}. For LP and CP responses, we respectively call these the phononic dipole moments of quantum metric and Berry curvature.
\subsection{Contributions from diagrams (c) and (d): ${\beta}^{\rm off,3}_{abc}$}
Here we analyze the remaining diagrams in \suppfigref{fig:feyn}c, d, corresponding to the terms given in Eqs.~\eqref{eq:chi-C} and \eqref{eq:chi-D}, which contribute to the susceptibility as
\begin{align}
    {\chi}^{\rm C}_{abc}(\omega) +{\chi}^{\rm D}_{abc}(\omega)=& \frac{-e^2}{2\omega^2} \sum_{\mathbf{p}} \Bigg \{ \sum_m f_m M^{mm}_{a;bc} + \sum_{mn} \frac{f_{mn} M^{mn}_a v^{nm}_{bc}}{i\hbar\Gamma - \varepsilon_{nm}} \Bigg \}.
\end{align}
Using the covariant derivative from Eq.~\eqref{eq.PhononCovDev} in the first term, expressing $M_a^{mn} = i \varepsilon_{mn} \mathbb{A}_a^{mn}$ and following some algebra, we rewrite the expression as
\begin{align}
    {\chi}^{\rm C}_{abc}(\omega) +{\chi}^{\rm D}_{abc}(\omega)
    =& \frac{-e^2}{2\omega^2} \sum_{\mathbf{p}} \Bigg \{ \sum_m f_m \partial_{u_a} v_{bc}^{mm} + \sum_{mn} f_{mn}\bigg[\frac{  M_a^{mn} v_{bc}^{nm}}{\varepsilon_{nm}}  + \frac{ M^{mn}_a v^{nm}_{bc}}{i\hbar\Gamma - \varepsilon_{nm}} \bigg] \Bigg\}.
\end{align}
Assuming a clean system far from band-touching (so that $i\hbar\Gamma$ can be neglected), the last two terms cancel, leaving
\begin{align}
    {\beta}^{\rm off,3}_{abc}(\omega) = 
     {\chi}^{\rm C}_{abc}(\omega) +{\chi}^{\rm D}_{abc}(\omega)=& \frac{-e^2}{2\omega^2} \sum_{\mathbf{p}} \sum_m f_m \partial_{u_a} v_{bc}^{mm}.
\end{align}
Expanding this using the symmetrized (in $p_b$ and $p_c$) momentum covariant derivative \cite{VenturaGaugeCov}, and substituting $\partial_{u_a} v_b^{mm} = \partial_{p_b} M_a^{mm}$ and $v_c^{mn} = -i \omega_{nm} \mathcal{A}_c^{mn}$, we obtain
\begin{align}
    {\beta}^{\rm off,3}_{abc}(\omega)
    = \frac{-e^2}{2\omega^2} \sum_{\mathbf{p}} \Bigg \{ \sum_m f_m \partial_{p_b} \partial_{p_c} M_{a}^{mm} + \frac{1}{2\hbar} \sum_{mn} f_{mn} \partial_{u_a} \bigg(\frac{v_b^{mn} v_{c}^{nm} + v_c^{mn} v_{b}^{nm}}{\omega_{nm}} \bigg) \Bigg \}.
\end{align}
In the first term, the $p$-derivatives can be moved onto $f_m$, while the quantum metric appears in the second, giving
\begin{align}
    {\beta}^{\rm off,3}_{abc}(\omega)
    = \frac{-e^2}{2\omega^2} \sum_{\mathbf{p}} \bigg \{\sum_m \big(\partial_{p_b} \partial_{p_c} f_m \big) M_{a}^{mm} - \frac{1}{\hbar}\sum_{mn} f_{mn} \Big(\frac{\Delta_a^{mn}}{\hbar} g_{bc}^{mn} + \omega_{mn} \partial_{u_a} g_{bc}^{mn} \Big) \bigg\}. \label{eq.IntraSusc}
\end{align}
The first term resembles a higher-order Drude response, which describes the Drude DC photocurrent if the electron-phonon coupling operator is replaced by the velocity operator (i.e., $\hat{M}_a \to \hat{v}_a$), and shows the usual divergence for $\omega\to0$ \cite{Watanabe_prx_2021}.  
The second term, meanwhile, includes the phononic dipole moment of the quantum metric, similar to that in Eq.~\eqref{eq.PVSusc}.
Since this expression is fully real, it contributes only to the response under LP light.
\begin{figure}[ht]
    \centering
\includegraphics[width=1\linewidth]{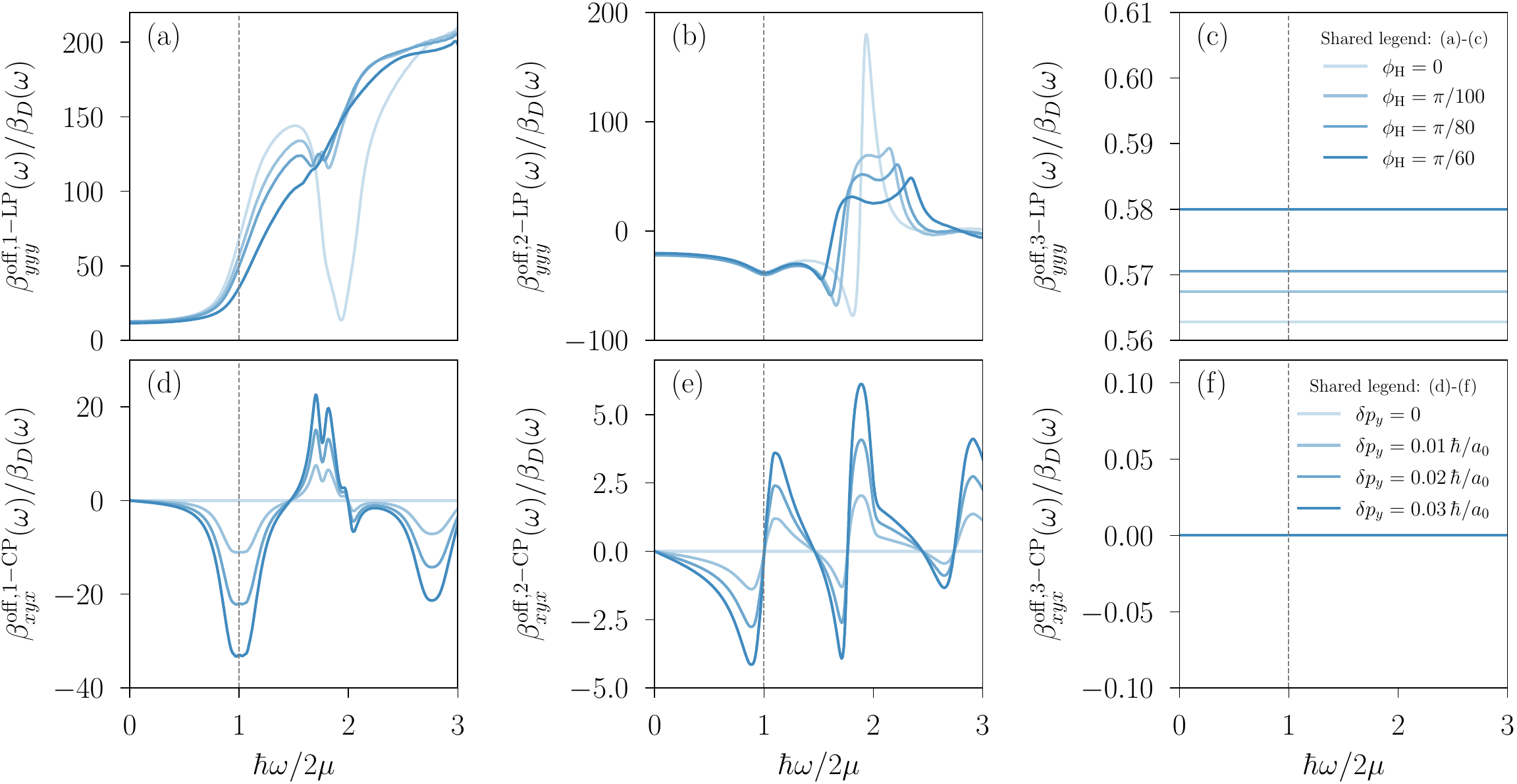}
\caption{
Spectra of the LP components (a--c) and CP components (d--f) of
${\beta}^{\rm off,1}_{abc}(\omega)$ [Eq.~\eqref{eq:OffRes1}],
${\beta}^{\rm off,2}_{abc}(\omega)$ [Eq.~\eqref{eq.PVSusc}], and
${\beta}^{\rm off,3}_{abc}(\omega)$ [Eq.~\eqref{eq.IntraSusc}],
respectively, in the bilayer Haldane model.
Each response is scaled by the frequency-dependent factor
\mbox{$\beta_D(\omega)=(\mu/\hbar\omega)^2\,\mathrm{nN}/\mathrm{V}^2$}.
The LP components (a--c) are obtained by varying the Haldane phase
$\phi_{\rm H}$ with no sublattice gap, $\delta=0$, whereas the CP
components (d--f) are obtained by varying $\delta p_y$, the electronic
momentum shift induced by a source--drain current along the $y$ direction,
with $\delta=0.05\,\mathrm{eV}$.
In panel (f), the result remains zero even when finite values of $\phi_{\rm H}=\pi/80$ and $\delta p_x=0.03\,\hbar/a_0$ are introduced to further enhance symmetry breaking.
The model parameters are listed in Fig.~4 of the main text.
}
    \label{fig:SM_OffRes}
\end{figure}

\suppfigref{fig:SM_OffRes} shows the LP (a--c) and CP (d--f)
components of the three off-resonant contributions in the
bilayer Haldane model. The responses are normalized to remove
the common $\omega^{-2}$ prefactor.
The first two contributions, $\beta^{\mathrm{off},1}_{abc}$
and $\beta^{\mathrm{off},2}_{abc}$, exhibit interband features
at the same transition energies as the resonant responses,
but with distinct spectral profiles.
The third contribution, $\beta^{\mathrm{off},3}_{abc}$,
scales as $\omega^{-2}$ in the LP channel, giving the
frequency-independent normalized curves in panel (c).
Its CP component vanishes identically [panel (f)], as
predicted by Eq.~\eqref{eq.IntraSusc}, even in the presence
of Haldane and sublattice gaps and a current-induced
momentum shift that breaks $C_3$ symmetry.

\subsection{Total off-resonant susceptibility and its two-band dielectric limit}

Here, we reorganize the principal-value terms of the first two off-resonant contributions obtained above into a form that can be directly compared with the length-gauge result. This is achieved using exact partial-fraction decompositions of the frequency-dependent kernels.
In particular,
\begin{align}
    \frac{\omega_{nm}}{\omega^2} \pv\left(\frac{1}{\omega - \omega_{nm}}\right)
    =& -\frac{1}{\omega^2}
    - \frac{1}{\omega_{nm}\omega}
    + \frac{1}{\omega_{nm}}
    \pv\left(\frac{1}{\omega-\omega_{nm}}\right),
    \\
    \frac{\omega_{nm}^2}{\omega^2}
    \pv\left(\frac{1}{(\omega - \omega_{nm})^2}\right)
    =& \frac{1}{\omega^2}
    + \frac{2}{\omega_{nm}\omega}
    - \frac{2}{\omega_{nm}}
    \pv\left(\frac{1}{\omega-\omega_{nm}}\right)
    + \pv\left(\frac{1}{(\omega-\omega_{nm})^2}\right).
\end{align}
Substituting these in, we can rewrite the first two off-resonant susceptibilities as
\begin{align}
    \beta^{\rm off,1}_{abc}(\omega) =& \frac{e^2}{2\hbar^2} \sum_{\mathbf p}\sum_{mn}f_{mn}\,\Delta_a^{mn}Q_{bc}^{nm}\left\{\frac{1}{\omega^2} + \frac{2}{\omega_{nm} \omega} - \frac{2}{\omega_{nm}} \pv\left(\frac{1}{\omega-\omega_{nm}}\right) + \pv\left(\frac{1}{(\omega-\omega_{nm})^2}\right) \right\} \\
    {\beta}^{\rm off,2}_{abc}(\omega)
    =& \frac{e^2}{2\hbar^2} \sum_{\mathbf{p}} \sum_{mn} f_{mn} (2 \Delta_a^{mn} Q_{bc}^{nm} - \hbar\omega_{nm} \partial_{u_a} Q_{bc}^{nm}) \left\{-\frac{1}{\omega^2} - \frac{1}{\omega_{nm}\omega} + \frac{1}{\omega_{nm}} \pv\left(\frac{1}{\omega-\omega_{nm}}\right)\right\}.
\end{align}
Adding these together, we see that many $\Delta_a^{mn}$-dependent terms will cancel, leaving
\begin{align}
    \beta^{\rm off,1+2}_{abc}(\omega) =& \frac{e^2}{2\hbar^2} \sum_{\mathbf p}\sum_{mn}f_{mn} \Bigg[\pv\!\left(\frac{\Delta_a^{mn}Q_{bc}^{nm}}{(\omega-\omega_{nm})^2}\right) \!-\! \pv\!\left(\frac{\hbar\partial_{u_a} Q_{bc}^{nm}}{\omega-\omega_{nm}}\right)\! + \frac{\hbar\partial_{u_a} Q_{bc}^{nm}}{\omega} - \frac{\Delta_a^{mn}Q_{bc}^{nm}}{\omega^2} + \frac{\hbar\omega_{nm} \partial_{u_a} Q_{bc}^{nm}}{\omega^2} \Bigg].
\end{align}
One can prove that the third term is purely Berry-curvature dependent and that the last two terms are purely quantum-metric dependent by splitting them in half and swapping $m$ and $n$ indices in one of the halves. Adding the third off-resonant contribution,
$\beta^{\rm off,3}_{abc}(\omega)$ from Eq.~\eqref{eq.IntraSusc},
then gives the total off-resonant Raman susceptibility:
\begin{align}
    \beta^{\rm off}_{abc}(\omega) =& \frac{e^2}{2} \sum_{\mathbf p}\sum_{mn} \Bigg[\frac{1}{\hbar^2}\pv\!\left(\frac{f_{mn} \Delta_a^{mn}Q_{bc}^{nm}}{(\omega-\omega_{nm})^2}\right) - \frac{1}{\hbar}\pv\!\left(\frac{f_{mn} \partial_{u_a} Q_{bc}^{nm}}{\omega-\omega_{nm}}\right) - \frac{if_{mn} \partial_{u_a} \Omega_{bc}^{nm}}{2\hbar\omega} - \frac{\delta_{nm}\big(\partial_{p_b} \partial_{p_c} f_m \big) M_{a}^{mm}}{\omega^2} \Bigg],
\end{align}
giving us squared and normal principal value contributions, a phononic Berry curvature dipole term and the nonlinear Drude contribution as the four off-resonant terms, just as one obtains from the length gauge \cite{Watanabe_prx_2021}.

Upon taking the two-band insulator with no Fermi surface, and where the EPC operator only contains momentum-independent intraband terms, which is the system we take when obtaining the conventional dielectric limit for the Raman force tensor in Supplementary Note \ref{app.dielectric_tensor}, only the first term from above will contribute. The final term vanishes as derivatives of the Fermi-Dirac function vanish.

For the second and third term, their vanishing is apparent from the length-gauge derivation of Supplementary Note \ref{app.length_gauge}, as these terms come from the second and third term of Eq.~\eqref{eq.FullSuscLengthGauge}, which require non-zero interband EPC elements. One can also check this by using Eq.~\eqref{eq.dpb_MT_conn} to rewrite the phononic quantum geometric dipole moment within our approximated system as 
\begin{align}
    \partial_{u_a} Q_{bc}^{nm} =&\hbar \left( \mathcal{A}_{c}^{mn}\partial_b \mathbb{A}_a^{nm}+ \mathcal{A}_{b}^{nm} \partial_c \mathbb{A}_a^{mn}\right), 
\end{align}
which when substituting back into the relevant terms of $\beta^{\rm off}_{abc}(\omega)$ gives zero for $n=m$ terms due to $f_{mn}$ and also gives zero for $m\neq n$ due to the identity $\mathbb{A}_a^{nm} = -iM_a^{nm}/\varepsilon_{nm}$, with $M_a^{nm}=M_a^{mm}\delta_{nm}$.
Therefore, in this approximated case, the off-resonant response is reduced to
\begin{align}
    \beta^{\rm off}_{abc}(\omega) \approx& -\frac{e^2 \Delta_a^{cv}}{2\hbar^2} \sum_{\mathbf p} \pv\!\left(\frac{Q_{bc}^{cv}}{(\omega-\omega_{cv})^2} + \frac{Q_{bc}^{vc}}{(\omega+\omega_{cv})^2}\right) = \frac{e^2 \Delta_a^{cv}}{2\hbar^2} \sum_{\mathbf p} \dv{\omega}\pv\!\left(\frac{Q_{bc}^{cv}}{\omega-\omega_{cv}} + \frac{Q_{bc}^{vc}}{\omega+\omega_{cv}}\right),
\end{align}
where the $\omega$-derivative form is what is used to obtain the dielectric-limit expression for the Raman force susceptibility. With $c=b$, one can finally reduce further to
\begin{align}
    \beta^{\rm off}_{abb}(\omega) \approx& \frac{e^2 \Delta_a^{cv}}{2\hbar^2} \sum_{\mathbf p} \dv{\omega}\pv\!\left(\frac{2\omega\mathcal{A}_b^{vc} \mathcal{A}_b^{cv}}{\omega^2-\omega_{cv}^2} \right) = \frac{e^2 \Delta_a^{cv}}{2\hbar^2} \sum_{\mathbf p} \dv{\omega}\pv\!\left(\frac{\omega}{\omega_{cv}}\frac{2\omega_{cv}\mathcal{A}_b^{vc} \mathcal{A}_b^{cv}}{\omega^2-\omega_{cv}^2} \right),
\end{align}
using $Q_{bb}^{cv} = Q_{bb}^{vc} = \mathcal{A}_b^{vc} \mathcal{A}_b^{cv}$, which contains exactly the derivative of $\omega/\omega_{cv}$ multiplied to the principal value part of linear electric susceptibility, $\chi_{bb}^{(1)}(\omega)$ of Eq.~\eqref{eq.chi_abb_chi_bb}. Therefore, the off-resonant part of the dielectric-limit Raman-force susceptibility evaluated in Supplementary Note \ref{app.dielectric_tensor} from the length-gauge results is consistent with our velocity-gauge results, along with the resonant term. The resulting expressions in terms of the dielectric tensor, $\epsilon_{bb}$ of Eq.~\eqref{eq.SM_dielectric_lim_final} and main text Eq.~(20) therefore follow directly from the velocity-gauge results.
 
\suppnote{Numerically-tractable form of the phononic shift vector}\label{app.shift_vector}
Here, we discuss how we obtain the phononic shift vector given by Eq.~\eqref{eq.PhSV} in our numerical computation. It is difficult to evaluate directly due to the derivative with respect to phonon displacement, so we find an alternative form. We note that this method is easily transferable to compute the usual shift vector describing the shift photocurrent.
We start with the numerically-accessible term $M_{ab}^{mn} v_b^{nm}$, which upon expansion with the phononic covariant derivative of Eq.~\eqref{eq.PhononCovDev} gives us, after use of the polar form for $v_b^{nm}$ and some algebra, that
\begin{align}
    M_{ab}^{mn} v_b^{nm} 
    =& (\partial_{u_a} \abs{v_{b}^{mn}}) \abs{v_b^{nm}} e^{-i(\vartheta_{b}^{mn} + \vartheta_{b}^{nm})} -i (\partial_{u_a} \vartheta_{b}^{mn}) v_{b}^{mn} v_b^{nm} - i v_{b}^{mn} v_b^{nm} (\mathbb{A}_a^{mm}  - \mathbb{A}_a^{nn}) 
   \nonumber\\ 
    -& i\sum_l \left\{\mathbb{A}_a^{ml} v_{b}^{ln} \Big|_{l \neq m} - \mathbb{A}_a^{ln} v_{b}^{ml} \Big|_{l \neq n}\right\} v_b^{nm},
\end{align}
where the phases (which are antisymmetric in $n$ and $m$) cancel in the first term, and we see the phononic shift vector as defined in Eq.~\eqref{eq.PhSV} appear between the second and third terms. Hence, we have
\begin{align}
    M_{ab}^{mn} v_b^{nm} =& \abs{v_b^{nm}}\partial_{u_a} \abs{v_{b}^{mn}}  -i \abs{v_{b}^{mn}}^2 \mathbb{R}_{a;b}^{mn} - i\sum_l \left\{\mathbb{A}_a^{ml} v_{b}^{ln} \Big|_{l \neq m} - \mathbb{A}_a^{ln} v_{b}^{ml} \Big|_{l \neq n}\right \} v_b^{nm}.
\end{align}
As the first term is fully real, the second is fully imaginary, while the third is, in general, complex, we can take just the imaginary part of both sides to get
\begin{align}
   \Im( M_{ab}^{mn} v_b^{nm}) =& - \abs{v_{b}^{mn}}^2 \mathbb{R}_{a;b}^{mn} - \Im\bigg\{i\sum_l (\mathbb{A}_a^{ml} v_{b}^{ln} \eval_{l \neq m} - \mathbb{A}_a^{ln} v_{b}^{ml} \eval_{l \neq n}) v_b^{nm}\bigg\}.
\end{align}
In the second term on the right-hand side, we can use $M_a^{mn} = i \varepsilon_{mn} \mathbb{A}_a^{mn}$ for $m \neq n$ to eliminate the non-Abelian MT connection.
Then, by simple rearrangement, we finally obtain the shift vector as
\begin{align}
    \mathbb{R}_{a;b}^{mn} =& - \frac{\Im( M_{ab}^{mn} v_b^{nm}) + \Im \sum_l \bigg \{\frac{M_a^{ml} v_{b}^{ln} v_b^{nm}}{\varepsilon_{ml}} \eval_{l \neq m} - \frac{M_a^{ln} v_{b}^{ml} v_b^{nm}}{\varepsilon_{ln}} \eval_{l \neq n} \bigg \} }{\abs{v_{b}^{mn}}^2},
\end{align}
which is easily calculable numerically.
We can perform similar analysis to obtain the chiral phononic shift vector of Eq.~\eqref{eq.ChiPhSV} in a numerically accessible form, which leads to
\begin{align}
     \mathbb{R}_{a;\pm}^{mn} =& - \frac{\Im(M_{a;\pm}^{mn} v_\mp^{nm}) + \Im \sum_l \bigg\{\frac{M_a^{ml} v_{\pm}^{ln} v_\mp^{nm}}{\varepsilon_{ml}} \eval_{l\neq m} - \frac{M_a^{ln}v_{\pm}^{ml} v_\mp^{nm}}{\varepsilon_{ln}} \eval_{l\neq n}\bigg\}}{\abs{v_{\pm}^{mn}}^2}.
\end{align}
Both the linear and chiral phononic shift vector expressions can be simplified further by canceling a factor of $v_b^{nm}$ and $v_{\mp}^{nm}$, respectively. The two can then be written in a unified form, given by
\begin{align}
    \mathbb{R}_{a;\ell}^{mn} =& -{\rm Im} \left\{\frac{M_{a\ell}^{mn}}{v_\ell^{mn}} + \sum_{l \neq m} \frac{M_a^{ml} v_{\ell}^{ln}}{\varepsilon_{ml} v_{\ell}^{mn}} - \sum_{l \neq n} \frac{M_a^{ln} v_{\ell}^{ml} }{\varepsilon_{ln} v_{\ell}^{mn}} \right\},
\end{align}
with $\ell \rightarrow b$ for the linear and $\ell \rightarrow \pm$ for the chiral case. This expression can be simplified for the low-energy two-band system which we use later in the main text, where sums over bands $l$ each give only one term, and where $\hat{M}_a$ has no $p$-dependence. In this case, we get
\begin{align}
    \mathbb{R}_{a;\ell}^{mn,\text {2-band}} =& {\rm Im} \left\{ \frac{M_a^{mn} (v_{\ell}^{mm}-v_{\ell}^{nn})}{\varepsilon_{mn} v_{\ell}^{mn}} \right\}.
\end{align}

\suppnote{Length-gauge derivation of the quantum-geometric Raman force}\label{app.length_gauge}

To begin, we write the length-gauge many-body Hamiltonian in terms of
fermionic field operators:
\begin{equation}
\hat{H}(t)
=
\int d^d\mathbf r\,
\hat{\Psi}^{\dagger}(\mathbf r)
\left[
\hat{H}_{e}
+
e\,\mathbf E(t)\cdot\mathbf r
\right]
\hat{\Psi}(\mathbf r),
\label{eq.Hamiltonian}
\end{equation}
where $\hat{\Psi}(\mathbf r)$ is the fermionic field operator,
$\hat{h}_{e}$ is the single-particle electronic Hamiltonian, and
$e>0$, such that the electronic charge is $-e$. The term
$e\,\mathbf E(t)\cdot\mathbf r$ is the electric-dipole light--matter
interaction in the length gauge.
Expanding the field operator in Bloch eigenstates,
\begin{equation}
\hat{\Psi}(\mathbf r)
=
\sum_{n,\mathbf p}
\phi_{n\mathbf p}(\mathbf r)\,
\hat{c}_{n\mathbf p},
\qquad
\phi_{n\mathbf p}(\mathbf r) = \langle {\bf r}|\phi_{n{\bf p}}\rangle  
=
e^{i\mathbf p\cdot\mathbf r/\hbar}
u_{n\mathbf p}(\mathbf r),
\end{equation}
the Hamiltonian in the band basis becomes
\begin{align}
\hat{H}(t)
={}&
\sum_{n,\mathbf p}
\varepsilon_n(\mathbf p) 
\hat{c}^{\dagger}_{n\mathbf p}
\hat{c}_{n\mathbf p}
+
e\,\mathbf E(t)\cdot
\sum_{nm}\sum_{\mathbf p,\mathbf p'}
\hat{c}^{\dagger}_{n\mathbf p}
\matrixel{\phi_{n\mathbf p}}{\hat{\mathbf r}}{\phi_{m\mathbf p'}}
\hat{c}_{m\mathbf p'} .
\label{eq.Hamiltonian_band}
\end{align}
Here, $\varepsilon_n(\mathbf p)$ is an eigenvalue of the field-free
electronic Hamiltonian $\hat H_e$, with the corresponding Bloch
eigenstate $\ket{\phi_{n\mathbf p}}$ satisfying
\begin{equation}
\hat H_e\ket{\phi_{n\mathbf p}}
=
\varepsilon_n(\mathbf p)\ket{\phi_{n\mathbf p}}.
\end{equation}
The index $n$ labels electronic band, while $\mathbf p$
denotes the crystal momentum. The position operator is not diagonal in crystal momentum
and must be treated through its full momentum-space kernel \cite{rostami2018}
\begin{equation}
\matrixel{\phi_{n\mathbf p}}{\hat{\mathbf r}}{\phi_{m\mathbf p'}}
=
i\hbar\,\delta_{nm}\,
\nabla_{\mathbf p}\delta(\mathbf p-\mathbf p')
+
\boldsymbol{\mathcal{A}}_{nm}(\mathbf p)\,
\delta(\mathbf p-\mathbf p'),
\label{eq.position_operator}
\end{equation}
where
\begin{align}
{\mathcal{A}}_a^{nm}(\mathbf p)
=
i\hbar
\left\langle
u_{n\mathbf p}
\middle|
\partial_{p_a}u_{m\mathbf p}
\right\rangle .
\end{align}
is the non-Abelian Berry connection component in the band basis. It is useful to consider the action of the position operator on an
arbitrary state $|\phi_f\rangle$, projected onto the Bloch state $\ket{\phi_n,\mathbf p}$:
\begin{align}
\langle \phi_{n\mathbf p}|\hat{\mathbf r}|\phi_f\rangle
&=
\sum_{m,\mathbf p'}
\langle \phi_{n\mathbf p}|\hat{\mathbf r}|\phi_{m\mathbf p'}\rangle
\langle \phi_{m\mathbf p'}|\phi_f\rangle
\nonumber\\
&=
\sum_{m,\mathbf p'}
\left[
i\hbar\,\delta_{nm}\nabla_{\mathbf p}
\delta(\mathbf p-\mathbf p')
+
\boldsymbol{\mathcal{A}}_{nm}(\mathbf p)
\delta(\mathbf p-\mathbf p')
\right]
\langle \phi_{m\mathbf p'}|\phi_f\rangle
\nonumber\\
&=
i\hbar\,\nabla_{\mathbf p}
\langle \phi_{n\mathbf p}|\phi_f\rangle
+
\sum_m
\boldsymbol{\mathcal{A}}_{nm}(\mathbf p)
\langle \phi_{m\mathbf p}|\phi_f\rangle
\nonumber\\
&=
\sum_m
\left[
i\hbar\,\delta_{nm}\nabla_{\mathbf p}
+
\boldsymbol{\mathcal{A}}_{nm}(\mathbf p)
\right]
\langle \phi_{m\mathbf p}|\phi_f\rangle .
\end{align}
Therefore, one may use the following notation for the band-basis momentum-diagonal notation for the position operator components \cite{Watanabe_prx_2021,Blount}:
\begin{align}\label{eq:Blount_r}
\mathbf{r}_{nm}(\mathbf{p})
\equiv
i\hbar\delta_{nm}\nabla_{\mathbf p}
+\boldsymbol{\mathcal{A}}_{nm}(\mathbf p). 
\end{align}

We now give the derivation of the injection and shift RF susceptibilities in the length gauge, following a similar process to \cite{Watanabe_prx_2021}. The atomic force is the electronic expectation value of the EPC operator, which can be written as a trace over the reduced density matrix (RDM), $\hat{\rho}(t)$, as
\begin{align}
    \mathbf{F}(t) =- \left\langle \frac{\partial H}{\partial \mathbf{u}} \right\rangle = - \Tr[\hat{\mathbf{M}}\hat{\rho}(t)] =- \sum_{\mathbf p}\sum_{nm} \mathbf{M}_{mn}(\mathbf{p}) \rho_{nm}(\mathbf{p},t), \label{eq.Force_RDM}
\end{align}
which is similar in meaning to Eq.~\eqref{eq:F} (written in terms of the RDM instead of Green's functions), and where the time-dependence of the RDM is determined by the Liouville equation:
\begin{align}
    i\hbar \pdv{\rho_{nm}(\mathbf{p},t)}{t} = \comm{\hat{H}(\mathbf{p},t)}{\hat{\rho}(\mathbf{p},t)}_{nm}. \label{eq.LiouvilleEq}
\end{align}
The time-dependence comes exclusively from the incident electric fields, $\mathbf{E}(t)$, and the Hamiltonian excludes the electron-phonon coupling term, as this would result in $\mathbf{F}\sim\order{\hat{\mathbf{M}}^2}$, leaving only the electronic and dipole components.
By splitting the two remaining components of the Hamiltonian, rearranging and Fourier transforming, we obtain the recursive equation for the RDM,
\begin{align}
    \left(\omega - \omega_{nm}\right)\rho^{(i+1)}_{nm}(\mathbf{p},\omega)=& \frac{e}{\hbar} \int \frac{\dd\omega_i}{2\pi} E_{b}(\omega-\omega_i) \mel**{\phi_{n\mathbf{p}}}{\comm{\hat{r}_{b}}{\hat{\rho}^{(i)}(\omega_i)}}{\phi_{m\mathbf{p}}} ,\label{eq.recursiveRDM}
\end{align}
where $\hat{\rho}^{(i)}(\omega_i)$ is the RDM term which is to the $i$th order in $\mathbf{E}$, where $\omega_i$ is the total of all $i$ electric field frequencies, and using the Blount position operator notation~\cite{Blount} defined in Eq.~\eqref{eq:Blount_r}. 
We solve Eq.~\eqref{eq.recursiveRDM} iteratively to obtain increasing orders of the RDM, using the zero-order term $\rho_{nm}^{(0)}(\mathbf{p},\omega) = 2\pi \delta(\omega) f_{n}(\mathbf{p}) \delta_{nm}$ as our starting point.
Following two iterations, and symmetrizing the result for the two electric fields, we obtain the second-order RDM term as
\begin{align}
\rho^{(2)}_{nm}(\mathbf{p},\omega_\Sigma) =&\,
\frac{e^2}{2\hbar^2} \int \frac{\dd\omega_1}{2\pi} 
\Bigg[
\Bigg( \frac{(i\hbar)^2 (\partial_c\partial_b f_m)\delta_{nm}}{(\omega_\Sigma+i\Gamma) (\omega_1+i\gamma)} \nonumber\\
&\quad
+\frac{i\hbar(\partial_b f_{mn})\mathcal{A}_c^{nm}}{(\omega_\Sigma+i\Gamma-\omega_{nm})(\omega_1+i\gamma)} \nonumber\\
&\quad
+\frac{i\hbar}{\omega_\Sigma+i\Gamma-\omega_{nm}}  \partial_c\left(\displaystyle \frac{f_{mn}\mathcal{A}_b^{nm}}{\omega_1+i\gamma-\omega_{nm}}\right) \nonumber\\
&\quad
+\sum_l \frac{1}{\omega_\Sigma+i\Gamma-\omega_{nm}}\left[\frac{f_{ml}\mathcal{A}_c^{nl}\mathcal{A}_b^{lm}}{\omega_1+i\gamma-\omega_{lm}} -\frac{f_{ln}\mathcal{A}_b^{nl}\mathcal{A}_c^{lm}}{\omega_1+i\gamma-\omega_{nl}}\right] \Bigg)\nonumber\\
&\qquad\times
E_b(\omega_1)E_c(\omega_2) + [b,\omega_1\leftrightarrow c,\omega_2]\Bigg],
\label{eq:rho2}
\end{align}
with $\omega_\Sigma=\omega_1+\omega_2$, and including the small imaginary components $\gamma$ and $\Gamma$ as we did before. Substituting this into Eq.~\eqref{eq.Force_RDM}, we obtain the full expression for the susceptibility of the Raman force: the EPC-induced atomic force second-order in $\mathbf{E}$, as
\begin{align}
\chi_{abc} (\omega_1,\omega_2)
=&\,
-\frac{e^2}{2\hbar^2}\sum_{\mathbf p}\Bigg[
\sum_m\frac{(i\hbar)^2 M_a^{mm}(\partial_c\partial_b f_m)}{(\omega_\Sigma+i\Gamma)(\omega_1+i\gamma)}
\nonumber\\
&\quad
+\sum_{mn}\frac{i\hbar(\partial_b f_{mn})M_a^{mn}\mathcal{A}_c^{nm}}{(\omega_\Sigma+i\Gamma-\omega_{nm})(\omega_1+i\gamma)}
\nonumber\\
&\quad
+\sum_{mn}\frac{i\hbar M_a^{mn}}{\omega_\Sigma+i\Gamma-\omega_{nm}} \partial_c\left(\displaystyle\frac{f_{mn}\mathcal{A}_b^{nm}}{\omega_1+i\gamma-\omega_{nm}}\right)
\nonumber\\
&\quad
-\sum_{mnl}\left[\frac{f_{mn}M_a^{nl}\mathcal{A}_c^{lm}\mathcal{A}_b^{mn}}{(\omega_\Sigma+i\Gamma-\omega_{ln})(\omega_1+i\gamma-\omega_{mn})} - \frac{f_{mn}M_a^{lm}\mathcal{A}_b^{mn}\mathcal{A}_c^{nl}}{(\omega_\Sigma+i\Gamma-\omega_{ml})(\omega_1+i\gamma-\omega_{mn})}
\right]
\nonumber\\
&\quad
+[b,\omega_1\leftrightarrow c,\omega_2]
\Bigg].
\label{eq.FullSuscLengthGauge}
\end{align}
Here, we consider only the third and fourth contributions from above, which give the resonant interband contributions in the clean limit of $\gamma,\Gamma\rightarrow 0$, leading to the injection and shift responses in the rectified case of $\omega_\Sigma\rightarrow 0$.

\subsection{Length-gauge injection force}

First, we consider the fourth and final term, taking $\omega_1=\omega$ and $\omega_2=\omega_\Sigma-\omega$, with rectification corresponding to $\omega_\Sigma\to0$, and keep only the terms in the sum with diagonal elements of $\hat{M}_a$ (which we denote with superscript d), which leads to
\begin{align}
\chi_{abc}^{(\text{IV};\text{d})} (\omega,\omega_\Sigma-\omega)
=&\,\frac{e^2}{2\hbar^2}\lim_{\omega_\Sigma\to0}\frac{1}{\omega_\Sigma+i\Gamma}\sum_{\mathbf p}\sum_{mn}
\Bigg[\frac{f_{mn}\Delta_a^{nm}\mathcal{A}_c^{nm}\mathcal{A}_b^{mn}}{\omega+i\gamma-\omega_{mn}} +\frac{f_{mn}\Delta_a^{nm}\mathcal{A}_b^{nm}\mathcal{A}_c^{mn}}{\omega_\Sigma-\omega+i\gamma-\omega_{mn}}\Bigg],
\end{align}
with $\Delta_a^{mn} = M^{mm}_a - M^{nn}_a$. We discuss the remaining off-diagonal portion later. 
As in the velocity gauge, we Taylor expand the second fraction in small $\omega_\Sigma$: 
\begin{align}\label{eq:taylor_inj}
  \lim_{\Gamma\to0}\frac{1}{\omega_\Sigma+i\Gamma}  \frac{1}{\omega_\Sigma-\omega+i\gamma-\omega_{nm}}
   \approx  \lim_{\Gamma\to0}\frac{1}{\omega_\Sigma+i\Gamma} 
   \frac{1}{-\omega+i\gamma-\omega_{nm}} - \lim_{\Gamma\to0}\frac{\omega_\Sigma}{\omega_\Sigma+i\Gamma}  \frac{1}{(-\omega+i\gamma-\omega_{nm})^2 } + {\cal O}(\omega_\Sigma). 
\end{align}
In the rectification limit, the zeroth- and first-order terms yield the singular and regular contributions, respectively, which we now consider in turn   

\subsubsection{Singular (conventional) injection contribution}
The zeroth-order term in the expansion of Eq.~\eqref{eq:taylor_inj} gives
\begin{align}
    \chi_{abc}^{(\text{IV};\text{d};0)}(\omega) =& \frac{e^2}{2\hbar^2}\lim_{\omega_\Sigma\to0} \lim_{\Gamma\to 0}\frac{1}{\omega_\Sigma+i\Gamma} \sum_{\mathbf p} \sum_{mn} \left[\frac{f_{mn} \Delta_a^{nm} \mathcal{A}_c^{nm} \mathcal{A}_b^{mn}}{\omega + i\gamma - \omega_{mn}} + \frac{f_{mn} \Delta_a^{nm} \mathcal{A}_b^{nm} \mathcal{A}_c^{mn}}{-\omega + i\gamma - \omega_{mn}}\right],
\end{align}
We first take the clean limit $\gamma,\Gamma\to0^+$ at fixed nonzero
$\omega_\Sigma$, and only then take the rectification limit
$\omega_\Sigma\to0$. Using
$1/(x+i0^+)=\pv(1/x)-i\pi\delta(x)$ and exchanging the band labels
$m\leftrightarrow n$ in one of the two terms, the principal-value
contributions cancel, whereas the resonant delta-function contributions
add. This yields
\begin{align}
\chi_{abc}^{(\text{IV};\text{d};0)}(\omega) =& \frac{-i\pi e^2}{\hbar^2} \lim_{\omega_\Sigma\to0}\frac{1}{\omega_\Sigma} \sum_{\mathbf p} \sum_{mn} f_{mn} \Delta_a^{nm} \mathcal{A}_c^{mn} \mathcal{A}_b^{nm} \delta(\omega - \omega_{nm}).
\end{align}

Multiplying by \(i\omega_\Sigma\) before taking the rectification limit
\(\omega_\Sigma\to 0\), and using the band-resolved quantum geometric tensor
\(Q_{bc}^{nm}=\mathcal A_b^{nm}\mathcal A_c^{mn}\), we obtain
\begin{equation}
    \alpha^{\rm inj}_{abc}(\omega) = \frac{\pi e^2}{\hbar^2} \sum_{\mathbf p} \sum_{mn} f_{mn} \Delta_a^{nm} Q_{bc}^{nm} \delta(\omega-\omega_{nm}) .
\end{equation}
Equivalently,
\begin{equation}
    \chi^{\rm IV;d}_{abc}(\omega) = \lim_{\omega_\Sigma\to 0} \frac{\alpha^{\rm inj}_{abc}(\omega)}{i\omega_\Sigma}.
\end{equation}
This is the exact resonant displacive susceptibility describing the injection
Raman force in the length gauge. Its linearly and circularly polarized components are given in Eqs.~(7) and (12) of the main text, respectively. 

\subsubsection{Regular injection contribution}
The first-order term in the expansion of Eq.~\eqref{eq:taylor_inj} gives the regular injection contribution,
\begin{align}
\chi_{abc}^{(\mathrm{IV};\mathrm{d};1)}(\omega)
=&\,\frac{e^2}{2\hbar^2}
\lim_{\omega_\Sigma\to0}
\frac{\omega_\Sigma}{\omega_\Sigma+i\Gamma}
\sum_{\mathbf p}\sum_{mn}
\partial_\omega
\Bigg[
\frac{
f_{mn}\Delta_a^{nm}
\mathcal{A}_b^{nm}\mathcal{A}_c^{mn}}
{\omega-i\gamma-\omega_{nm}}
\Bigg].
\end{align}
As above, we first take the clean limit $\gamma,\Gamma\to0^+$ at fixed nonzero
$\omega_\Sigma$, followed by the rectification limit $\omega_\Sigma\to0$.
Retaining the resonant delta-function contribution and introducing the
band-resolved quantum geometric tensor then gives
\begin{align}
\chi_{abc}^{(\mathrm{IV};\mathrm{d};1)}(\omega)
=&\,\frac{i\pi e^2}{2\hbar^2}
\sum_{\mathbf p}\sum_{mn}
\partial_\omega
\Big[
f_{mn}\Delta_a^{nm}
Q_{bc}^{nm}
\delta(\omega-\omega_{nm})
\Big].
\end{align}
Using the definition of the injection susceptibility, this can be written as
\begin{align}
\beta_{abc}^{\mathrm{reg-inj}}(\omega)
=
\chi_{abc}^{(\mathrm{IV};\mathrm{d};1)}(\omega)
=&\,\frac{i}{2}\,
\partial_\omega\alpha^{\rm inj}_{abc}(\omega),
\end{align}
in agreement with the corresponding velocity-gauge result.

\subsection{Length-gauge shift force}
We now consider the third term in the full expression for \(\chi_{abc}\). In
the rectification limit \(\omega_\Sigma\to 0\), we integrate by parts to move
the momentum derivative, use the off-diagonal ($m\neq n$) identity
\(M_a^{mn}=i\hbar\omega_{mn}\mathbb A^a_{mn}\) (the diagonal sector does not contribute, since \(f_{mn}=0\) for \(m=n\)), and then take
\(\Gamma\to 0\). This leaves
\begin{align}
    \chi_{abc}^{(\text{III})}(\omega) =& -\frac{e^2}{2} \sum_{\mathbf p} \sum_{mn} \left[\frac{f_{mn} \mathcal{A}_b^{nm} (\partial_c \mathbb{A}_a^{mn})}{\omega + i\gamma - \omega_{nm}} + \frac{f_{mn} \mathcal{A}_c^{nm} (\partial_b \mathbb{A}_a^{mn})}{-\omega + i\gamma - \omega_{nm}} \right].
\end{align}
From here, we substitute
\begin{align}
    \hbar\partial_b \mathbb{A}_a^{mn} = \partial_{u_a} \mathcal{A}_b^{mn} + i\mathcal{A}_b^{mn} (\mathbb{A}_a^{nn} - \mathbb{A}_a^{mm}) + \sum_l \left(\frac{\mathcal{A}_b^{ml} M_a^{ln}}{\hbar\omega_{ln}} \eval_{l\neq n} - \frac{M_a^{ml} \mathcal{A}_b^{ln}}{\hbar\omega_{ml}} \eval_{l\neq m}\right), \label{eq.dpb_MT_conn}
\end{align}
which can be derived by expanding $\mathbb{A}_a^{mn}$ in Dirac notation, using the product rule to act $\partial_b$ on this and also to move $\partial_{u_a}$. 
Keeping only the first two terms of this substituting expression, we obtain the effective diagonal (d) portion of this susceptibility term, given by
\begin{align}
    \chi_{abc}^{(\text{III};\text{d})}(\omega)=&- \frac{e^2}{2\hbar}\sum_{\mathbf p} \sum_{mn} \left[\frac{f_{mn} \mathcal{A}_b^{nm} (\partial_{u_a} \mathcal{A}_c^{mn} + i\mathcal{A}_c^{mn} (\mathbb{A}_a^{nn} - \mathbb{A}_a^{mm}))}{\omega + i\gamma - \omega_{nm}} + \frac{f_{mn} \mathcal{A}_c^{nm} (\partial_{u_a} \mathcal{A}_b^{mn} + i\mathcal{A}_b^{mn} (\mathbb{A}_a^{nn} - \mathbb{A}_a^{mm}))}{-\omega + i\gamma - \omega_{nm}} \right],
\end{align}
with the remaining off-diagonal portion canceling with the off-diagonal term from $\chi_{abc}^{(\text{IV})}$ in exactly the way shown in Supplementary Note~\ref{app.cancel} for the velocity gauge derivation. Turning back to the diagonal component, the resonant (delta function-containing) contribution from taking the clean limit is
\begin{align}
    \chi_{abc}^{(\text{III};\text{d},\delta)}(\omega) =& \frac{i\pi e^2}{2\hbar} \sum_{\mathbf p} \sum_{mn} f_{mn} \Big(\big(\abs{\mathcal{A}_b^{nm}} \partial_{u_a} \abs{\mathcal{A}_c^{mn}} - \abs{\mathcal{A}_c^{mn}} \partial_{u_a} \abs{\mathcal{A}_b^{nm}} \big) e^{-i(\vartheta_b^{nm} - \vartheta_c^{nm})} \!- \!i \left(\mathbb{R}_{a;c}^{mn} + \mathbb{R}_{a;b}^{mn}\right) Q_{bc}^{nm}  \Big)  \delta(\omega - \omega_{nm}) ,
\end{align}
upon writing in terms of the quantum geometric tensor and using the definition of the phononic shift vector, $\mathbb{R}_{a;c}^{mn}$, of Eq.~\eqref{eq.PhSV} and Eq.~(10) of the main text (and where the Berry connections in polar form are $\mathcal{A}_b^{nm}=\abs{\mathcal{A}_b^{nm}}e^{-i\vartheta_b^{nm}}$).
Comparing with Eq.~\eqref{eq.full_shift_term}, we identify the expression
above as the resonant impulsive shift-force susceptibility,
\begin{equation}
    \chi_{abc}^{({\rm III};{\rm d},\delta)}(\omega) = \beta^{\rm shi}_{abc}(\omega).
\end{equation}
Considering this, one can follow the exact derivation of the ``Shift force'' subsection of
Supplementary Note~\ref{app.rect_force} to obtain the linear and circular shift responses described respectively by Eqs.~(7) and (13) of the main text. 

Thus, we conclude that both the shift and injection-force susceptibilities are obtained
consistently within the length-gauge formulation.

\suppnote{Derivation of the conventional dielectric-function limit} \label{app.dielectric_tensor}

Here, we take our expression for the Raman-force susceptibility derived in the length gauge and apply the approximations and assumptions from Ref. \cite{Stevens2002}, which they use to be able to write in terms of the dielectric tensor.
First, we assume that there is no Fermi surface and that we are at zero temperature (so that derivatives of $f_n$ vanish), and that only intraband matrix elements of the EPC tensor exist ($M_a^{nm}=M_a^{nn}\delta_{mn}$). In this scenario, Eq.~\eqref{eq.FullSuscLengthGauge} is heavily simplified to
\begin{align}
    \chi_{abc} \approx& \frac{e^2}{2\hbar^2}\frac{1}{\omega_\Sigma + i\Gamma} \sum_{\mathbf{p}} \sum_{mn} \Bigg(
    \frac{f_{mn} (M_a^{nn} - M_a^{mm}) \mathcal{A}_b^{mn} \mathcal{A}_c^{nm}}{\omega_1 + i\gamma - \omega_{mn}} + \frac{f_{mn} (M_a^{nn} - M_a^{mm})\mathcal{A}_b^{mn} \mathcal{A}_c^{nm}}{\omega_2 + i\gamma - \omega_{nm}} \Bigg).
\end{align}
We now make a further approximation, taking a two-band insulator model with one valence band, $v$ and conduction band $c$, with $f_v=1$ and $f_c=0$, and that $M_a^{cc} - M_a^{vv} = \overline{\Delta}_a$, which we take as momentum-independent. In this case, we get
\begin{align}
    \chi_{abc} \approx& \frac{e^2 \overline{\Delta}_a}{2\hbar^2} \frac{1}{\omega_\Sigma + i\Gamma} \sum_{\mathbf{p}} \Bigg(
    \frac{\mathcal{A}_b^{vc} \mathcal{A}_c^{cv}}{\omega_1 + i\gamma - \omega_{vc}} + \frac{\mathcal{A}_b^{cv} \mathcal{A}_c^{vc}}{\omega_1 + i\gamma - \omega_{cv}} + \frac{\mathcal{A}_b^{vc} \mathcal{A}_c^{cv}}{\omega_2 + i\gamma - \omega_{cv}} + \frac{\mathcal{A}_b^{cv} \mathcal{A}_c^{vc}}{\omega_2 + i\gamma - \omega_{vc}} \Bigg).
\end{align}
Under the same assumptions, the linear electric susceptibility (which determines the polarization \cite{SipeSOOR}) is given by 
\begin{align}
    \chi_{ab}^{(1)}(\omega) \approx- \frac{e^2}{\hbar\epsilon_0} \sum_{\mathbf{p}} \left(\frac{\mathcal{A}_a^{vc} \mathcal{A}_b^{cv}}{\omega + i\gamma - \omega_{cv}} - \frac{\mathcal{A}_a^{cv} \mathcal{A}_b^{vc}}{\omega + i\gamma - \omega_{vc}}\right).
\end{align}

We consider a linearly polarized input, \(c=b\), and take the clean limit for the total-frequency damping, \(\Gamma\to0\), in which case the Raman susceptibility becomes
\begin{align}
    \chi_{abb} \approx \frac{e^2 \overline{\Delta}_a}{2\hbar^2} \frac{1}{\omega_\Sigma} \sum_{\mathbf{p}} \Bigg(\frac{2\bar{\omega}_1 \mathcal{A}_b^{vc} \mathcal{A}_b^{cv}}{\bar{\omega}_1^2 - \omega_{cv}^2} + \frac{2\bar{\omega}_2 \mathcal{A}_b^{vc} \mathcal{A}_b^{cv}}{\bar{\omega}_2^2 - \omega_{cv}^2} \Bigg), \qquad
    \chi_{bb}^{(1)}(\omega) \approx -\frac{e^2}{\hbar\epsilon_0} \sum_{\mathbf{p}} \left(\frac{2\omega_{cv} \mathcal{A}_b^{vc} \mathcal{A}_b^{cv}}{\bar{\omega}^2 - \omega_{cv}^2}\right), \label{eq.chi_abb_chi_bb}
\end{align}
also writing the diagonal elements of the linear susceptibility, and where $\bar{\omega}_j=\omega_j+i\gamma$. From here, we can see a correspondence between the terms in the integrands of the Raman and linear susceptibilities. Neglecting the imaginary $i\gamma$ part of $\omega_i$ when appearing in the numerator, and setting $\omega_1=\omega$ and
$\omega_2=\omega_\Sigma-\omega$, we obtain
\begin{align}
\chi_{abb}(\omega,\omega_\Sigma-\omega) \approx -\frac{\epsilon_0\overline{\Delta}_a}{2\hbar\omega_\Sigma} \sum_{\mathbf p}\frac{1}{\omega_{cv}} \Big[\omega\chi^{(1)}_{bb}(\omega,\mathbf p) +(\omega_\Sigma-\omega) \chi^{(1)}_{bb}(\omega_\Sigma-\omega,\mathbf p) \Big].
\end{align}
Expanding the second term in small $\omega_\Sigma$ at fixed $\omega$ and retaining terms up to $O(\omega_\Sigma^0)$ gives
\begin{align}
\chi_{abb}(\omega,\omega_\Sigma-\omega) \approx \frac{\epsilon_0\overline{\Delta}_a}{2\hbar} \sum_{\mathbf p}\frac{1}{\omega_{cv}} \left[\frac{2\omega\Im\chi^{(1)}_{bb}(\omega,\mathbf p)}{i\omega_\Sigma} - \partial_\omega [\omega \chi^{(1)*}_{bb}(\omega,\mathbf p)] \right] +O(\omega_\Sigma),
\end{align}
noting that $\chi^{(1)}_{bb}(-\omega,\mathbf p) =\chi^{(1)*}_{bb}(\omega,\mathbf p)$. Writing the dielectric tensor as $\epsilon_{ab}(\omega)=\epsilon_0[\delta_{ab}+\chi^{(1)}_{ab}(\omega)]$, and for a system with approximately momentum-independent gap $\hbar\omega_{cv}(\mathbf p)\approx\varepsilon_g$, performing the momentum sum yields Eq.~(19) of the main text:
\begin{align}
\chi_{abb}(\omega) \approx \frac{\overline{\Delta}_a}{\varepsilon_g}  \left[\frac{\omega\Im\epsilon_{bb}(\omega)}{i\omega_\Sigma} - \frac{1}{2} \partial_\omega \big\{\omega (\epsilon^{*}_{bb}(\omega)-\epsilon_0)\big\} \right].
\end{align}
The nonsingular term separates into
\begin{align}
\partial_\omega
\big[\omega(\epsilon_{bb}^{*}(\omega)-\epsilon_0)\big]
={}&-i\partial_\omega
\big[\omega\operatorname{Im}\epsilon_{bb}(\omega)\big]
+\omega\partial_\omega\operatorname{Re}\epsilon_{bb}(\omega)
+\operatorname{Re}\epsilon_{bb}(\omega)-\epsilon_0.
\label{eq.SM_dielectric_lim_final}
\end{align}
The imaginary part gives the regular injection correction,
while the real part gives the total off-resonant response.

The result contains the absorptive injection pole and the
dispersive dielectric derivative appearing in the second
line of Eq.~(9) of Ref.~\cite{Stevens2002}.
Near resonance, $\hbar\omega\simeq\varepsilon_g$, the factor
$\overline{\Delta}_a\omega/\varepsilon_g$ approaches
$\overline{\Delta}_a/\hbar$, with the relative signs and
numerical coefficients retained as written above.
The full expression also includes the regular injection
correction and the contribution
$\operatorname{Re}\epsilon_{bb}(\omega)-\epsilon_0$.
Neglecting these additional terms recovers the conventional
form after matching coupling and normalization conventions.
Throughout, we take the clean limit at fixed nonzero
$\omega_\Sigma$ before expanding in the output frequency.

\suppnote{Symmetry constraints on Raman forces in the bilayer Haldane model}
\label{app.sym}
At finite Haldane gap, the equilibrium bilayer model retains
threefold rotational symmetry, $C_3$, while mirror covariance
relates opposite Haldane gaps through
$M_x:(x,\delta_{\rm H})\to(-x,-\delta_{\rm H})$.
We parameterize the Haldane gap as
$\delta_{\rm H}\propto\sin\phi_{\rm H}$, where the dimensionless
hopping phase $\phi_{\rm H}$ is proportional to the magnetic
flux $\Phi_{\triangle}=\int_{\triangle}\mathbf{B}(\mathbf{r})
\cdot d\mathbf{a}$ through a triangular plaquette.
The fictitious magnetic field $\mathbf{B}(\mathbf{r})$
has zero net flux per unit cell.
Like $\bf B$, the gap $\delta_{\rm H}$ is an axial quantity and thus it is odd under  $x$-reflection,
\begin{align}
M_x \delta_{\rm H}  = -\delta_{\rm H}.   
\end{align}
In other words, the positive and negative $\delta_{\rm H}$ stand for  two opposite chiral counterparts which map to each other under $M_x$ reflection. Operating the time-reversal symmetry following Table~II of the main text implies that displacive injection susceptibility $\alpha^{\rm inj}_{abc}$ and impulsive shift susceptibility $\beta^{\rm shi}_{abc}$ are even and odd under $\delta_{\rm H} \to -\delta_{\rm H}$, respectively:
\begin{align}
\alpha^{\rm inj}_{abc}(-\delta_{\rm H})=\alpha^{\rm inj}_{abc}(\delta_{\rm H}),
\qquad
\beta^{\rm shi}_{abc}(-\delta_{\rm H})=-\beta^{\rm shi}_{abc}(\delta_{\rm H}).
\end{align}
Note that the frequency dependence of the Raman responses are assumed implicit for the sake of shorthand notation. The regular injection correction,
$\beta^{\rm reg\text{-}inj}=(i/2)\partial_\omega\alpha^{\rm inj}$,
inherits the parity of $\alpha^{\rm inj}$ under
$\delta_{\rm H}\to-\delta_{\rm H}$.
The off-resonant contributions are outside the scope of
the following symmetry analysis.
\subsection{$M_x$: mirror covariance}
The injection and shift contributions to the Raman force read
\begin{align}
    {\cal F}_a={\cal F}^{\rm inj}_a+{\cal F}^{\rm shi}_a
    =\sum_{bc}\left[
    \frac{\alpha^{\rm inj}_{abc}(\delta_{\rm H})}{i\omega_\Sigma}
    +\beta^{\rm shi}_{abc}(\delta_{\rm H})
    \right]E_bE_c.
\end{align}
Frequency arguments and the frequency integral
(Eq.~(2) of the main text) are suppressed for brevity.
Applying the $x$-reflection, $M_x$, we obtain 
\begin{align}\label{eq:MxFMx}
    (M_x \bm{\mathcal  F})_a = \sum_{bc} \left\{
    \frac{[M_x\cdot\bm\alpha^{\rm inj}(\delta_{\rm H}):M^{-1}_xM^{-1}_x]_{abc}}{i\omega_\Sigma} + [M_x\cdot \bm\beta^{\rm shi}(\delta_{\rm H}):M^{-1}_xM^{-1}_x]_{abc}
    \right\} (M_x {\bf E})_b  (M_x {\bf E})_c 
\end{align}
where 
\begin{align}
     M_x \bm{\mathcal  F} = \begin{bmatrix}-{\cal F}_x\\{\cal F}_y\end{bmatrix},\qquad
    M_x {\bf E} = \begin{bmatrix}-E_x\\E_y\end{bmatrix}
\end{align}
Here we should note that, although the Raman force and electric field vectors transform similarly under the mirror operation $M_x$, they do not transform in the same way under inversion, which also sends $y \to -y$ and $z \to -z$. The Raman force remains invariant under inversion, since $z \to -z$ exchanges the two layers, whereas the electric field is an ordinary vector and is odd under inversion. Moreover, the mirror covariance implies that
\begin{align}
    &[M_x\cdot\bm\alpha^{\rm inj}(\delta_{\rm H}):M^{-1}_xM^{-1}_x]_{abc}=
    \alpha^{\rm inj}_{abc}(M_x\delta_{\rm H})=\alpha^{\rm inj}_{abc}(-\delta_{\rm H})=\alpha^{\rm inj}_{abc}(\delta_{\rm H}),
    \\
    &[M_x\cdot\bm\beta^{\rm shi}(\delta_{\rm H}):M^{-1}_xM^{-1}_x]_{abc}
    =\beta^{\rm shi}_{abc}(M_x\delta_{\rm H})=\beta^{\rm shi}_{abc}(-\delta_{\rm H})=-\beta^{\rm shi}_{abc}(\delta_{\rm H}).
\end{align}
Considering the above symmetry argument, Eq.~\eqref{eq:MxFMx} can be written for each component: 
\begin{align}
    -\mathcal{F}_x =& \left(\frac{\alpha^{\rm inj}_{xxx}}{i\omega_\Sigma}-\beta^{\rm shi}_{xxx}\right) E_x E_x + \left(\frac{\alpha^{\rm inj}_{xyy}}{i\omega_\Sigma}-\beta^{\rm shi}_{xyy}\right) E_y E_y - \left(\frac{\alpha^{\rm inj}_{xxy}}{i\omega_\Sigma}-\beta^{\rm shi}_{xxy}\right) E_x E_y - \left(\frac{\alpha^{\rm inj}_{xyx}}{i\omega_\Sigma}-\beta^{\rm shi}_{xyx}\right) E_y E_x \\
    \mathcal{F}_y =& \left(\frac{\alpha^{\rm inj}_{yyy}}{i\omega_\Sigma}-\beta^{\rm shi}_{yyy}\right) 
    E_y E_y + \left(\frac{\alpha^{\rm inj}_{yxx}}{i\omega_\Sigma}-\beta^{\rm shi}_{yxx}\right) E_x E_x - \left(\frac{\alpha^{\rm inj}_{yyx}}{i\omega_\Sigma}-\beta^{\rm shi}_{yyx}\right) E_y E_x - \left(\frac{\alpha^{\rm inj}_{yxy}}{i\omega_\Sigma}-\beta^{\rm shi}_{yxy}\right) E_x E_y.
\end{align}
Equating expressions for each of $\mathcal{F}_x$ and $\mathcal{F}_y$ before and after $M_x$ reflection, we obtain
\begin{align} \alpha^{\rm inj}_{xxx}=\alpha^{\rm inj}_{xyy}=\alpha^{\rm inj}_{yyx}=\alpha^{\rm inj}_{yxy}=0; \hspace{2cm} \beta^{\rm shi}_{yyy}=\beta^{\rm shi}_{yxx}=\beta^{\rm shi}_{xxy}=\beta^{\rm shi}_{xyx}=0,
\end{align}
with remaining elements surviving.
\subsection{$C_3$: rotational symmetry}
Moving onto three-fold rotation, we consider components of $\alpha^{\rm inj}_{abc}$ and $\beta^{\rm shi}_{abc}$ separately. The former of these, considering the results from above, gives force components of
\begin{align}\label{eq:Finj_x}
   \mathcal{F}^{\rm inj}_x =& \frac{1}{i\omega_\Sigma} \left\{\alpha^{\rm inj}_{xxy} E_x E_y + \alpha^{\rm inj}_{xyx} E_y E_x\right\} \\\label{eq:Finj_y}
   \mathcal{F}^{\rm inj}_y =&\frac{1}{i\omega_\Sigma} \left\{\alpha^{\rm inj}_{yyy} E_y E_y + \alpha^{\rm inj}_{yxx} E_x E_x\right\}.
\end{align}
Acting once with the $C_3$ operator, vectors transform as 
\begin{align}
 C_3 
 \begin{bmatrix}
   \hat{x} \\[5pt] \hat{y}  
 \end{bmatrix}
= 
 \begin{bmatrix}
     -\frac{1}{2}\hat{x} + \frac{\sqrt{3}}{2}\hat{y}
     \\[5pt]
     -\frac{\sqrt{3}}{2}\hat{x} - \frac{1}{2}\hat{y}
 \end{bmatrix}, \quad
 C_3 
 \begin{bmatrix}
   {\cal F}_x \\[5pt] {\cal F}_{y} 
 \end{bmatrix}
=
 \begin{bmatrix}
     -\frac{1}{2}{\cal F}_x + \frac{\sqrt{3}}{2}{\cal F}_y
     \\[5pt]
     -\frac{\sqrt{3}}{2}{\cal F}_x - \frac{1}{2}{\cal F}_y
 \end{bmatrix}, \quad
 C_3 
 \begin{bmatrix}
   E_x \\[5pt] E_{y} 
 \end{bmatrix}
= 
 \begin{bmatrix}
     -\frac{1}{2}E_x + \frac{\sqrt{3}}{2}E_y
     \\[5pt]
     -\frac{\sqrt{3}}{2}E_x - \frac{1}{2}E_y
 \end{bmatrix},      
\end{align}
and pseudovectors transform as regular vectors, as rotation is a proper symmetry operation, meaning $\delta_{\rm H}$ is invariant under $C_3$. Moreover, the Raman-force susceptibility tensors are invariant due to the three-fold rotational symmetry:  
\begin{align}
    &[C_3\cdot\bm\alpha^{\rm inj}(\delta_{\rm H}):C^{-1}_3C^{-1}_3]_{abc}=
    \alpha^{\rm inj}_{abc}(\delta_{\rm H}),
    \\
    &[C_3\cdot\bm\beta^{\rm shi}(\delta_{\rm H}):C^{-1}_3C^{-1}_3]_{abc}
    = \beta^{\rm shi}_{abc}(\delta_{\rm H}).
\end{align}
Looking just at $\mathcal{F}^{\rm inj}_y$, rotation yields
\begin{align}
     -\frac{\sqrt{3}}{2}\mathcal{F}^{\rm inj}_x - \frac{1}{2}\mathcal{F}^{\rm inj}_y =& \frac{3\alpha^{\rm inj}_{yyy} + \alpha^{\rm inj}_{yxx}}{4(i\omega_\Sigma)} E_x E_x + \frac{\sqrt{3}(\alpha^{\rm inj}_{yyy} - \alpha^{\rm inj}_{yxx})}{4(i\omega_\Sigma)} E_x E_y + \frac{\sqrt{3}(\alpha^{\rm inj}_{yyy} - \alpha^{\rm inj}_{yxx})}{4(i\omega_\Sigma)} E_y E_x + \frac{\alpha^{\rm inj}_{yyy} + 3\alpha^{\rm inj}_{yxx}}{4 (i\omega_\Sigma)} E_y E_y.
\end{align}
Substituting in $\mathcal{F}^{\rm inj}_x$ and $\mathcal{F}^{\rm inj}_y$ from Eqs.~\eqref{eq:Finj_x} and \eqref{eq:Finj_y}, and equating terms with the same $E^bE^c$, we obtain simultaneous equations with solutions
\begin{align}\label{eq:alpha_sym}
    -\alpha^{\rm inj}_{yyy} = \alpha^{\rm inj}_{yxx} = \alpha^{\rm inj}_{xxy} = \alpha^{\rm inj}_{xyx} \neq 0.
\end{align}
Looking at $\beta^{\rm shi}_{abc}$ now, our remaining terms give forces of
\begin{align}
\label{eq:Fshi_x}
    \mathcal{F}^{\rm shi}_x =&\beta^{\rm shi}_{xxx} E_x E_x + \beta^{\rm shi}_{xyy} E_y E_y \\ \label{eq:Fshi_y}
    \mathcal{F}^{\rm shi}_y =&\beta^{\rm shi}_{yxy} E_x E_y + \beta^{\rm shi}_{yyx} E_y E_x.
\end{align}
Acting the $\mathcal{C}_3$ operator on $\mathcal{F}^{\rm shi}_x$, we obtain
\begin{align}
    -\frac{1}{2} \mathcal{F}^{\rm shi}_x + \frac{\sqrt{3}}{2} \mathcal{F}^{\rm shi}_y =& \frac{\beta^{\rm shi}_{xxx} + 3\beta^{\rm shi}_{xyy}}{4} E_x E_x - \frac{\sqrt{3}(\beta^{\rm shi}_{xxx} - \beta^{\rm shi}_{xyy})}{4} E_x E_y - \frac{\sqrt{3}(\beta^{\rm shi}_{xxx} - \beta^{\rm shi}_{xyy})}{4} E_y E_x + \frac{3\beta^{\rm shi}_{xxx} + \beta^{\rm shi}_{xyy}}{4} E_y E_y.
\end{align}
Substituting in $\mathcal{F}^{\rm shi}_x$ and $\mathcal{F}^{\rm shi}_y$ from Eqs.~\eqref{eq:Fshi_x} and \eqref{eq:Fshi_y}, then solving the resulting simultaneous equations, we obtain
\begin{align}\label{eq:beta_sym}
    -\beta^{\rm shi}_{xxx} = \beta^{\rm shi}_{xyy} = \beta^{\rm shi}_{yxy} = \beta^{\rm shi}_{yyx} \neq 0.
\end{align}
The final symmetry results, summarized in Eqs.~\eqref{eq:alpha_sym} and \eqref{eq:beta_sym} and identifying the non-vanishing tensor elements of the injection and shift responses, are collected in Eq.~(21) of the main text and are in exact agreement with our analytical and numerical calculations. 

\suppnote{Temperature dependence of linearly-polarized Raman forces }\label{app.Temp_LP}
\begin{figure}[b]
    \centering  \includegraphics[width=1\linewidth]{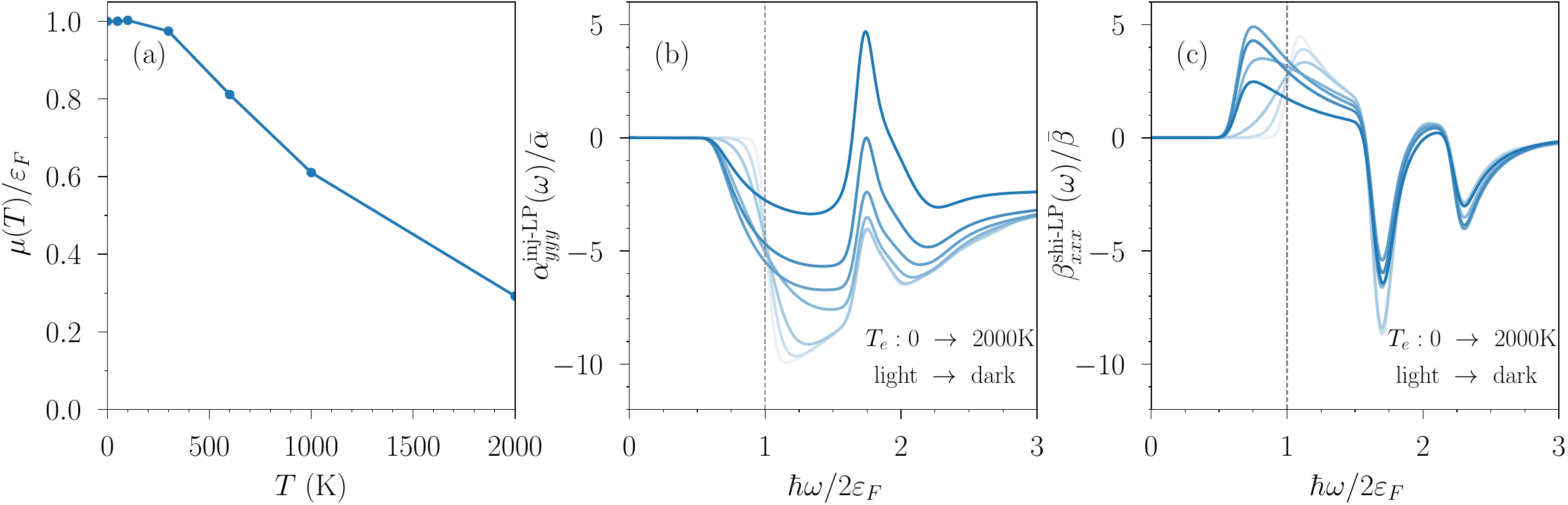}
    \caption{(a) Chemical potential, $\mu$ as a function of temperature, $T$ in the bilayer Haldane model relative to the Fermi energy, $\epsilon_F = 0.1$ eV. (b, c) Spectra for LP (b) injection (c) shift shear Raman-force susceptibility  at different electronic temperatures with $\phi=\pi/80$ and $\delta=0$, where we account for both population spreading and changing of chemical potential due to increasing temperature. Note that $\bar\alpha = 100\, \mathrm{nN}/(\mathrm{V}^2 \, \mathrm{ps})$ and $\bar\beta = \mathrm{nN}/\mathrm{V}^2$. The same units and parameter values are used for the LP Raman-force responses presented in Fig.~4 of the main text. From light to dark, the curves correspond to
$T=0$, $50$, $100$, $300$, $600$, $1000$, and $2000\,\mathrm{K}$.
    }
    \label{fig:SM_VaryingT}
\end{figure}
Here, we briefly provide and discuss the numerical results for the resonant Raman shear force susceptibilities due to LP light at various temperatures. We first compute the chemical potential at different temperatures and then obtain the numerical result for the Raman-force susceptibilities. 
The chemical potential is obtained by numerically solving \mbox{$n=n_e(T)-n_h(T) = n_e(0)-n_h(0)$} for $\mu$, where $n_{e(h)}(T)$ is the density of electrons (holes) at temperature, $T$. We calculate the differences in density using 
\begin{align}\label{eq:mu}
    n=n_e(T)-n_h(T) = \int_{\rm BZ} \frac{d^2{\bf p}}{(2\pi)^2}
    \left \{ \sum_{m\in \text{conduction}} f_m-\sum_{m\in \text{valence} } (1-f_m)
    \right\} =\int_{\rm BZ} \frac{d^2{\bf p}}{(2\pi)^2} \left \{ \sum_{m\in \text{all bands}} f_m- N_{\rm val}\right\} ,
\end{align}
where $N_{\rm val}=2$ is the number of valence bands, zero-temperature chemical potential is the Fermi energy $\mu(0) = \varepsilon_F$ which is set by the carrier density $n$, and Fermi distribution function of band $m$ reads
\begin{align}\label{eq:fm}
    f_m = \frac{1}{1+\exp\left \{\frac{\varepsilon_m({\bf p})-\mu(T)}{k_B T}\right\}}.
\end{align}
By solving Eq.~\eqref{eq:mu}, for a given Fermi energy $\varepsilon_F$ we obtain the chemical potential at different temperature.  
The dotted points in \suppfigref{fig:SM_VaryingT}a indicate the values of chemical potential we calculated at various temperatures. As seen it stays nearly constant until the temperature exceeds 100 K, when it starts to decrease quickly, with this descent slowing down as one increases temperature further. 
We use the chemical potential values shown in panel (a) and the finite-temperature Fermi distribution, Eq.~\eqref{eq:fm}, to compute the injection and shift Raman-force responses at various temperatures, shown in panels (b) and (c), respectively. 

Beyond simple thermal broadening, raising the electron temperature depletes carriers in the UVB and populates the LCB. This redistribution is partly compensated by the downward shift of the chemical potential with increasing temperature. Together, these effects in the electronic distribution account for most of the temperature dependence observed in the response functions.
At low temperature, the absorption edge at $2\epsilon_F$, arising from UVB–LCB transitions, gradually flattens and, above $300$K, shifts entirely to lower energies. This behaviour is driven by the reduced LCB occupation near the Dirac point, caused both by thermal spreading to higher energies and by the lowering of the chemical potential. As a result, transitions near the band (Haldane) gap become increasingly allowed and are no longer Pauli blocked.

Remarkably, the increased LCB population enhances LCB–UCB transitions, which appear as a sharp peak opposite in sign to the absorption step in the injection response. In contrast, the corresponding contribution to the shift response is reduced, since this response is maximal near the Dirac point and the thermally broadened LCB population shifts resonances away from it. Interestingly, beyond $600$K the shift-response peak grows again, due to the competition between (i) the detrimental effect of population spreading in the LCB and (ii) the lowering chemical potential, which creates additional vacancies in the UVB and enables previously forbidden LVB–UVB transitions.
\bibliography{refs}